\documentclass[sigconf]{acmart}
\AddToHook{package/amssymb/before}{}
\usepackage{popets}

\setcopyright{popets}
\copyrightyear{2027}

\acmYear{2027}
\acmVolume{2027}
\acmNumber{1}
\acmDOI{XXXXXXX.XXXXXXX}
\acmISBN{}
\acmConference{Proceedings on Privacy Enhancing Technologies}
\usepackage[dvipsnames]{xcolor}
\usepackage{graphicx}
\usepackage{amssymb}
\usepackage{booktabs,multirow}
\usepackage{enumitem}
\usepackage{colortbl}
\newcommand{\sigA}[1]{\cellcolor{teal!42}#1}
\newcommand{\sigB}[1]{\cellcolor{teal!22}#1}
\newcommand{\sigC}[1]{\cellcolor{teal!9}#1}
\newcommand{\sigkeyA}{\colorbox{teal!42}{\,$p{<}0.001$\,}}
\newcommand{\sigkeyB}{\colorbox{teal!22}{\,$p{<}0.01$\,}}
\newcommand{\sigkeyC}{\colorbox{teal!9}{\,$p{<}0.05$\,}}
  \definecolor{linrecon}{gray}{0.88}
\definecolor{takeawayframe}{HTML}{3A4E85}
\definecolor{takeawaytitle}{HTML}{2C3D6E}
\definecolor{takeawayfill}{HTML}{EEF9FE}
\usepackage{tcolorbox, fontawesome5}
\tcbuselibrary{skins}
\usepackage{orcidlink}
\usepackage{algorithm}
\usepackage[noend]{algpseudocode}
\usepackage{placeins}
\algrenewcommand\algorithmicrequire{\textbf{Input:}}
\algrenewcommand\algorithmicensure{\textbf{Output:}}

\begin{document}

\title[Reconstruction Attaxonomy]{SoK: Reconstruction Attacks on Synthetic Tabular Data \\ (Insights from Winning the NIST CRC)}

\author{%
  Steven Golob\orcidlink{0009-0008-4442-2762}$^{1}$,\enspace
  Sikha Pentyala\orcidlink{0000-0001-7486-6016}$^{1}$,\enspace
  Martine De Cock\orcidlink{0000-0001-7917-0771}$^{1,2}$%
}
\affiliation{%
  \institution{%
    $^{1}$School of Engineering and Technology, University of Washington Tacoma, Tacoma, WA, USA\\[2pt]
    $^{2}$Department of Mathematics, Computer Science, and Statistics, Ghent University, Gent, Belgium%
  }
  \city{}
  \state{}
  \country{}
}
\email{golobs@uw.edu}

\renewcommand{\shortauthors}{Golob et al.}

\begin{abstract}
Synthetic data is increasingly promoted as a privacy-preserving substitute for releasing sensitive tabular records, yet its central adversarial threat (\emph{reconstruction}, the recovery of an individual's hidden attribute values from a synthetic release and a handful of known quasi-identifiers) has been studied only in scattered, hard-to-compare settings. We systematize reconstruction (equivalently, attribute inference) attacks on de-identified and synthetic tabular data. We contribute a taxonomy that organizes attacks by the structure they exploit; a broad, controlled empirical evaluation, pitting fourteen attacks against thirteen synthetic data generation (SDG) methods across five benchmark datasets; and a set of new attacks that fill gaps in the taxonomy, one of which (CoBP-RA) is the strongest attack we measure. We also introduce a methodology for interpreting what attack success means: a memorization test that distinguishes reconstruction of the population distribution from memorization of training records, and a reduction that places reconstruction and membership inference on a single comparable scale. Our findings: the choice of SDG method governs risk far more than the choice of attack; differential privacy reduces reconstruction steadily up to $\varepsilon{\approx}10$, above which reconstruction risk levels off for all six DP mechanisms we sweep, bounded by what each synthesizer can represent rather than by its noise; de-identification methods are the most exposed, with diffusion the only synthesizer close behind; and most reconstruction reflects distributional structure rather than memorization, concentrating individual risk on atypical records. Our attacks and infrastructure are externally validated by our first-place finish in the 2025 \textit{National Institute of Standards and Technology} (NIST) Collaborative Research Cycle.
\end{abstract}

\keywords{privacy, synthetic data, de-identification, reconstruction attack, attribute inference}

\maketitle

\section{Introduction}\label{sec:intro}

Synthetic data promises to resolve a central tension in data sharing: produce a dataset that reproduces the statistical structure of sensitive records,
and publish it for downstream users to analyze without exposing any real individual~\cite{hu2023sok}. The promise has drawn substantial institutional investment~\cite{jordon2022synthetic}. National statistical agencies~\cite{benedetto2018creation}, hospital systems~\cite{walonoski2018synthea}, and the financial sector~\cite{assefa2020generating} are evaluating synthetic data release as a successor to traditional disclosure-limitation techniques such as $k$-anonymization~\cite{sweeney2002k} and record suppression~\cite{samarati2001protecting}, and as a path to compliance under frameworks like HIPAA Expert Determination~\cite{hhs2012deid}.

\begin{figure}
    \centering
    \includegraphics[width=\columnwidth]{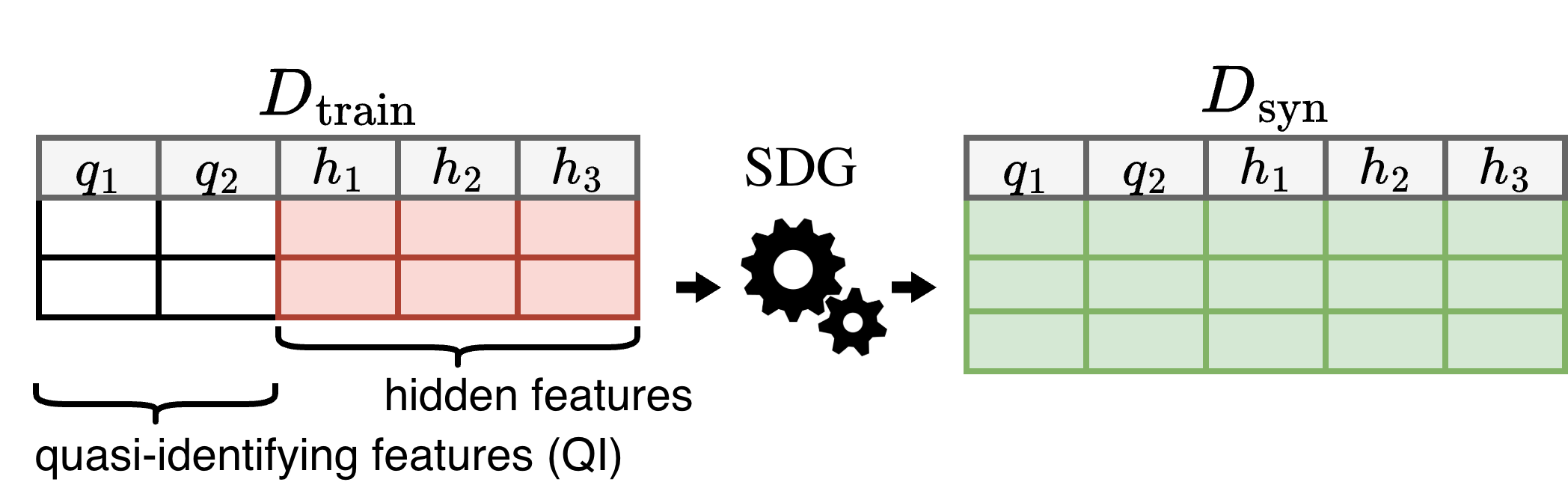}
    \caption{The reconstruction threat model (synthetic-data branch of Figure~\ref{fig:taxonomy}; the de-identified branch is identical, with \textsc{sdg} replaced by de-identification). The \textsc{sdg} process turns the private training table into the synthetic release (\emph{green} rows). Every training record is a target: the adversary sees the records' quasi-identifiers (\emph{white} columns) plus the synthetic data, and fills in their hidden attributes (\emph{red} columns).}
    \Description{A schematic of the reconstruction threat model. On the left, the private training table $D_\text{train}$ is drawn as a grid whose first two columns, headed $q_1$ and $q_2$, are white and bracketed below as ``quasi-identifying features (QI)'', while its remaining three columns, headed $h_1$ to $h_3$, are shaded red and bracketed as ``hidden features''. An arrow leads rightward through a gear icon labelled SDG to a second table, $D_\text{syn}$, carrying the same five column headings but with every cell shaded green. The adversary is assumed to hold the white QI cells of a target row together with the whole green synthetic table, and must infer that row's red cells.}
    \label{fig:threat_model}
\end{figure}

Whether that promise of privacy survives adversarial scrutiny is an empirical question, and the sharpest test of it is the \textbf{reconstruction attack}. Given a released synthetic dataset and a partial view of a target individual, the adversary attempts to recover that individual's remaining, hidden attributes. The partial view is a set of \emph{quasi-identifying} features, or \textbf{quasi-identifiers} (QI): attributes such as age, sex, or ZIP code that an adversary can plausibly obtain from public records, voter rolls, or social-media profiles~\cite{sweeney2000simple,narayanan2008robust} (Figure~\ref{fig:threat_model}: the adversary knows the \emph{white} quasi-identifier columns and must fill in the \emph{red} hidden columns). Reconstruction thus completes a partial record, predicting sensitive values such as income, disease status, or household wealth.\footnote{Prior work sometimes reserves ``attribute inference'' for predicting a single withheld attribute~\cite{yeom2018privacy}; under our threat model (Section~\ref{sec:threat}) it is the same problem as reconstructing all hidden attributes, and we use the terms interchangeably.} It is more consequential than \emph{membership inference}~\cite{shokri2017membership} (the de facto privacy audit for synthetic data~\cite{stadler2022synthetic,allard2023snake,golob_privacy_2025}), which asks only \emph{whether} a record was used in training  and assumes the adversary already knows \emph{what} it contains.

\begin{figure*}[t]
    \centering
    \includegraphics[width=\textwidth]{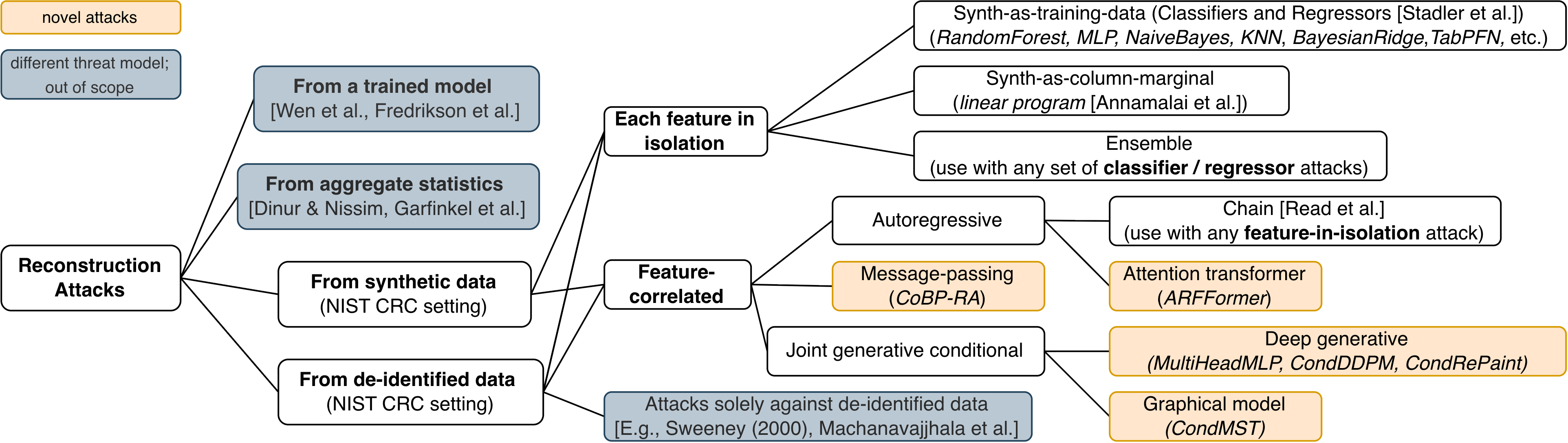}
    \caption{Taxonomy of reconstruction attacks. The two lower branches (from synthetic data and from de-identified microdata) share the NIST CRC quasi-identifier threat model, so we merge their attack subtrees; the de-identified branch additionally admits attacks on \emph{retained real records} (Section~\ref{sec:related}). Within the merged subtree, \emph{in-isolation} attacks fit one model per hidden feature, while \emph{feature-correlated} attacks exploit dependencies among the hidden features, the category our new attacks target (Section~\ref{sec:attack_taxonomy}). The top two branches assume a different release artifact and are out of scope.}
    \Description{A tree diagram drawn left to right. The root, ``Reconstruction Attacks'', branches into four release settings. Two are shaded grey and marked by the legend as a different threat model and out of scope: ``From trained model'' (Fredrikson et al., Wen et al.) and ``From aggregate statistics'' (Dinur \& Nissim, Garfinkel et al.). Two are in scope and both labelled ``NIST CRC setting'': ``From synthetic data'' and ``From de-identified data''; a further grey out-of-scope leaf, ``Attacks solely against de-identified data'' (Sweeney, Machanavajjhala et al.), hangs off the latter. The two in-scope branches converge on one shared subtree that splits in two. ``Each feature in isolation'' leads to three leaves: synth-as-training-data classifiers and regressors (RandomForest, MLP, NaiveBayes, KNN, BayesianRidge, TabPFN), synth-as-column-marginal (linear program, Annamalai et al.), and Ensemble. ``Feature-correlated'' leads to ``Autoregressive'', which yields Chain (Read et al.) and an attention transformer (ARFFormer); to ``Message-passing'' (CoBP-RA); and to ``Joint generative conditional'', which splits into ``Deep generative'' (MultiHeadMLP, CondDDPM, CondRePaint) and ``Graphical model'' (CondMST). Nodes shaded orange, identified by the legend as ``novel attacks'', are ARFFormer, CoBP-RA, the deep-generative leaf, and CondMST.}
    \label{fig:taxonomy}
\end{figure*}

Reconstruction attacks differ in the \emph{artifact} the adversary holds (Figure~\ref{fig:taxonomy}): a trained model (\emph{model inversion}~\cite{fredrikson2015model}), released aggregate statistics (\emph{reconstruction from tabulations}~\cite{dinur2003revealing}), or a published de-identified or synthetic table. We study the last: the setting of the NIST Collaborative Research Cycle,\footnote{\label{fn:nistcrc}\url{https://pages.nist.gov/privacy_collaborative_research_cycle/}} where both pose one threat model and attack surface (Section~\ref{sec:threat}).

\paragraph{Why systematize?}
Despite its centrality to disclosure risk, reconstruction from synthetic tabular data has, to our knowledge, not yet been systematized. The literature is a patchwork: studies fix a single attack method \citep{stadler2022synthetic,annamalai2024linear}
or a single privacy metric~\cite{houssiau2022tapas,giomi2023unified}, and report numbers that cannot be placed side by side. Basic questions therefore remain open: which attacks work against which SDG methods, whether the adversary or the generator drives risk, whether a successful attack has memorized a record or learned the population, and how reconstruction compares with membership inference. This paper answers them through the following \textit{main contributions:} 

\begin{itemize}[leftmargin=*,noitemsep,topsep=3pt]
  \item \textbf{A taxonomy of reconstruction attacks on de-identified and synthetic tabular data} (Section~\ref{sec:taxonomy}, Figure~\ref{fig:taxonomy}). It organizes attacks by the structure they exploit, separating those that predict each hidden feature \emph{in isolation}~\cite{stadler2022synthetic,annamalai2024linear} from those that exploit \emph{correlations among hidden features} (autoregressive, message-passing, and joint-generative).
  \item \textbf{A broad, controlled empirical evaluation} (Section~\ref{sec:conditions}): fourteen attacks against thirteen SDG methods (the seven from the NIST CRC, a GAN- and a diffusion-based generator, and four more differentially private (DP) synthesizers) across a wide range of privacy budgets and five datasets, with quasi-identifier sets varied in composition and size (3 to 16 features). This substantially extends the NIST CRC, which fixed a single dataset and QI configuration.
  \item \textbf{New attacks that populate sparse branches} of the taxonomy, offered as an exploration of the design space rather than an exhaustive enumeration: \textbf{CoBP-RA}, which runs belief propagation of per-feature random-forest posteriors over a learned graphical model and is the single strongest attack in our study; and a family that recasts reconstruction as QI-conditioned synthesis: \textbf{CondMST}~\cite{mckenna2019graphical} and the diffusion-based \textbf{CondDDPM}~\cite{kotelnikov2023tabddpm} and \textbf{CondRePaint}, the latter porting image-inpainting \emph{RePaint}~\cite{lugmayr2022repaint} to tables. CoBP-RA improves on the strongest prior attack across every categorical dataset we test; the graphical-model and diffusion attacks are competitive but do not consistently beat well-tuned per-feature classifiers (Section~\ref{sec:taxonomy}).
  \item \textbf{A methodology for interpreting attack success} (Section~\ref{sec:interrogating}). Measured only against training targets (as in the literature and the NIST CRC), a high score is ambiguous: it may reflect memorization of training records (a breach) or merely a well-modeled population distribution. Our memorization test separates the two by running each attack against training and held-out targets from the same population (an attribute analogue of the train-versus-holdout gap behind membership advantage~\cite{yeom2018privacy}).
  We also propose a reduction that casts a reconstruction attack as a membership inference attack (\emph{RA-as-MIA}), placing both paradigms on a single axis.
  \item \textbf{Policy-relevant guidance} for data holders and auditors, synthesizing the above into recommendations (Sections~\ref{sec:conditions} and~\ref{sec:policy}).
\end{itemize}

\paragraph{External validation.} Our attacks and infrastructure placed first in the 2025 NIST Privacy Collaborative Research Cycle (CRC)\textsuperscript{\ref{fn:nistcrc}}, a structured red-team/blue-team program whose privacy-measurement results inform policy at federal statistical agencies; the US Census Bureau, for instance, adopted differential privacy for its 2020 census data releases~\cite{abowd2018us}.

\paragraph{Main findings:}
\begin{itemize}[leftmargin=*,noitemsep,topsep=3pt]
  \item \textbf{The generator, not the adversary, is the dominant lever on risk.} Reconstruction accuracy varies far more with the SDG method than with the attack; a more protective generator lowered risk more than any attack refinement raised it (Section~\ref{sec:attack_taxonomy}).
  \item \textbf{Differential privacy works, but the budget stops mattering early.} Protection accrues up to $\varepsilon{\approx}10$; beyond it, two more orders of magnitude of budget move reconstruction by about a point for most of the six DP synthesizers we sweep, and by at most ${\approx}3.5$~pp for any, because they are limited not by noise but by the structure they can represent (Section~\ref{sec:whichgenerator}).
  \item \textbf{De-identification is the most exposed family; among synthesizers, only diffusion comes close.} Cell suppression and rank swapping release real records that attacks read back, and TabDDPM is the one synthesizer that likewise memorizes individuals (Sections~\ref{sec:whichgenerator} and~\ref{sec:memorization}). 
  \item \textbf{Adversary knowledge quality beats quantity.} The \emph{composition} of the quasi-identifying features shapes risk more than its size (Section~\ref{sec:whichadversary}).
  \item \textbf{Most reconstruction is distributional, not memorization.} Our memorization test (Section~\ref{sec:memorization}) separates the two: attacks succeed largely by reproducing population structure the release preserves rather than memorizing individuals. Two releases that reconstruct non-members equally well can differ eightfold in how much they memorize, so the test complements $R_{adv}$ rather than replacing it.
  \item \textbf{Reconstruction and membership inference agree in aggregate but diverge per record.} The two rank SDG methods almost identically, yet succeed on substantially different individuals (about a third of records give discordant signals), so neither subsumes the other and a thorough audit needs both (Section~\ref{sec:mia}).
  \item \textbf{Averages hide who is at risk.} Diffusion concentrates risk on atypical, demographically rare records, cell suppression on the majority, and DP spreads it evenly (Section~\ref{sec:disparate}).
\end{itemize}

\subsection{Formal Problem Description}
\label{sec:formulation}

Let $D_{\text{train}} = \{x_1, \ldots, x_n\}$ be a tabular dataset over $n$ individuals, each record a vector of $p$ features, each feature either categorical or continuous. We write $\mathcal{F}$ for the set of features, $|\mathcal{F}| = p$. A \textbf{synthetic data generator} (SDG) $\mathcal{G}$ takes $D_{\text{train}}$ as input and produces a release $D_{\text{syn}} = \mathcal{G}(D_{\text{train}})$ (we write $D_{\text{syn}}$ for both synthetic and de-identified releases, since we treat them identically).

We partition features into two disjoint sets: the quasi-identifiers $Q \subset \mathcal{F}$ comprising features assumed observable by the adversary from auxiliary sources~\cite{sweeney2000simple,narayanan2008robust} (the \emph{white} columns of Figure~\ref{fig:threat_model}), and the \textbf{hidden features} $H = \mathcal{F} \setminus Q$ that the adversary wishes to infer (the \emph{red} columns). A \textbf{reconstruction attack} is a function
\begin{equation}\label{eqn:attack}
  f:\bigl(D_{\text{syn}},\; x^Q_i\bigr) \;\longmapsto\; \hat{x}^H_i,
\end{equation}
mapping the synthetic release and a target's observed quasi-ident\-ifi\-ers to a prediction of the target's hidden attributes.

\textbf{Scoring: rarity-weighted reconstruction advantage ($R_{adv}$).}
Na\"{i}ve accuracy rewards predicting the mode, a trivially achievable but uninformative baseline. We use a \emph{rarity-weighted} score: a correct prediction of rare value $v$ earns weight $n/\text{count}(v, D_{\text{train}})$ (where $\text{count}(v, D_{\text{train}})$ is the number of training records in $D_{\text{train}}$ taking value $v$ on feature $h$), reflecting the greater harm of inferring information the adversary could not have guessed. The score for inferring hidden feature $h$ is:
\begin{equation}\label{eqn:scoring}
  R_{adv}(f, h) \;=\;
  \frac{\sum_{i}\;
        \mathbf{1}[\hat{x}^h_i = x^h_i]\cdot
        n/{\text{count}(x^h_i,\,D_{\text{train}})}}
       {\sum_{i}\;
        n/{\text{count}(x^h_i,\,D_{\text{train}})}}
  \times 100.
\end{equation}

Under this measure, a mode-baseline attack scores exactly $100/C$\% regardless of the distribution's shape, where $C$ is the number of distinct values; a perfect attack scores $100$\%. This was also the official scoring metric for the NIST Privacy CRC competition. We report $R_{adv}$ on this 0--100 scale and write it with a `\%' sign as shorthand, though it is a weighted accuracy rather than a share of records. Headline scores are the mean $R_{adv}$ across all hidden features (per-feature results are in Sections~\ref{sec:whatdata} and~\ref{sec:disparate}), and each table caption states what else its cells average over, depending on the dimension under study (training samples, attacks, generators, or QI sets).

\subsection{Differential Privacy}
\label{sec:dp}

Differential privacy (DP)~\cite{dwork2006calibrating,dwork2014algorithmic} is the principal formal privacy guarantee studied here. A randomized mechanism $\mathcal{M}$ is \textbf{$\varepsilon$-dif\-fe\-ren\-tial\-ly private} if, for neighboring datasets $D, D'$ (i.e., datasets that differ in one record) and all subsets $S$ of the range of $\mathcal{M}$: $\Pr[\mathcal{M}(D) \in S] \le e^\varepsilon \Pr[\mathcal{M}(D') \in S]$. Smaller $\varepsilon$ provides stronger privacy; an $\varepsilon$-DP SDG guarantees that no individual's presence in $D_{\text{train}}$ shifts the distribution of $D_{\text{syn}}$ by more than $e^\varepsilon$. The bound is worst-case and does not directly specify empirical reconstruction accuracy; we measure $R_{adv}$ across $\varepsilon \in \{10^{-1}, \dots, 10^3\}$ for six DP SDG methods, MST \cite{mckenna2021winning}, AIM \cite{mckenna2022aim}, PrivBayes~\cite{zhang2017privbayes}, PrivSyn~\cite{zhang2021privsyn}, MWEM-PGM~\cite{hardt2012simple,mckenna2019graphical}, and PrivateGSD~\cite{liu2023generating}, in Section~\ref{sec:conditions}.

\subsection{Threat Model}
\label{sec:threat}

\textbf{Setting and adversary.} 
The adversary targets a training individual $x_i$ and observes \emph{only} $D_{\text{syn}}$ and $x^Q_i$: never the original data, the generator's parameters, seed, or weights (the attack is fully black-box), nor any other individual's record. The QI set $Q$ sets how much the adversary knows; we vary it in size and composition (Section~\ref{subsec:qi}), so a larger or more informative $Q$ is a stronger adversary.

\textbf{Scope.}
We study the lower two branches of Figure~\ref{fig:taxonomy}, treated identically. Out of scope are model inversion and reconstruction-from-tabulations (which assume a different release artifact: a trained model or aggregate statistics), along with attacks that jointly exploit multiple target records, model extraction, and white-box attacks on SDG internals. Ours is the black-box, single-record setting of published releases and compliance audits.

\paragraph{Paper organization.}
Section~\ref{sec:taxonomy} presents the attack taxonomy (our ``attaxonomy'', borrowing the wordplay of~\cite{cummings2024attaxonomy}) and our main cross-method benchmark. Section~\ref{sec:conditions} examines how the SDG method and its privacy budget, the data domain, the dataset size, and quasi-identifier design shape reconstruction risk. Section~\ref{sec:interrogating} interrogates what attack success \emph{means}: memorization versus distributional inference, reconstruction versus membership inference, and disparate impact. Section~\ref{sec:policy} translates these findings into auditing and policy guidance. The appendix adds preliminaries on the datasets, quasi-identifiers, and generators; a notation table (Table~\ref{tab:notation}); pseudocode for the new attacks; the ensembling and chaining oracle analyses; and full results for the remaining datasets.

\section{
A Taxonomy of Reconstruction Attacks}
\label{sec:taxonomy}
\subsection{The Reconstruction Attack Landscape}
\label{sec:related}
We survey the whole family of reconstruction attacks, distinguished by the \emph{artifact} the adversary holds (Figure~\ref{fig:taxonomy}, top level), before narrowing to the two branches we study; the same map shows where attacks are missing, gaps the new attacks of Section~\ref{sec:attack_taxonomy} are built to fill.

\paragraph{Model inversion (out of scope).}
The first branch attacks a \emph{trained model}: model inversion~\cite{fredrikson2015model} exploits a classifier's confidence scores to recover sensitive training attributes, and \citet{wen2025sok} systematize this branch for vision with a taxonomy, evaluation metrics, and a benchmark. 
\citet{zhao2021feasibility} argue that exact attribute inference is largely infeasible (only an approximate, imputation-like recovery succeeds), and \citet{jayaraman2022attribute} reduce many attribute-inference attacks to ordinary imputation; leverage comes from feature correlations, not a membership signal.
The federated-learning \emph{attribute reconstruction attack} (ARA), despite its name, likewise needs model internals: it recovers attributes from shared gradients~\cite{lyu2021novel,chen2024practical}. We borrow the imputation lens of these attacks, but our adversary never observes a model or its gradients.

\paragraph{Reconstruction from aggregate statistics (out of scope).}
A second branch attacks \emph{released tabulations}. \citet{dinur2003revealing} proved that answering too many counting queries too accurately permits near-complete reconstruction, the result that motivated differential privacy~\cite{dwork2014algorithmic,dwork2017exposed}. \citet{garfinkel2019understanding} traced this threat to U.S.\ Census releases, informing the Bureau's adoption of DP for 2020; \citet{dick2023confidence} reconstructed and confidence-ranked individual census records; \citet{cohen2024data} characterized when reconstruction from aggregates is feasible; \citet{dwork2017exposed} survey the landscape, and \citet{muralidhar2023database} caution that practice is often harder than the theory implies. Our results are the generative-setting counterpart: reconstruction is consistently above chance, but its ceiling is set by the structure the release preserves (echoing \citeauthor{cohen2024data}'s feasibility boundary), not by adversary effort.

\paragraph{Reconstruction from synthetic data (in scope).}
This branch attacks \emph{synthetic data}: records drawn from a generative model fit to the original. \citet{stadler2022synthetic} framed attribute inference as a game in which a machine learning (ML) adversary trained on the synthetic release predicts withheld attributes of real records, concluding that synthetic data rarely provides strong anonymization. 
\citet{houssiau2022tapas} (TAPAS) and \citet{giomi2023unified} (Anonymeter) assemble attack toolkits for auditing synthetic data (TAPAS spanning membership and attribute inference, Anonymeter the GDPR risks of singling-out, linkability, and attribute inference), but both are evaluation frameworks rather than systematic comparisons of which generators leak most.
\citet{ganev_inadequacy_2025} show that the similarity-based privacy scores marketed by synthetic-data vendors badly understate real risk. Each of these works fixes a single attacker, threat model, or metric; none organizes the attack space or compares attacks on a common footing: the gap this paper addresses.

\paragraph{Reconstruction from de-identified microdata (in scope).}
This branch attacks \emph{de-identified record-level data}: a table whose identifiers have been removed or perturbed by suppression, generalization, or swapping. The re-identification literature established that such releases leak. \citet{sweeney2000simple} showed that quasi-identifiers alone re-identify most individuals and proposed $k$-anonymity; \citet{machanavajjhala2007diversity} then showed $k$-anonymity still permits \emph{attribute} disclosure when a quasi-identifier group is homogeneous, motivating $\ell$-diversity. We attack de-identified releases in this spirit, but score attribute reconstruction; the retained-record attacks unique to de-identification (re-identification and these homogeneity attacks on $k$-anonymity) have no synthetic analog and are out of scope (Figure~\ref{fig:taxonomy}).

\paragraph{Two branches, one attack surface.}
Because de-identified and synthetic releases present the adversary with the same information (a published record-level table plus the target's QI), their attack methodologies coincide, and we merge the two subtrees of Figure~\ref{fig:taxonomy}. A data holder choosing between de-identifying and synthesizing a table faces the same adversary either way, so scoring both with the same attacks measures how the two differ instead of assuming it. Both were the setting of the NIST CRC, where our submission placed first (Appendix~\ref{sec:nist_appendix}). We attack thirteen generators spanning three paradigms. Seven are those evaluated in the NIST CRC: the differentially private marginal synthesizers MST~\cite{mckenna2021winning} and AIM~\cite{mckenna2022aim}; the deep generative models TVAE~\cite{xu2019modeling} and ARF~\cite{watson2023adversarial}; the non-private statistical synthesizer Synthpop~\cite{nowok2016synthpop}; and the de-identification methods rank swapping~\cite{moore1996rankswap} and cell suppression~\cite{templ2015sdcmicro}. To broaden coverage, we add two more from the dominant deep-generative families, the GAN-based CTGAN~\cite{xu2019modeling} and the diffusion-based TabDDPM~\cite{kotelnikov2023tabddpm}, and four DP synthesizers with distinct mechanisms (PrivBayes, PrivSyn, MWEM-PGM, and PrivateGSD) that join MST and AIM in the privacy-budget sweep of Section~\ref{sec:whichgenerator}. Table~\ref{tab:ra_mean_adult} covers the first nine. Full descriptions are deferred to Appendix~\ref{sec:sdg_descriptions}.

\paragraph{Reconstruction as imputation.} Reconstruction has a cooperative twin in \emph{missing-data imputation}, a routine step in clinical and epidemiological analysis~\cite{sterne2009,demirtas2018flexible}. \citet{jayaraman2022attribute} show that attribute inference against trained models largely reduces to such imputation, and the same holds for a released table: MissForest~\cite{stekhoven2012missforest}, fit on the release, is essentially our random-forest attack, and MICE~\cite{van2011mice} is a chained per-feature regression. Only the intent differs (restoring values a data holder lost vs.\ inferring values it withheld), so every better imputer is a better attack. Because the missing cells here are whole columns, present in the release and absent for every target, imputers reduce to the per-feature and chained predictors we benchmark, and we do not evaluate them separately.

\begin{table*}
  \centering
  \small
  \caption{Mean reconstruction accuracy ($R_{adv}$, \%): each cell averages over all hidden features and 5 disjoint training sets, with rows grouped by taxonomy branch (Figure~\ref{fig:taxonomy}). The three reference points are not attacks, and the \emph{Best ensemble}/\emph{Best chain} rows give each enhancement's strongest configuration (Appendix~\ref{sec:ensembling} and \ref{sec:chaining}). $\dagger$~= novel attack. (Adult, 10k rows, $\text{QI}_{\text{demo}}$.)}
  \label{tab:ra_mean_adult}
  \resizebox{\textwidth}{!}{%
\begin{tabular}{lrrrrrrrrrrrrr>{\bfseries}r}
\toprule
Attack &
\rotatebox{60}{RankSwap} &
\rotatebox{60}{Cell Supp.} &
\rotatebox{60}{Synthpop} &
\rotatebox{60}{TVAE} &
\rotatebox{60}{CTGAN} &
\rotatebox{60}{ARF} &
\rotatebox{60}{TabDDPM} &
\rotatebox{60}{MST $\varepsilon{=}0.1$} &
\rotatebox{60}{MST $\varepsilon{=}1$} &
\rotatebox{60}{MST $\varepsilon{=}10$} &
\rotatebox{60}{MST $\varepsilon{=}100$} &
\rotatebox{60}{MST $\varepsilon{=}1000$} &
\rotatebox{60}{AIM $\varepsilon{=}1$} &
\rotatebox{60}{Avg.} \\
\midrule
\multicolumn{15}{l}{\emph{Reference points (not attacks)}}\\
Mode & 10.5 & 10.5 & 10.5 & 10.5 & 10.5 & 10.5 & 10.5 & 10.3 & 10.3 & 10.3 & 10.3 & 10.3 & 10.5 & 10.4 \\
Random & 10.5 & 10.5 & 10.5 & 10.5 & 10.3 & 10.5 & 10.5 & 10.3 & 10.3 & 10.3 & 10.3 & 10.2 & 10.5 & 10.4 \\
Copy & 49.3 & 10.3 & 10.5 & 10.4 & 10.3 & 10.4 & 10.5 & 10.2 & 10.2 & 10.4 & 10.2 & 10.3 & 10.5 & 13.3 \\
\midrule
\multicolumn{15}{l}{\emph{Each feature in isolation: classifiers}}\\
\textsc{knn} & 25.5 & 55.7 & 26.9 & 22.1 & 19.3 & 18.6 & 39.3 & 15.4 & 20.3 & 21.7 & 21.5 & 21.7 & 15.4 & 24.9 \\
Naive Bayes & 22.2 & 46.3 & 31.1 & 18.6 & 18.4 & 18.8 & 31.5 & 15.5 & 17.6 & 23.5 & 25.0 & 24.4 & 17.1 & 23.8 \\
Logistic Regression & 15.6 & 17.3 & 16.7 & 16.1 & 15.3 & 14.7 & 16.9 & 12.3 & 14.3 & 15.6 & 15.6 & 15.7 & 12.6 & 15.3 \\
\textsc{svm} & 21.5 & 26.0 & 25.5 & 21.1 & 20.6 & 19.5 & 26.1 & 15.0 & 19.4 & 21.0 & 21.2 & 21.3 & 14.4 & 21.0 \\
Random Forest & 25.7 & 52.3 & 28.8 & 24.1 & 21.2 & 20.5 & 38.9 & 16.0 & 22.1 & 24.1 & 25.1 & 25.4 & 20.9 & 26.5 \\
LightGBM & 23.7 & 34.2 & 27.9 & 23.1 & 22.6 & 22.7 & 23.8 & 16.0 & 23.1 & 23.0 & 22.1 & 26.9 & 21.1 & 23.9 \\
\textsc{mlp} & 23.3 & 33.9 & 30.1 & 27.0 & 24.3 & 22.9 & 34.8 & 16.2 & 24.2 & 27.2 & 27.3 & 27.0 & 23.8 & 26.3 \\
TabPFN & 23.0 & 46.3 & 27.6 & 23.6 & 21.2 & 19.7 & 34.7 & 15.5 & 20.6 & 23.1 & 24.2 & 23.3 & 14.7 & 24.4 \\
Best ensemble & 25.3 & 42.1 & 29.7 & 25.2 & 23.3 & 22.4 & 36.0 & 15.9 & 23.3 & 25.8 & 25.7 & 26.5 & 22.0 & 26.4 \\
\midrule
\multicolumn{15}{l}{\emph{Feature-correlated: autoregressive}}\\
ARFFormer$^\dagger$ & 22.7 & 30.5 & 30.1 & 27.6 & 25.0 & 23.0 & 30.6 & 15.4 & 23.5 & 26.4 & 26.3 & 26.3 & 23.5 & 25.5 \\
Best chain & 25.5 & 54.3 & 28.1 & 23.3 & 20.0 & 19.8 & 39.2 & 16.0 & 22.4 & 24.9 & 25.3 & 25.3 & 20.0 & 26.5 \\
\midrule
\multicolumn{15}{l}{\emph{Feature-correlated: row-wise message passing}}\\
CoBP-RA$^\dagger$ & 26.7 & 53.1 & 30.7 & 27.6 & 24.9 & 23.8 & 40.4 & 16.1 & 23.5 & 25.7 & 26.2 & 26.1 & 22.4 & 28.2 \\
\midrule
\multicolumn{15}{l}{\emph{Feature-correlated: joint generative conditioning}}\\
MultiHeadMLP$^\dagger$ & 23.4 & 39.2 & 30.1 & 26.4 & 23.5 & 22.3 & 33.5 & 15.8 & 24.0 & 26.7 & 27.2 & 26.9 & 23.8 & 26.4 \\
CondMST$^\dagger$ & 20.9 & 27.9 & 27.4 & 23.0 & 21.4 & 18.2 & 27.1 & 15.5 & 21.7 & 25.5 & 23.0 & 24.9 & 22.1 & 23.0 \\
CondDDPM$^\dagger$ & 22.8 & 28.0 & 26.7 & 22.3 & 18.5 & 18.0 & 29.1 & 13.9 & 20.0 & 18.6 & 18.6 & 18.7 & 21.2 & 21.3 \\
CondRePaint$^\dagger$ & 14.8 & 18.2 & 18.5 & 15.9 & 13.4 & 12.8 & 19.3 & 11.3 & 14.0 & 13.0 & 12.5 & 13.0 & 15.1 & 14.8 \\
\midrule
\textbf{Avg.} & \textbf{22.5} & \textbf{31.8} & \textbf{24.1} & \textbf{20.6} & \textbf{18.9} & \textbf{18.1} & \textbf{26.9} & \textbf{14.2} & \textbf{18.8} & \textbf{20.4} & \textbf{20.4} & \textbf{20.7} & \textbf{17.6} & \\
\bottomrule
\end{tabular}
  }
\end{table*}

\subsection{Attacks on Released Tables, by the Structure They Exploit}
\label{sec:attack_taxonomy}

Every attack we study maps a release $D_{\text{syn}}$ and a target's QI to a guess $\hat{x}^H$ of the hidden attributes  (Eq.~\ref{eqn:attack}); they differ in \emph{how much joint structure they exploit} (Figure~\ref{fig:taxonomy}, lower subtree). One family reconstructs each hidden feature \emph{in isolation}; the other exploits the \emph{correlations among hidden features}.

\paragraph{Reference points (not attacks).}
Three trivial procedures fix a scale for the rarity-weighted score $R_{adv}$, which (unlike the AUC of a binary test, where $0.5$ is always chance) shifts with the dataset, the QI, and the generator. For each hidden feature $h$ we (i)~impute the \textbf{mode} (categorical) or \textbf{mean} (continuous) of $D_{\text{syn}}[h]$; (ii)~draw a \textbf{random} value from $D_{\text{syn}}[h]$; or (iii)~\textbf{copy} $D_{\text{syn}}[h]$ row-for-row (using $D_{\text{syn}}[h]_{i \bmod |D_{\text{syn}}|}$ when $|D_{\text{syn}}| \neq |D_{\text{train}}|$). These are not attacks; they calibrate what a given feature's $R_{adv}$ score means, and under rarity weighting all three sit near $100/C$ for a $C$-valued feature. The one revealing exception is \textbf{copy} against RankSwap ($49.3$ in Table~\ref{tab:ra_mean_adult}): RankSwap leaves most columns untouched, so copying the release returns the originals verbatim, which is a property of the RankSwap technique, not of any attack.

\paragraph{Reconstructing each feature in isolation.}
The dominant paradigm, due to \citet{stadler2022synthetic}, treats the release ($D_{\text{syn}}$) as \emph{training data}: for each hidden feature $h$, fit a supervised model with the QI columns as inputs and $D_{\text{syn}}[h]$ as the label, then apply it to the target's QI. Any classifier serves; we evaluate logistic regression, naive Bayes, $k$-nearest neighbors, an SVM, random forests~\cite{breiman2001random}, gradient-boosted trees~\cite{ke2017lightgbm}, a multilayer perceptron (MLP), and (new to this setting) \textbf{TabPFN}~\cite{hollmann2023tabpfn}, a pre-trained transformer that produces a tabular predictor in a single forward pass. For a \emph{continuous} hidden feature the same recipe substitutes a regressor for the classifier (we report a random-forest regressor and Bayesian ridge regression). Among these per-feature classifiers the random forest is the strongest in our benchmark and the MLP a close second (Table~\ref{tab:ra_mean_adult}). \emph{Ensembling} belongs to the same branch: any set of these classifiers can be combined by voting or averaging.

A different approach replaces the classifier with a \emph{column-mar\-gi\-nal} model: the linear-program (LP) attack of \citet{annamalai2024linear} assigns each record a soft membership score so that the release's implied counting-query answers are matched, then rounds to a prediction. The LP handles one low-cardinality column at a time and grows intractable as the value count rises; on single binary targets it does not beat a random forest or MLP, and it is infeasible for the high-cardinality, many-feature NIST CRC setting. We therefore report it only on binary targets (Table~\ref{tab:linear_sweep_summary}) and defer extended results to Appendix~\ref{sec:lp_appendix}.

\paragraph{Exploiting correlations among hidden features.}
Per-feature attacks discard a real signal: the hidden features are themselves correlated, and the release encodes those correlations. Three designs exploit them, and (apart from classifier chaining \cite{read2011classifier}) each is our contribution.

\textbf{Autoregressive} attacks predict hidden features in sequence, feeding each prediction back into the next. \emph{Chaining}~\cite{read2011classifier} (see Appendix \ref{sec:chaining}) wraps any in-isolation attack; \textbf{ARFFormer} (\emph{autoregressive feature transformer}) instead learns the dependencies directly, encoding the QI with multi-head self-attention and predicting the hidden features autoregressively. New to this setting, it tracks the MLP's success closely.

\begin{figure*}[t]
  \centering
  \includegraphics[width=\textwidth]{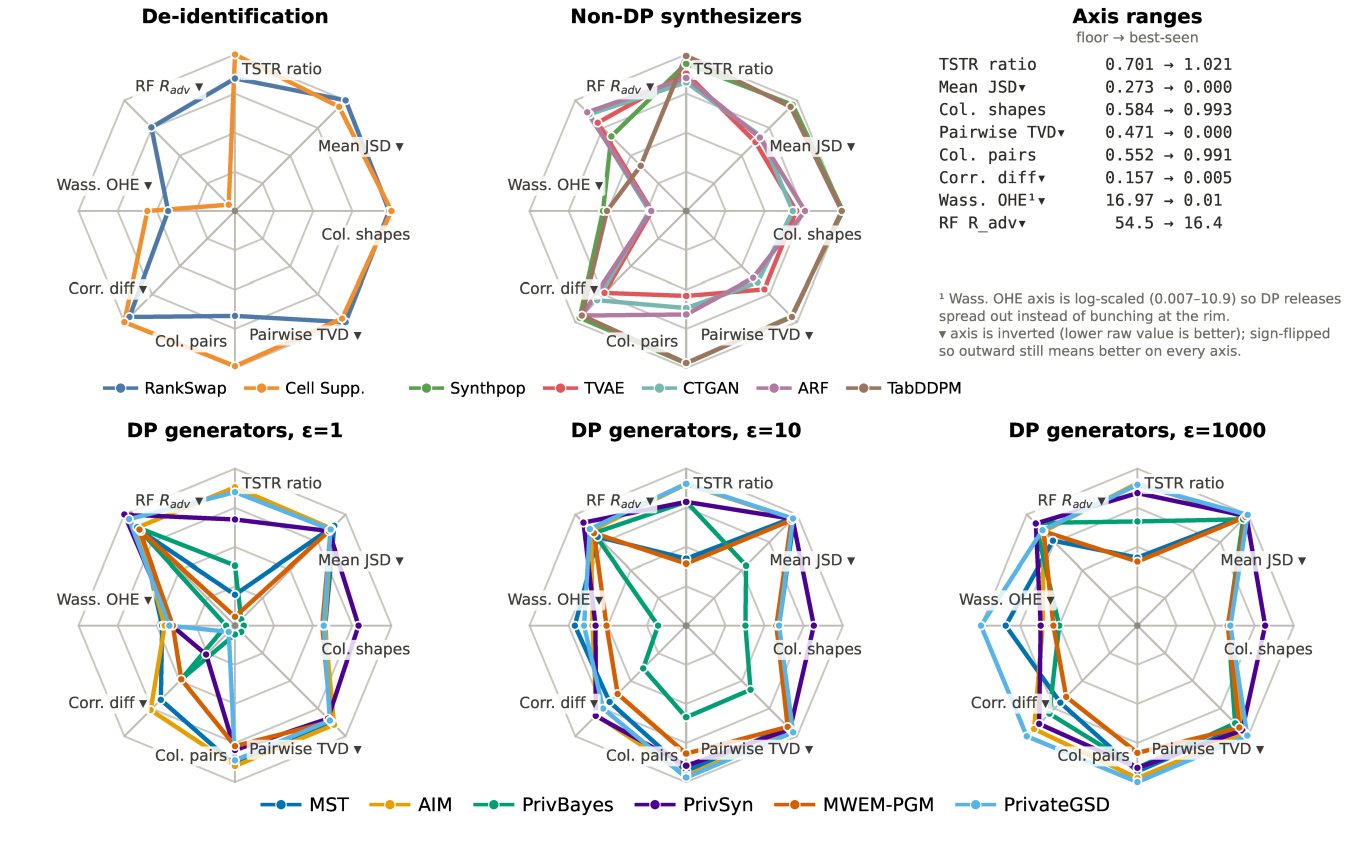}
  \caption{Fidelity, utility, and reconstruction risk of every release; Table~\ref{tab:quality_overview} gives the exact values and the \textit{Train--Train} real-vs-real benchmark. \textbf{TSTR ratio}~\cite{esteban2017real}, F1-macro of a classifier trained on synthetic data and tested on real, relative to one trained on real ($1$ is ideal); \textbf{Mean JSD}~\cite{lin1991divergence}, Jensen--Shannon divergence over marginals; \textbf{Col.\ shapes} and \textbf{Col.\ pairs}~\cite{patki2016synthetic}, SDV marginal and bivariate fidelity ($1$ = indistinguishable from real); \textbf{Pairwise TVD}, mean TV distance over pairwise marginals; \textbf{Corr.\ diff}, mean absolute error between correlation matrices; \textbf{Wass.\ OHE}~\cite{golob_privacy_2025}, summed per-feature $L_1$ marginal distance; \textbf{RF $R_{adv}$}, a Random Forest attack's reconstruction accuracy. Outward is always better; $\blacktriangledown$ marks a sign-flipped axis. Each axis spans worst-to-best plotted, so a release's \emph{shape} compares, not its position. (Adult, 10k rows, $\text{QI}_{\text{demo}}$.)}
  \label{fig:quality_overview}
\end{figure*}

\begin{figure}[t]
  \centering
  \includegraphics[width=\columnwidth]{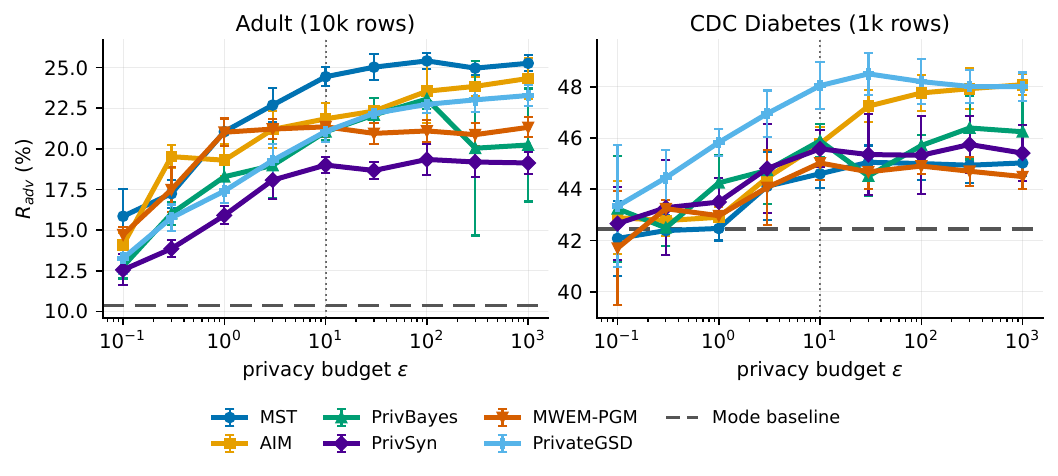}
  \caption{Reconstruction risk against privacy budget, mean of Random Forest, Naive Bayes and CoBP-RA; bars are 95\% confidence intervals. The dotted line marks $\varepsilon{=}10$, beyond which the curves change little; the dashed rule is the Mode baseline. AIM and (on Adult) PrivateGSD keep climbing, owing most of the rise to Naive Bayes; Figure~\ref{fig:eps_curves_perattack} breaks the curves out by attack. (Adult 10k and CDC Diabetes 1k, $\text{QI}_{\text{demo}}$.)}
  \label{fig:eps_curves}
\end{figure}

\textbf{Row-wise message passing} is our strongest design.
\textbf{CoBP-RA} (\emph{correction with belief propagation}, the ``cobra'' attack) keeps the per-feature random-forest posteriors as unary beliefs but couples them on a graph: it takes the maximum-spanning-tree of pairwise mutual information over the hidden features, attaches to each edge a local pointwise-mutual-information table estimated from the synthetic neighbors nearest the target in QI space, and runs one pass of sum--product belief propagation. The local tables are the crux: because the random forest already conditions on the QI, a global co-occurrence table would double-count it, whereas a neighbor-local table isolates the \emph{residual} dependence between two hidden features that survives that conditioning. Belief propagation then lets a confident feature correct an uncertain neighbor: if the synthetic records around a target make the joint value $(B,C)$ far more common than $(A,C)$, the belief for the first feature shifts from $A$ toward $B$, a correction no independent classifier can make. By default CoBP-RA also adds the QI values as observed graph nodes and damps messages from high-entropy senders, the best configuration we found and the one we report throughout. (A final \emph{column-wise} step nudges predictions toward the synthetic marginals; it moves accuracy by about a point at most.) For continuous data we discretize each hidden feature into equal-frequency bins of the synthetic column, run the same message passing on the bin labels, and decode each prediction back to its bin midpoint (Table~\ref{tab:memorization_california}).

\textbf{Joint generative conditioning} turns a generative model into the attack: instead of predicting features one by one, it samples all hidden attributes \emph{at once} from an SDG model conditioned on the target's QI, so the sample respects the full joint distribution. We explore this region of the attack taxonomy with four new attacks. The deep-generative form includes \textbf{MultiHeadMLP} (a single network with one head per feature, predicting all hidden features jointly), \textbf{CondDDPM} (a tabular diffusion model whose reverse process is conditioned on the QI during training), and \textbf{RePaint}/\textbf{CondRePaint}, which port the image-inpainting RePaint scheme~\cite{lugmayr2022repaint} to tables: at each denoising step the QI columns are overwritten with a forward-noised copy of their true values, anchoring the generated hidden columns to the observed QI with no change to the trained model~\cite{ho2020denoising}. The graphical-model form is \textbf{CondMST}: we fit an MST-style graphical model~\cite{mckenna2019graphical} on the release and condition it on the QI to sample the hidden features jointly. CondMST and CondDDPM are named for the generators they adapt, but the construction is general (each is run against \emph{all nine} generators of Table~\ref{tab:ra_mean_adult}, not only its namesake), and the details appear in Appendices~\ref{sec:partial_sdg_mst} and~\ref{sec:partial_sdg_tabddpm}, with pseudocode for all six new attacks in Appendix~\ref{sec:pseudocode}.

\paragraph{What the benchmark shows.}
Table~\ref{tab:ra_mean_adult} pits every attack against nine generators (MST at five privacy budgets) on the Adult dataset (the five datasets are summarized in Table~\ref{tab:datasets} and described in Appendix~\ref{app:datasets}). Every reported score is averaged over 5 disjoint training samples.
Two patterns dominate.

First, \emph{the generator matters far more than the attack}. Down any column, the three strongest attacks lie within three points of each other; across any row the same attack swings much more: a random forest ranges from $16.0$ against MST at $\varepsilon{=}0.1$ to $52.3$ against CellSuppression. The vertical \textbf{Avg.}\ row makes the ordering plain: de-identification (CellSuppression $31.8$, RankSwap $22.5$) and the highest-fidelity synthesizers (TabDDPM $26.9$, Synthpop $24.1$) are the exposed regimes, while DP marginal synthesis collapses toward the Mode reference as the budget tightens (MST $\varepsilon{=}0.1$: $14.2$). Section~\ref{sec:conditions} dissects this axis and why it, not the attack, is decisive.

Second, \emph{no attack wins everywhere, and sophistication buys little}. KNN leads on CellSuppression, where retained rows are real records and the attack is effectively a lookup; naive Bayes is the best on Synthpop, whose sequential generation mirrors its column-wise independence; on TVAE, CoBP-RA, ARFFormer, and MLP cluster at the top ($27.6$, $27.6$, $27.0$), separated by under a point. \textbf{CoBP-RA is the best attack on average} ($28.2$) and is best or near-best in almost every column, but its margin over a plain random forest ($26.5$) is under two points, and its belief-propagation correction helps most on the deep generators TVAE, CTGAN, and ARF (${+}3.3$ to ${+}3.7$ over the random forest) and least against heavy DP noise (${+}0.1$ at MST $\varepsilon{=}0.1$). The joint-generative attacks are a cautionary case: CondDDPM and CondRePaint trail simple classifiers even on TabDDPM-generated data, plausibly because $10$k synthetic rows are too few to pin down their parameters.

\paragraph{Can attack combinations increase success rates?}
Two leaves of the taxonomy combine attacks rather than introduce new ones: \emph{ensembling} (voting across in-isolation classifiers) and \emph{chaining} (autoregressive prediction). Both helped our NIST CRC submission, yet in controlled experiments neither beats CoBP-RA alone, and in each case an oracle analysis shows that the limit lies in the release, not in how the attacks are combined. Ensembled attacks are complementary (an oracle picking the right attack per record would gain over eleven points), but nothing in the QI marks which records a given attack will get right, per-feature selection recovers almost none of the gain, and a shared core of hard records defeats every attack, so even weights tuned on the test labels do not beat the best single attack. \emph{Chaining} fails for the opposite reason: an oracle fed the \emph{true} earlier hidden values gains only a few points ($+4.0$ for a random forest and $+5.8$ for an MLP, on TabDDPM), so error propagation is \emph{not} the bottleneck; the release does not preserve the higher-order conditional structure autoregression would exploit, even under TabDDPM (Appendices~\ref{sec:ensembling} and~\ref{sec:chaining}).

\begin{tcolorbox}[
    enhanced,
    title={\faLightbulb\quad Takeaway},
colbacktitle=takeawaytitle,
coltitle=white,
colback=takeawayfill,
colframe=takeawayframe,
    fonttitle=\bfseries,
    arc=3pt,
    boxrule=1.2pt,
    left=6pt, right=6pt, top=4pt, bottom=4pt
]
On a fixed release, switching \emph{attacks} moves risk by a few points; switching \emph{generators}, by tens. CoBP-RA is our strongest attack, yet the more striking pattern is how close simple methods come: a plain random forest lands within a couple of points of it, and the best attack is often mechanism-specific (naive Bayes rivals far richer models on column-sequential Synthpop). Sharper attacks may yet emerge, but on every release, the generator, not the adversary, drives exposure. 
\end{tcolorbox}

\begin{table}[t]
\centering
\small
\setlength{\tabcolsep}{4pt}
\caption{Datasets used in this study, all real and publicly available. Feature counts are categorical/continuous; `$\checkmark$' marks data of a sensitive nature. $^{\ast}$NIST Arizona is IPUMS 1940 census microdata from the NIST Privacy CRC; we use its 25-feature setting.}
\label{tab:datasets}
\begin{tabular}{lcclcc}
\toprule
\textbf{Dataset} & \textbf{Feat.} & \textbf{Rows} & \textbf{Domain} & \textbf{Sens.} & \textbf{Ref.} \\
\midrule
Adult                 & 15 (9/6)    & 48k  & Demographic & \checkmark & \cite{adult1996} \\
NIST Arizona$^{\ast}$ & 25 (19/6)   & 235k & Demographic & \checkmark & \cite{nist_arizona} \\
California Housing     & 9 (0/9)     & 22k  & Housing     &            & \cite{california_housing} \\
NIST SBO              & 130 (125/5) & 1.6M & Finance     & \checkmark & \cite{nist_sbo} \\
CDC Diabetes          & 22 (19/3)   & 200k & Healthcare  & \checkmark & \cite{cdc_diabetes} \\
\bottomrule
\end{tabular}
\end{table}

\begin{table*}[htbp]
\centering
\caption{Reconstruction accuracy on binary hidden features. $R_{adv}^{\mathrm{tr}}$ = accuracy on training-set targets; $\Delta = R_{adv}^{\mathrm{tr}} - R_{adv}^{\mathrm{NT}}$ is the gap to held-out (non-training, NT) targets from the same population, the individual-level (memorization) leakage (Section~\ref{sec:memorization}). Best $R_{adv}^{\mathrm{tr}}$ per column in \textbf{bold}; the \colorbox{linrecon}{shaded} row is the LP attack proposed as state-of-the-art by \citet{annamalai2024linear}. The lower block adds two attacks new to this setting, TabPFN~\cite{hollmann2023tabpfn} and CoBP-RA$^\dagger$. $\dagger$~= novel attack. (One binary hidden feature per column, all other features known; mean over all SDG methods.)}
\label{tab:linear_sweep_summary}
\begin{tabular}{l c c c c c c c c c c c c c c c c}
\toprule
 & \multicolumn{4}{c}{Adult 1k} & \multicolumn{4}{c}{Adult 10k} & \multicolumn{2}{c}{Arizona 10k} & \multicolumn{6}{c}{CDC 1k} \\
\cmidrule(lr){2-5} \cmidrule(lr){6-9} \cmidrule(lr){10-11} \cmidrule(lr){12-17}
 & \multicolumn{2}{c}{income} & \multicolumn{2}{c}{sex} & \multicolumn{2}{c}{income} & \multicolumn{2}{c}{sex} & \multicolumn{2}{c}{sex} & \multicolumn{2}{c}{Diabetes} & \multicolumn{2}{c}{HighBP} & \multicolumn{2}{c}{Stroke} \\
\cmidrule(lr){2-3} \cmidrule(lr){4-5} \cmidrule(lr){6-7} \cmidrule(lr){8-9} \cmidrule(lr){10-11} \cmidrule(lr){12-13} \cmidrule(lr){14-15} \cmidrule(lr){16-17}
Attack & $R_{adv}^{\mathrm{tr}}$ & $\Delta$ & $R_{adv}^{\mathrm{tr}}$ & $\Delta$ & $R_{adv}^{\mathrm{tr}}$ & $\Delta$ & $R_{adv}^{\mathrm{tr}}$ & $\Delta$ & $R_{adv}^{\mathrm{tr}}$ & $\Delta$ & $R_{adv}^{\mathrm{tr}}$ & $\Delta$ & $R_{adv}^{\mathrm{tr}}$ & $\Delta$ & $R_{adv}^{\mathrm{tr}}$ & $\Delta$ \\
\midrule
Random & 50.0 & +0.0 & 50.0 & +0.0 & 50.0 & +0.0 & 50.0 & +0.0 & 50.0 & +0.0 & 50.0 & +0.0 & 50.0 & +0.0 & 50.0 & +0.0 \\
KNN & 63.7 & +5.3 & 67.3 & +5.7 & 69.5 & +4.6 & 75.2 & +4.0 & 66.6 & +4.2 & \textbf{60.7} & +7.1 & 64.3 & +7.4 & \textbf{59.1} & +7.5 \\
Random Forest & 67.0 & +5.5 & \textbf{75.1} & +3.9 & 69.6 & +4.1 & 80.4 & +3.3 & 70.8 & +3.8 & 58.4 & +6.4 & \textbf{68.1} & +6.2 & 56.0 & +5.7 \\
\textsc{mlp} & \textbf{70.1} & +2.5 & 75.1 & +2.8 & \textbf{71.8} & +2.8 & \textbf{80.6} & +1.3 & \textbf{71.1} & +1.3 & 60.2 & +5.0 & 65.2 & +2.9 & 57.5 & +4.1 \\
TabPFN & 60.8 & +3.3 & 71.2 & +2.4 & 61.8 & +0.2 & 73.4 & +0.0 & 69.9 & +0.1 & 53.8 & +2.7 & 64.3 & +2.9 & 51.8 & +1.6 \\
CoBP-RA$^\dagger$ & 67.8 & +5.2 & 75.0 & +4.3 & 70.5 & +3.8 & 80.2 & +3.4 & 70.8 & +3.9 & 58.2 & +6.2 & 67.2 & +5.5 & 55.7 & +5.4 \\
\rowcolor{linrecon}Linear Program & 63.3 & +4.0 & 66.1 & +3.2 & 67.9 & +2.4 & 76.1 & +2.3 & 63.8 & +1.2 & 59.4 & +6.1 & 63.0 & +5.5 & 58.8 & +7.3 \\
\midrule
\textit{Avg.} & \textit{63.1} & \textit{+3.3} & \textit{68.1} & \textit{+2.8} & \textit{65.8} & \textit{+2.3} & \textit{73.8} & \textit{+1.7} & \textit{65.8} & \textit{+1.8} & \textit{57.4} & \textit{+4.2} & \textit{63.0} & \textit{+3.9} & \textit{55.6} & \textit{+4.1} \\
\bottomrule
\end{tabular}%
\end{table*}

\section{
How Does the Scenario Shape Risk?}
\label{sec:conditions}

Section~\ref{sec:taxonomy} fixed almost everything but the attack; here we hold the attack family roughly fixed and sweep the \emph{scenario}: the generator and its privacy budget (Section~\ref{sec:whichgenerator}), the data domain, rows, and features (Section~\ref{sec:whatdata}), and the adversary's quasi-identifier (Section~\ref{sec:whichadversary}). The release again governs how much can be reconstructed; the interesting structure is in how each dimension moves that ceiling.

\subsection{SDG Method and Privacy Budget}\label{sec:whichgenerator}

As stated in Section \ref{sec:taxonomy},
the SDG method is the single largest lever on reconstruction risk, and the ordering follows how much each method transforms the data: the \emph{least} transformed releases leak the most. De-identification leads: cell suppression enforces $k$-ano\-ny\-mity~\cite{sweeney2002k} by dropping records whose QI combination is too rare and releasing the rest unmodified, so KNN's $55.7\%$ on Adult data (Table~\ref{tab:ra_mean_adult}) is little more than a lookup against surviving real records; rank swapping permutes values within blocks, preserving marginals while leaving enough joint structure for a random forest to reach $25.7\%$.

In general, as seen in Figure~\ref{fig:quality_overview}, vulnerability loosely tracks the quality of the release. Across the nine generators of Table~\ref{tab:ra_mean_adult} (DP ones at $\varepsilon{=}1$), downstream utility (TSTR: the F1-score of a classifier trained on the synthetic data to predict \texttt{income} and evaluated on real data) and RF reconstruction are rank-correlated at Spearman $\rho{=}0.63$ ($p{=}0.07$), a positive trend that nine generators are too few to make significant.

No single fidelity metric reads out risk, however. Every metric in Figure~\ref{fig:quality_overview} correlates with RF reconstruction in the expected direction ($|\rho|$ between $0.59$ and $0.79$), and every one of them is broken by some pair of releases. TVAE and MST at $\varepsilon{=}0.1$ match almost exactly on marginal distance, yet TVAE is reconstructed half again as accurately; conversely PrivBayes at $\varepsilon{=}1000$ has marginals as faithful as a second real sample and is still \emph{safer} than three deep generators whose marginals are far worse. AIM and MST break the utility trend the same way. A data holder cannot read privacy off a fidelity score: how closely a release matches the real data says little about how reconstructable its records are, which only an attack measures (per-dataset breakdowns in Appendix~\ref{sec:quality_appendix}, Tables~\ref{tab:quality_arizona}--\ref{tab:quality_california}).

Differential privacy helps mainly below $\varepsilon{\approx}10$. For MST on Adult, RF reconstruction drops sharply at $\varepsilon{=}0.1$ (to $16.0\%$); from $\varepsilon{=}10$ to $\varepsilon{=}1000$ it barely stirs ($24.1\%\!\to\!25.4\%$; Table~\ref{tab:ra_mean_adult}),
and the full nine-budget sweep rises steeply to $\varepsilon{\approx}10$ and then flattens for all six mechanisms, the few curves still rising being carried almost entirely by one attack, Naive Bayes (Figure~\ref{fig:eps_curves}; two further QIs and per-attack results in Appendix~\ref{sec:eps_appendix}, per-feature curves in Appendix~\ref{sec:feature_eps_appendix}). Averaged over our three attacks on Adult, moving from $\varepsilon{=}0.1$ to $\varepsilon{=}10$ buys the adversary $5.7$--$11.4$~pp; moving from $\varepsilon{=}10$ to $\varepsilon{=}1000$ buys between $-0.9$ and $+3.4$~pp, and is statistically indistinguishable from zero for three of the six mechanisms (Appendix~\ref{sec:statistics}). Above $\varepsilon{\approx}10$ the binding constraint is thus what the synthesizer can \emph{represent}, not its noise. All six DP synthesizers share this ceiling, though they share little machinery: MST, AIM, and MWEM-PGM fit graphical models to noisy marginals, PrivBayes fits a Bayesian network, PrivSyn nudges a record set toward noisy pairwise marginals with no model at all, and PrivateGSD evolves a synthetic population by genetic search. They do not even get there the same way: from $\varepsilon{=}1$ to $\varepsilon{=}10$, MWEM-PGM gains only $0.3$~pp ($p{=}0.50$) while PrivateGSD gains $3.7$~pp ($p{<}0.001$), yet both level off. Where each lands is another matter: the ordering is mechanism- and data-specific (MST leaks most on Adult, PrivateGSD and AIM on CDC), and it follows what each mechanism elects to preserve (Appendix~\ref{sec:eps_appendix}). This is a clean qualitative break from \emph{membership} inference, which on MST and PrivateGSD keeps strengthening with $\varepsilon$ out to $1000$~\cite{golob_privacy_2025}. The two tasks ask different amounts of a release. Membership needs one bit per record, and a release can keep surrendering it long after its marginals have stopped moving. Reconstruction needs a joint assignment over many features at once, and the number of candidate assignments grows combinatorially with them, so an attack exhausts the structure a synthesizer actually represents while the budget still has orders of magnitude left to give. Reconstruction therefore meets its ceiling early in $\varepsilon$, where membership does not.

That ceiling is measured, not promised, and the two should not be confused: at large $\varepsilon$ the worst-case guarantee is nearly vacuous. Randomized response is $\varepsilon$-DP yet preserves over $90\%$ of a sensitive bit once $\varepsilon\ge\ln 9\approx2.2$~(\emph{blatant non-privacy}~\cite{dinur2003revealing}), and real deployments run far higher still, the 2020 US Census at $\varepsilon\approx19.61$~\cite{abowd2022topdown}.

\subsection{Domain, Features, and Scale}
\label{sec:whatdata}

Raw $R_{adv}$ is not comparable across our five datasets (Table~\ref{tab:datasets}): both the baseline and the cardinality ceiling shift with the data. Two structural drivers explain most of the per-feature variation. The first is \textbf{cardinality}: on Adult (the RF attack on the TabDDPM release), high-cardinality targets sit near the floor (\texttt{fnlwgt}, $C{=}8{,}477$, is reconstructed at $0.9\%$; \texttt{capital-gain}, $C{=}98$, $9.7\%$) while low-cardinality ones are exposed (\texttt{income}, $C{=}2$, $77.9\%$). The second, and more telling, is \textbf{predictability from the QI}: \texttt{education-num} ($C{=}16$) scores $90.0\%$ (above even binary \texttt{income}) because it is essentially a re-encoding of QI education, whereas \texttt{income} is genuinely harder to guess from demographics. Once cardinality is low, what the QI \emph{determines} sets the score. The baseline moves for the same reason: CDC Diabetes' mostly-binary features yield a Mode baseline of $42.5\%$ (on a binary feature, Mode scores $100/C = 50\%$), against Adult's $10.5\%$, where high-cardinality columns earn Mode almost nothing. Aggregate scores must therefore be read with the per-feature view: reconstructing \texttt{income} and \texttt{occupation} is real harm even when \texttt{fnlwgt} is unrecoverable.

\paragraph{Across domains.} Mostly the ceiling shifts across datasets, though the attack ranking moves too: on CDC, Naive Bayes overtakes CoBP-RA on most DP releases, but not on the de-identified ones, where CoBP-RA stays well ahead (Appendix~\ref{sec:eps_appendix}) (QI memberships in Appendix~\ref{sec:qi_appendix}). On Adult the best attack averages $28.2\%$ across generators, $18$~pp above Mode; the other four datasets each add a distinct lesson.

\textbf{CDC Diabetes} data (full per-generator results in Appendix~\ref{sec:cdc_appendix}, Table~\ref{tab:ra_mean_cdc}) is the most exposed categorical dataset under de-ident\-ifi\-ca\-tion (KNN hits $83.5\%$ on rank swapping), and TabDDPM, at $75.7\%$, rivals it despite being fully synthetic, since diffusion captures binary health dependencies cleanly even from 1k rows. DP is correspondingly tight: across the entire MST sweep KNN moves only $42.2\!\to\!44.2\%$, because binary marginals already carry most of the signal and leave little joint structure beyond them for an attack to use. CoBP-RA again leads or ties the field on most CDC and Arizona data generators.

\textbf{Feature count} matters through how unevenly risk spreads across features, not the average. On NIST SBO (130 mostly-binary financial features) the aggregate gap above Mode is a modest $6$~pp (Appendix Table~\ref{tab:ra_mean_nist_sbo}), but it splits into two tiers: a handful of employee-benefit flags (health insurance, retirement, holidays) leak $9$--$12$~pp because they are predictable from firm sector and size, while noisy high-cardinality financials (receipts, payroll) leak under $2$~pp: the \texttt{fnlwgt} pattern at scale. Only two of 123 hidden features clear $10$~pp, but they are among the most sensitive business facts. NIST Arizona (25 census features) behaves much like Adult, with the best attack at $22.7\%$. 

\textbf{Continuous data} (California Housing, scored by NRMSE) shifts the quality landscape, not the conclusion. Attacks still beat mean-imputation on every release but MST at $\varepsilon{=}0.1$ (Table~\ref{tab:memorization_california}), but the trade-off sharpens into a cliff: MST and AIM at $\varepsilon{\le}1$ discretize the columns so aggressively that a regressor trained on their output predicts worse than the mean (TSTR $R^2{\le}0$; Appendix Table~\ref{tab:quality_california}), recovering to about half the real-data utility from $\varepsilon{=}10$, whereas non-DP Synthpop and TabDDPM deliver real utility at real reconstruction risk. For continuous data, no budget buys both.

\paragraph{Scale cuts both ways.} The effect of dataset size is not monotone: it is set by the mechanism (Table~\ref{tab:cdc_100k}). Growing CDC Diabetes from 1k to 100k makes \emph{de-identification} markedly more exposed (RF on cell suppression $58.6\!\to\!87.1\%$, as more QI groups clear the $k$-anonymity threshold and ship verbatim) yet makes \emph{diffusion} safer (TabDDPM $75.1\!\to\!47.7\%$, as abundant data dilutes per-record memorization); Adult shows the same diffusion effect ($48.6\!\to\!38.9\%$, 1k$\to$10k). Attacker-side scale interacts with this: TabPFN, a prior-fitted transformer built for small tables, matches or beats RF at 1k (on Adult, Cell Supp.\ $51.1$ vs.\ $49.6$; TabDDPM $48.9$ vs.\ $48.6$) but falls behind by 10k (Cell Supp. $46.3$ vs.\ $52.3$; MST $\varepsilon{=}1$ $20.6$ vs.\ $22.1$), where its small-sample prior stops paying off. More data is not uniformly more or less private; it depends on what consumes it. Which \emph{attack} leads, by contrast, follows the dataset and not its size (Appendix~\ref{sec:eps_appendix}).

\subsection{Quasi-Identifying Features}
\label{sec:whichadversary}
\label{subsec:qi}

The quasi-identifier (the features the adversary is assumed to know) is the last design axis. Our default $\text{QI}_{\text{demo}}$ is a set of standard demographic attributes per dataset (for Adult: age, sex, race, native-country, education, marital-status; full per-dataset QI definitions in Appendix~\ref{sec:qi_appendix}, Tables~\ref{tab:qi_def_adult}--\ref{tab:qi_def_sbo}); an adversary can obtain such auxiliary knowledge easily, and a few attributes go a long way, as \citet{sweeney2000simple} re-identified $87\%$ of Americans from \{ZIP, birth date, sex\} and \citet{rocher2019estimating} estimate $99.98\%$ from 15 attributes. Table~\ref{tab:qi_analysis_adult} varies QI \emph{size} and, at a matched size, QI \emph{composition}, each scored over a fixed hidden-feature intersection for comparability (a parallel CDC analysis is in Appendix~\ref{sec:cdc_qi_appendix}).

\paragraph{Composition beats size.} Adding features helps only until the QI captures the main predictors: on Adult, RF climbs $+15.5$~pp from $\text{QI}_{\text{tiny}}$ (3 features) to $\text{QI}_{\text{demo}}$ (6), then just $+2.7$~pp more to $\text{QI}_{\text{large}}$ (10). \emph{Which} features are known matters more. At a matched $|\text{QI}|{=}6$, a behavioral QI (employment, hours) gives RF $27.8\%$ versus $24.2\%$ for the demographic one: a $+3.6$~pp swing from composition alone, larger than the gain from four extra features, and of similar size for KNN and the MLP (Naive Bayes reverses it). This mirrors \citet{narayanan2008robust}: behavioral side-channels re-identify where demographics cannot.

\paragraph{The most revealing QI depends on the data.} On CDC Diabetes (Appendix~\ref{sec:cdc_qi_appendix}), a 10-feature \emph{behavioral} QI scores \emph{below} the same-size demographic one across every attack (RF $33.6$ vs.\ $29.6\%$), the opposite of Adult: on this health data the demographic attributes carry the predictive structure. Size matters even less there (RF gains $1.6$~pp from 4 to 16 known features), so on both datasets what the adversary knows matters more than how much.

\begin{tcolorbox}[
    enhanced,
    title={\faLightbulb\quad Takeaway},
colbacktitle=takeawaytitle,
coltitle=white,
colback=takeawayfill,
colframe=takeawayframe,
    fonttitle=\bfseries,
    arc=3pt,
    boxrule=1.2pt,
    left=6pt, right=6pt, top=4pt, bottom=4pt
]
Risk is set by the release, then shaped at the margins:

by the \emph{data}'s cardinality and QI-predictability, by \emph{scale} (which exposes de-identification yet dilutes a diffusion model's memorization), and by the \emph{composition}, not the size, of what the adversary knows. The biggest surprise is the shape of differential privacy. Across six mechanistically distinct DP synthesizers, protection accrues up to $\varepsilon{\approx}10$ and then \emph{flattens}: beyond it, reconstruction is capped by what each synthesizer can \emph{represent}, not by its noise. In practice, a budget of $\varepsilon{=}10$, in the range real deployments use, already exposes nearly as much as a release at $\varepsilon{=}1000$, whose guarantee is nominal.
\end{tcolorbox}

\section{What Does ``Success'' Actually Mean?
}
\label{sec:interrogating}

The previous two sections measured \emph{how much} can be reconstructed and \emph{under what conditions}. But a single number, $R_{adv}$, conflates phenomena that carry very different privacy meaning. A record can be reconstructed because the generator memorized it, or because the release merely preserves population structure that would predict the same value for anyone like them. We ask three questions of every success: is it memorization or inference (\S\ref{sec:memorization})? Does the signal that drives reconstruction also reveal \emph{membership} (\S\ref{sec:mia})? And is the risk it measures shared evenly across people, or concentrated on a few (\S\ref{sec:disparate})? The answers show that releases with similar average scores can pose very different risks to individuals.

\subsection{Memorization or Generalization?}
\label{sec:memorization}

Two mechanisms produce a correct reconstruction. A generator may \emph{memorize} an individual training record, letting the attacker read it back; or the release may faithfully preserve \emph{distributional} structure, letting the attacker \emph{infer} a hidden value that any similar person would share. The distinction is the crux of what ``leakage'' means: distributional inference is intrinsic to any useful release (the same signal that makes synthetic data valuable), whereas memorization is a failure of the privacy mechanism, exposing a specific person in the data.

\citet{cohen2024data} formalize exactly this boundary as \emph{narcissus resiliency}: a reconstruction counts as genuine only if it could not have been produced by a process that never saw the target; otherwise the attacker is, fittingly for the name, merely \emph{admiring a reflection} of the population rather than recovering the individual. Our \emph{memorization test} is an empirical proxy motivated by it. For each generator we run the attack against training targets ($R_{adv}^{\mathrm{tr}}$) and against disjoint \emph{non-training} targets from the same distribution ($R_{adv}^{\mathrm{NT}}$). The gap $\Delta = R_{adv}^{\mathrm{tr}} - R_{adv}^{\mathrm{NT}}$ is the \emph{memorization gap}: $\Delta\!\approx\!0$ means every success would have occurred without the target in the data, while a large $\Delta$ isolates individual memorization. Table~\ref{tab:memorization_and_ds_risk} reports it for the binary \texttt{income} feature (RF, KNN, MLP averaged), alongside the classical disclosure score discussed below.

\begin{table}[t]
  \centering
  \small
  \caption{Reconstruction accuracy by quasi-identifier setting. Within each section, cells average over the \emph{intersection} of hidden features across that section's QI variants, so columns are comparable within a section (per-variant QI memberships in Table~\ref{tab:qi_def_adult}); bold = row maximum within a section. $\Delta_{\text{size}}{=}\text{QI}_{\text{large}}{-}\text{QI}_{\text{tiny}}$, $\Delta_{\text{comp}}{=}\text{QI}_{\text{beh.}}{-}\text{QI}_{\text{demo}}$, and $\Delta^{*}_{t{\to}d}$ is the gain from adding 3 demographic features, scored on the broader 9-feature two-way intersection. (Adult, 10k rows, mean over all SDG methods.)}
  \label{tab:qi_analysis_adult}
  \footnotesize
  \setlength{\tabcolsep}{3pt}
  \begin{tabular}{lrrrrr|rrr}
    \toprule
     & \multicolumn{5}{c}{QI size sweep ($\nearrow$ adversary knowledge)} & \multicolumn{3}{c}{Composition (matched $|\text{QI}|$)} \\
    \cmidrule(lr){2-6}\cmidrule(l){7-9}
    Attack & $\text{QI}_{\text{tiny}}$ & $\text{QI}_{\text{demo}}$ & $\Delta^{*}_{t{\to}d}$ & $\text{QI}_{\text{large}}$ & $\Delta_{\text{size}}$ & $\text{QI}_{\text{demo}}$ & $\text{QI}_{\text{beh.}}$ & $\Delta_{\text{comp}}$ \\
     & (3 kn.) & (6 kn.) &  & (10 kn.) &  & (6 kn.) & (6 kn.) &  \\
    \midrule
    Random & 11.49 & 11.48 & $-$0.01 & \textbf{11.49} & $+$0.00 & 17.05 & \textbf{17.07} & $+$0.02 \\
    \midrule
    \textsc{knn} & 12.84 & \textbf{25.84} & $+$11.90 & 25.17 & $+$12.33 & 24.82 & \textbf{28.06} & $+$3.24 \\
    NB & 13.74 & 26.14 & $+$9.50 & \textbf{27.10} & $+$13.36 & \textbf{24.55} & 20.97 & $-$3.58 \\
    RF & 12.46 & 27.96 & $+$13.36 & \textbf{30.69} & $+$18.23 & 24.20 & \textbf{27.80} & $+$3.61 \\
    LGBM & 12.35 & 24.82 & $+$11.07 & \textbf{24.91} & $+$12.56 & 21.95 & \textbf{21.95} & $+$0.00 \\
    \textsc{mlp} & 12.52 & 29.52 & $+$13.24 & \textbf{30.71} & $+$18.19 & 23.49 & \textbf{27.69} & $+$4.20 \\
    \bottomrule
  \end{tabular}
\end{table}

\begin{table}[t]
  \centering
  \small
  \caption{Memorization test: training vs.\ non-training accuracy ($R_{adv}^{\mathrm{tr}}$, $R_{adv}^{\mathrm{NT}}$) and gap $\Delta$, alongside a simplified El~Emam disclosure score ($d_S$, higher = more  risk)~\cite{el_emam2020identity}. Bold $\Delta$ marks a gap significantly greater  than zero (one-sample $t$-test, $p\le0.0001$); no other row separates from zero, including ARF at $+2.3$ ($p{=}0.31$). Per-row intervals in Appendix~\ref{sec:statistics}. (Adult, 1k rows, $\text{QI}_{\text{linear}}$, whose single hidden feature is binary \texttt{income}; RF+KNN+MLP with round-robin holdout.)}
  \label{tab:memorization_and_ds_risk}
  \begin{tabular}{@{}lrrrr@{}}
    \toprule
    \textbf{SDG method} & $R_{adv}^{\mathrm{tr}}$ & $R_{adv}^{\mathrm{NT}}$ & $\Delta$ & $d_S$ \\
    \midrule
\multicolumn{5}{@{}l}{\emph{De-identification}} \\
Cell Supp. & 92.8 & 72.8 & $\mathbf{+20.0}$ & 0.417 \\
RankSwap & 81.4 & 68.3 & $\mathbf{+13.1}$ & 0.089 \\
\midrule
\multicolumn{5}{@{}l}{\emph{Non-DP synthesizers}} \\
TabDDPM & 86.4 & 72.9 & $\mathbf{+13.5}$ & 0.281 \\
Synthpop & 73.7 & 72.1 & $+1.6$ & 0.148 \\
CTGAN & 51.2 & 50.7 & $+0.5$ & 0.055 \\
TVAE & 63.3 & 63.4 & $-0.1$ & 0.104 \\
ARF & 65.9 & 63.6 & $+2.3$ & 0.045 \\
\midrule
\multicolumn{5}{@{}l}{\emph{DP synthesizers}} \\
AIM $(\varepsilon{=}1)$ & 57.0 & 56.6 & $+0.4$ & 0.019 \\
MST $(\varepsilon{=}1)$ & 54.9 & 54.7 & $+0.2$ & 0.018 \\
MST $(\varepsilon{=}10)$ & 62.6 & 61.6 & $+1.0$ & 0.011 \\
MST $(\varepsilon{=}1000)$ & 59.9 & 58.9 & $+1.0$ & 0.010 \\
PrivBayes $(\varepsilon{=}1)$ & 53.5 & 53.6 & $-0.1$ & 0.021 \\
PrivBayes $(\varepsilon{=}1000)$ & 60.6 & 59.8 & $+0.8$ & 0.040 \\
PrivSyn $(\varepsilon{=}1)$ & 56.3 & 55.7 & $+0.6$ & 0.007 \\
PrivSyn $(\varepsilon{=}1000)$ & 63.1 & 62.6 & $+0.5$ & 0.025 \\
MWEM-PGM $(\varepsilon{=}1)$ & 51.0 & 50.7 & $+0.4$ & 0.036 \\
MWEM-PGM $(\varepsilon{=}1000)$ & 63.3 & 62.5 & $+0.7$ & 0.038 \\
PrivateGSD $(\varepsilon{=}1)$ & 56.5 & 54.4 & $+2.1$ & 0.022 \\
PrivateGSD $(\varepsilon{=}1000)$ & 59.8 & 58.6 & $+1.2$ & 0.053 \\
    \bottomrule
  \end{tabular}
\end{table}

The generators split into three regimes. \textbf{De-identification leaks individuals.} CellSuppression and RankSwap carry the largest gaps ($\Delta = {+}13$ to ${+}20$~pp): they edit records rather than generate new ones, so original rows survive and stay individually traceable: the high de-identification scores of Section~\ref{sec:taxonomy} are, in substantial part, memorization rather than inference. \textbf{Most generative and DP methods do not.} Synthpop, CTGAN, TVAE, ARF, AIM, and MST, like the four further DP synthesizers, all sit at $|\Delta|\!\le\!2.3$~pp, none significantly above zero; whatever their magnitude, their reconstruction success is mostly distributional. The DP gaps (${-}0.1$ to ${+}2.1$ across all six mechanisms, at $\varepsilon{=}1$ and $\varepsilon{=}1000$ alike) are within sampling noise and consistent with the DP guarantee, and they do not grow meaningfully with the budget even where aggregate risk does (Section~\ref{sec:whichgenerator}).

\textbf{Diffusion memorizes.} TabDDPM's gap ($\Delta = {+}13.5$~pp) is several times that of every other synthesizer (Synthpop ${+}1.6$, CTGAN ${+}0.5$, TVAE ${-}0.1$, ARF ${+}2.3$) and rivals RankSwap, which copies records outright: a \emph{fully synthetic} generator is memorizing individuals. The mechanism mirrors what \citet{carlini2021extracting,carlini2023extracting} document for language and image diffusion models: high-capacity generators interpolate between generalization and memorization, and for records with rare QI combinations they fall back on encoding the individual. For TabDDPM, the \emph{same} capacity that earns it the highest aggregate risk among synthesizers in Table~\ref{tab:ra_mean_adult} is what makes it memorize atypical people. The continuous domain agrees: on California Housing the three-regime structure reappears with the sign reversed (TabDDPM RF $\Delta = -0.034$, MST $|\Delta|\!\le\!0.001$; Appendix~\ref{sec:cal_memorization_appendix}, Table~\ref{tab:memorization_california}), so the signature is structural, not an artifact of categorical scoring.

\textbf{Same score, different risk.} 
Synthpop and TabDDPM are the two highest-fidelity synthesizers (Figure~\ref{fig:quality_overview}), and on people outside the training data an attack reconstructs them equally well ($R_{adv}^{\mathrm{NT}}$ $72.1$ vs.\ $72.9$, Table~\ref{tab:memorization_and_ds_risk}). On the people they were trained on, TabDDPM pulls ahead ($86.4$ vs.\ $73.7$): its memorization gap is eightfold Synthpop's (${+}13.5$ vs.\ ${+}1.6$). The gap shows that Synthpop exposes a target only as far as its population statistics would expose anyone like them, whereas TabDDPM also exposes the people it was trained on. Neither number replaces the other. $R_{adv}$ measures what an adversary actually learns about a person, which harms them whether or not they were in the data, and needs no held-out sample to compute; the gap says whether the release, rather than the population, is to blame.

\paragraph{Classical disclosure risk.}
\label{sec:disclosure_risk}
The rightmost column of Table~\ref{tab:memorization_and_ds_risk} places our adversary against the bottom rung of the classical ladder: a simplified form of the El~Emam \emph{et al.}\ identity-disclosure-risk model~\cite{el_emam2020identity}, $d_S = \frac{1}{n}\sum_s E_s I_s R_s$ (the subscript $S$ labels the synthetic release; the indicators are per record $s$), which credits a match only when a record sits in a QI equivalence class of size $\le k$ ($E_s$), its sensitive value is correctly inferred ($I_s$), and that value is rare ($R_s$): a lookup attacker with no learning (we drop their error/verification adjustment). $d_S$ agrees with $R_{adv}$ at the extremes (CellSuppression most exposed, DP releases among the least), but misorders the middle:
it rates RankSwap ($0.089$), whose retained records carry a ${+}13.1$ memorization gap, below Synthpop ($0.148$) and TVAE ($0.104$), which carry none, because ML attacks generalize \emph{across} equivalence classes rather than requiring exact QI matches. Its one independent signal echoes our memorization finding: TabDDPM's $d_S$ ($0.28$) exceeds every other synthesizer's: even this model-free lookup attacker finds its copied records.

\subsection{Relationship to Membership Inference}
\label{sec:mia}

Membership inference (MIA) asks a strictly easier question than reconstruction (i.e. was this record in the training set?), yet it is the standard currency of privacy auditing~\cite{shokri2017membership,carlini2022membership}, including on synthetic data~\cite{hilprecht2019monte,van2023membership,golob_privacy_2025}.
The two have been scored on incomparable scales, which has kept them apart. We bridge them with a \emph{reduction} in the complexity-theoretic sense: a procedure that solves membership inference by calling reconstruction as a subroutine. Given a candidate record, the adversary runs the full reconstruction attack on it and returns \emph{how well it reconstructs} as the membership score, on the hypothesis that a generator encodes members' conditional structure more faithfully. The reduction is fair to both paradigms: it consumes nothing a MIA adversary lacks (only the synthetic data and the candidate's QI), so any success it achieves is a success MIA could have claimed. We can then ask whether the two paradigms flag the same \emph{releases} and the same \emph{people}. We compare it against two standard distance baselines~\cite{hayes2017logan,stadler2022synthetic}: \emph{SynthDistance} (negative distance to the nearest synthetic row) and \emph{NNDR} (that distance taken relative to the nearest leave-one-out training neighbor). Both are standard in synthetic-data audits and already separate the generators sharply (AUC $0.50$ to $1.00$); we ask whether RA-as-MIA tracks that signal, not which MIA is strongest. Table~\ref{tab:mia_comparison} spans the full vulnerability spectrum across three datasets.

\begin{table}[t]
\centering
\small
\caption{AUC of MIA (SynthDistance, NNDR) vs.\ RA-as-MIA, which uses RF $R_{adv}$ as the membership signal. Members are one training sample, non-members a disjoint holdout sample from the same population; bold = the stronger MIA.}
\label{tab:mia_comparison}
\resizebox{\columnwidth}{!}{%
\begin{tabular}{lcccccccc}
  \toprule
  & \rotatebox{55}{Cell Supp.} & \rotatebox{55}{RankSwap} & \rotatebox{55}{TabDDPM} & \rotatebox{55}{Synthpop} & \rotatebox{55}{MST $(\varepsilon{=}1)$} & \rotatebox{55}{MST $(\varepsilon{=}1000)$} & \rotatebox{55}{PrivateGSD $(\varepsilon{=}1)$} & \rotatebox{55}{PrivateGSD $(\varepsilon{=}1000)$} \\
  \midrule
  \multicolumn{9}{l}{\textbf{Adult} data ($n=10{,}000$, 15 features, $\text{QI}_{\text{demo}}$)} \\[2pt]
  SynthDistance   & 0.96 & 0.82 & 0.77 & 0.55 & 0.50 & 0.50 & 0.50 & 0.51 \\
  NNDR            & \textbf{0.97} & \textbf{0.87} & \textbf{0.80} & \textbf{0.58} & \textbf{0.50} & \textbf{0.50} & \textbf{0.50} & \textbf{0.51} \\
  \cmidrule(l){1-9}
  RA-as-MIA       & 0.65 & 0.57 & 0.62 & 0.52 & 0.49 & 0.50 & 0.51 & 0.50 \\
  \midrule
  \multicolumn{9}{l}{\textbf{CDC Diabetes} data ($n=1{,}000$, 22 features, $\text{QI}_{\text{demo}}$)} \\[2pt]
  SynthDistance   & 0.62 & 0.99 & 0.80 & 0.57 & 0.51 & 0.52 & 0.51 & 0.54 \\
  NNDR            & \textbf{0.64} & \textbf{1.00} & \textbf{0.81} & \textbf{0.60} & \textbf{0.51} & \textbf{0.53} & \textbf{0.51} & \textbf{0.55} \\
  \cmidrule(l){1-9}
  RA-as-MIA       & 0.60 & 0.77 & 0.71 & 0.52 & 0.51 & 0.51 & 0.50 & 0.51 \\
  \midrule
  \multicolumn{9}{l}{\textbf{NIST Arizona} data ($n=10{,}000$, 25 features, $\text{QI}_{\text{medium}}$)} \\[2pt]
  SynthDistance   & 0.93 & 0.84 & 0.72 & 0.51 & 0.50 & 0.50 & 0.50 & 0.51 \\
  NNDR            & \textbf{0.96} & \textbf{0.91} & \textbf{0.80} & \textbf{0.56} & \textbf{0.52} & \textbf{0.52} & \textbf{0.52} & \textbf{0.52} \\
  \cmidrule(l){1-9}
  RA-as-MIA       & 0.62 & 0.55 & 0.57 & 0.51 & 0.50 & 0.50 & 0.50 & 0.50 \\
  \bottomrule
\end{tabular}}
\end{table}

\begin{table*}[!t]
  \centering
  \small
  \caption{Disparate impact. \textbf{Mean} and the per-race/per-sex columns are mean row-level $R_{adv}$ (\%); \textbf{Outlier~$\times$} is the \emph{unitless} ratio of that score on the outlier $10\%$ (the most QI-atypical individuals; Isolation Forest on the public QI, see text) to the typical $90\%$ (${>}1$ = outliers more exposed).
  The `\%' beside each subgroup header is its population share. AI/AN = Amer.-Indian/Alaska Native, API = Asian-Pacific Islander. Per-attack and per-feature breakdowns in Appendix~\ref{sec:disparity_perfeature_appendix}. (Adult, 10k rows, $\text{QI}_{\text{demo}}$; RF.)}
  \label{tab:disparate_impact}
  \begin{tabular}{l r r r r r r r r r}
    \toprule
    & & & \multicolumn{5}{c}{Mean row $R_{adv}$ by race (\%)} & \multicolumn{2}{c}{By sex (\%)} \\
    \cmidrule(lr){4-8}\cmidrule(lr){9-10}
    \textbf{SDG} & \textbf{Mean (\%)} & \textbf{Outlier $\times$} & \textbf{AI/AN} {\footnotesize(1.1\%)} & \textbf{API} {\footnotesize(3.1\%)} & \textbf{Black} {\footnotesize(9.9\%)} & \textbf{White} {\footnotesize(85.3\%)} & \textbf{Other} {\footnotesize(0.7\%)} & \textbf{Female} {\footnotesize(32.1\%)} & \textbf{Male} {\footnotesize(67.9\%)} \\
    \midrule
    Cell Supp.              & 33.1 & $\times$2.3 & 4.1 & \textbf{52.7} & \textbf{61.0} & 29.8 & 0.4 & \textbf{42.2} & 28.8 \\
    TabDDPM                 & 1.4  & $\times$3.4 & \textbf{8.4} & 3.9  & 1.9  & 1.2  & 0.9 & 1.7  & 1.3  \\
    Synthpop                & 1.6  & $\times$1.0 & 2.2  & 1.2  & 1.2  & \textbf{1.7}  & 0.4 & 1.5  & \textbf{1.7}  \\
    MST $(\varepsilon{=}1)$    & 0.25 & $\times$1.0 & 0.17 & \textbf{0.27} & 0.19 & \textbf{0.25} & 0.16 & 0.24 & \textbf{0.25} \\
    MST $(\varepsilon{=}1000)$ & 0.28 & $\times$1.3 & 0.20 & \textbf{0.40} & 0.21 & \textbf{0.29} & \textbf{0.31} & 0.27 & \textbf{0.29} \\
    \bottomrule
  \end{tabular}
\end{table*}

Two things hold across all three datasets. First, every paradigm draws the same map of exposure: de-identification is wide open (CellSuppression and RankSwap NNDR up to $1.00$), TabDDPM is meaningfully but less exposed, and both MST budgets sit at chance ($\approx0.5$) irrespective of $\varepsilon$. So do PrivateGSD, PrivBayes, PrivSyn, and MWEM-PGM (AUC $0.49$--$0.56$ on every dataset and budget). Their flat membership scores cast no doubt on the reconstruction plateau of Section~\ref{sec:whichgenerator}, which reconstruction attacks measure directly. The distance attacks are not blind, either: on CDC's 1k rows, PrivateGSD's NNDR climbs from $0.51$ to $0.55$ with the budget. Second (and this is the point of the reduction), RA-as-MIA traces that same map using reconstruction alone: where dedicated MIA succeeds it succeeds, and where MIA fails (MST) it fails too. Reconstruction success and membership read the same leakage, which is why the reduction works at all.

What it does \emph{not} read is the same \emph{individuals}. The agreement is aggregate, not per-record: for TabDDPM training records, RA-as-MIA and SynthDistance correlate at only Spearman $\rho{=}0.41$, and of $10{,}000$ members, $3{,}269$ are flagged \emph{discordantly}: ranked above the median by one attack and below it by the other. The two attacks succeed on overlapping but distinct subpopulations. Records flagged only by SynthDistance are typical people ($2\%$ QI outliers, against $10\%$ overall) with rare hidden values and a near-exact copy in the release; records flagged only by RA-as-MIA are atypical ($19\%$ outliers), more often women and racial minorities, with common hidden values and no exact copy. Distance attacks find the records a generator copied, reconstruction the unusual people it modeled without copying, so an audit that runs only one will systematically miss the people the other would catch; neither subsumes the other. RA-as-MIA's discriminative power is itself concentrated on outliers (AUC $0.71$ on QI outliers vs.\ $0.63$ on the rest), the same concentration we turn to next.

\subsection{Disparate Impact: Who Is Most at Risk?}
\label{sec:disparate}
A mean hides its tail, and the tail is what matters once the question turns from \emph{whether} a release leaks (the memorization gap, Section~\ref{sec:memorization}) to \emph{whom} it exposes.\footnote{The memorization gap is itself a mean, deliberately: it asks whether a release memorizes \emph{systematically}, for which averaging over members and non-members is the natural estimator. It can absorb a few memorized individuals without trace, so we ask who is exposed here, from the row-level scores of the same attack runs.} Risk concentrates on atypical records (\citet{meeus2023achilles} call them ``Achilles' heels'' in synthetic data, and \citet{yaghini2022disparate} show disparate membership-inference vulnerability across demographic groups), and we ask how that concentration behaves across the reconstruction spectrum. We score each individual by feature-averaged row-level $R_{adv}$ and label as \emph{outliers} the $10\%$ whose quasi-identifier profile is most atypical: those an Isolation Forest, fit on the public QI columns alone, most readily isolates from the rest of the population. Outlier status is therefore a property of how unusual a person looks to the adversary \emph{before} any attack, not of the hidden values or whether reconstruction succeeds. The \textbf{outlier penalty} (the ``Outlier~$\times$'' column of Table~\ref{tab:disparate_impact}) is then the mean row-level $R_{adv}$ on the outlier $10\%$ divided by the mean on the typical $90\%$ ($1$ = even; above $1$ = unusual individuals reconstructed more accurately). The remaining columns break the same score down by race and sex. One scale caveat applies: row-level $R_{adv}$ weights a single person's values by rarity, so most rows score far below the population aggregates of Section~\ref{sec:taxonomy}; compare \emph{within} the table, not against Table~\ref{tab:ra_mean_adult}.

The disparity is real, large, and \emph{opposite in direction} depending on the mechanism, which is why we report it per generator rather than per attack: the protection mechanism, not the attacker's algorithm, decides who is exposed. Under high-fidelity synthesis, rarity is a liability. TabDDPM reconstructs outliers $3.4\times$ as well as typical records, and Amer.-Indian/Alaska Native individuals ($1.1\%$ of the data) at $7\times$ the rate of the White majority ($8.4\%$ vs.\ $1.2\%$): few synthetic records share a minority QI pattern, so each maps back to a single real person. Suppression inverts the sign: CellSuppression \emph{removes} the most heavily suppressed minority records (AI/AN individuals score $4.1\%$, against $29.8\%$ for White) while leaving the majority and partially-retained groups exposed (Black $61.0\%$, API $52.7\%$). Differential privacy alone spreads risk evenly: both MST variants keep every subgroup within $1.5\times$ of their mean, the equity its guarantee promises.

This concentration is a property of the \emph{release}, not of our attacker: from simple in-isolation classifiers to CoBP-RA, the outlier penalty holds near ${\times}2.2$ for CellSuppression and ${\times}2.6$--$2.9$ for TabDDPM, while Synthpop stays between ${\times}0.9$ and ${\times}1.3$. A stronger adversary recovers more on \emph{average} but does not change \emph{who} is exposed. Per-attack and per-feature breakdowns are in Appendix~\ref{sec:disparity_perfeature_appendix} (Tables~\ref{tab:disparity_perattack}~and~\ref{tab:disparity_perfeature}).

The lesson is not that one mechanism is fairer but that \emph{mean $R_{adv}$ is silent on this axis}: diffusion endangers the rare, suppression the common, and only per-subgroup, per-feature reporting tells them apart.

\begin{tcolorbox}[
    enhanced,
    title={\faLightbulb\quad Takeaway},
colbacktitle=takeawaytitle,
coltitle=white,
colback=takeawayfill,
colframe=takeawayframe,
    fonttitle=\bfseries,
    arc=3pt,
    boxrule=1.2pt,
    left=6pt, right=6pt, top=4pt, bottom=4pt
]
A single $R_{adv}$ hides three different things. Most reconstruction is not a breach but \emph{distributional inference} (true of anyone like the target); genuine memorization is confined to record-retaining de-identification and, among synthesizers, diffusion. Reconstruction ranks releases as \emph{membership inference} does, yet flags different people: distance attacks catch the records a generator copied, reconstruction the unusual people it modeled. And the same score can endanger \emph{opposite} individuals: diffusion concentrates risk on the rare, suppression on the common. An aggregate score is where an audit starts, never where it ends.
\end{tcolorbox}

\section{
Implications for Auditing and Policy}\label{sec:policy}

Across three sections one ordering held: the release governs how much can be reconstructed, and the attacker only realizes it. For a practitioner that is reassuring (risk is chosen at synthesis time, not imposed later by an adversary's ingenuity), but the metrics in widest use see neither how much a learning adversary recovers nor the \emph{shape} of the risk: who is exposed, which attribute leaks, and whether the leakage is memorization a generator could remove or distributional inference no useful release can avoid. We close by translating these findings into auditing practice and marking what they leave open.

\subsection{Guidance for Privacy Auditing}

\paragraph{Audit with a learning adversary, not a matching one.}
The standard of practice for disclosure control remains matching-based: equivalence-class risk in the El~Emam tradition~\cite{el_emam2011evaluating} and the similarity/distance scores shipped with synthetic-data tooling. Our calibrated ladder (Section~\ref{sec:disclosure_risk}) shows what these miss: the El~Emam score ranks RankSwap, whose retained records measurably leak individuals, below two synthesizers that do not, because learning attacks generalize \emph{across} equivalence classes rather than requiring exact QI matches; \citet{ganev_inadequacy_2025} reach the same verdict for similarity metrics even under differential privacy. The recommendation (already operationalized by the NIST CRC) is to audit with the strongest available attribute-inference attack, not a distributional-distance proxy. Frameworks are converging on attack-based evaluation~\cite{stadler2022synthetic,houssiau2022tapas,giomi2023unified}; our taxonomy and benchmark contribute a ranked attack suite, with CoBP-RA a strong, inexpensive default.

\paragraph{Report the distribution of risk, not its mean.}
A single mean $R_{adv}$ conceals the two things a regulator most needs. The first is \emph{mechanism}: our memorization test (motivated by \emph{narcissus resiliency}~\cite{cohen2024data}) separates individual memorization, which a better generator can remove, from distributional inference, which no informative release can. The second is \emph{incidence}: diffusion-based synthesis can post a reassuring mean while exposing demographically rare individuals several-fold (Section~\ref{sec:disparate}), and because that concentration is a property of the release, it cannot be audited away with a weaker attacker. Both diagnostics (the train-versus-holdout gap and per-subgroup, per-feature scores) fall out of the same attack run at no extra cost; we recommend reporting both as standard.

\paragraph{Reconstruction and membership are complementary.}
Synthetic-data auditing has standardized on membership inference~\cite{shokri2017membership,carlini2022membership}, but the two answer different questions and, as Section~\ref{sec:mia} shows, flag different individuals: roughly a third of records are scored discordantly. Membership asks whether a person was in the data; reconstruction asks what can be recovered about them, and EU anonymisation guidance names this \emph{inference} risk, alongside singling out and linkability, as one an anonymised release must resist~\cite{wp29_2014anonymisation}. A complete audit should run both.

\subsection{Implications for Release Mechanisms}

That de-identification carries no worst-case bound is not new~\cite{sweeney2002k,machanavajjhala2007diversity}; our contribution is making that failure \emph{measurable} and characterizing its shape against synthetic data's. The leakage of CellSuppression and RankSwap is record-retention memorization (visible as a large train--holdout gap), borne unevenly: it falls on the majority and partially-retained minority groups while protecting the smallest, most heavily suppressed ones. Our diagnostics let an auditor see this in a specific release, not infer it from first principles.

Differential privacy is the only family in our study that is at once memorization-free (all six mechanisms, at every budget we test, Table~\ref{tab:memorization_and_ds_risk}) and equitable across subgroups (the privacy mirror of \citet{bagdasaryan2019differential}'s disparate-\emph{utility} finding: the same noise that degrades a rare subgroup's utility also denies an attacker its sharp conditional structure). Those two properties, not any particular level of $R_{adv}$, are what DP reliably buys, at $\varepsilon{=}1$ and $\varepsilon{=}1000$ alike. Reconstruction protection differs: it is bought almost entirely below $\varepsilon{\approx}10$; past that, \emph{mechanism} matters more than budget, since at matched $\varepsilon$ the six spread by several points and which leaks most changes with the dataset (Section~\ref{sec:whichgenerator}). The practical order follows: choose the mechanism and measure it on the data at hand, spend budget only where it still moves the measured curve, and treat anything far above $\varepsilon{\approx}10$ as a formal guarantee bought for risk that barely moves.

\subsection{Limitations and Open Problems}\label{sec:limitations}

\paragraph{Scope of the threat model.} Our findings cover the two branches we study, not model inversion (the adversary holds the trained model) or reconstruction from aggregate statistics, the setting of the Census reconstruction theorem~\cite{dinur2003revealing,garfinkel2019understanding}.

\paragraph{The plateau is empirical.} The $\varepsilon$ plateau of Section~\ref{sec:whichgenerator} is a regularity of the marginal-based mechanisms and datasets we test, not a law: DP-SGD deep generators and DP language-model synthesis remain untested, and it holds where membership risk keeps climbing~\cite{golob_privacy_2025}.

\paragraph{An open taxonomy.} Figure~\ref{fig:taxonomy} is a snapshot of the attacks that exist, not a closed system. We populate several of its leaves ourselves, but new leaves will grow as existing ideas are refined, and new methodologies may earn branches we have not drawn.

\paragraph{Continuous-feature synthesis.} The continuous case (California, Section~\ref{sec:conditions}) is evaluated with a subset of attacks under NRMSE. It raises distinct challenges: DP discretization destroys utility at low $\varepsilon$, GANs and VAEs struggle with sharp continuous marginals, and there is no clean rarity-weighting analog for real-valued outputs. A complete treatment remains open.

\paragraph{LLM-based synthesis.} Large language models now generate tables too, fine-tuned on serialized records~\cite{borisov2023great}, fine-tuned under differential privacy~\cite{tran2024dpllmtgen}, or prompted through an API alone~\cite{swanberg2025api}. Our attacks evaluate them unchanged, since they consume only the published table (a foundation model, TabPFN, already sits on the attack side). What would change is the risk profile. A non-private fine-tuned LLM is a high-capacity model trained directly on records, the profile that made TabDDPM memorize individuals (Section~\ref{sec:memorization}) and underlies training-data extraction from language models~\cite{carlini2021extracting}; the memorization gap is thus the natural first audit. Under DP the evidence is mixed~\cite{tran2024dpllmtgen,swanberg2025api}; whether the reconstruction plateau survives a mechanism that never reduces data to noisy marginals is the sharpest open test of that finding.

\paragraph{Access, quasi-identifiers, and scale.} Three axes we held fixed each bound our optimism. All our attacks are \emph{black-box}; white-box access to model weights or training metadata (MST's selected marginals, TabDDPM's score function) would enable stronger attacks. Each experiment assumes the auditor knows the adversary's \emph{QI set} exactly, yet Section~\ref{subsec:qi} shows its composition strongly shapes risk, so a mis-specified audit QI can mis-estimate exposure; auditing when the adversary's QI is unknown is open. Our largest benchmark has $130$ features, while genomics, EHR, and transaction data reach thousands; several of our strongest attacks (CoBP-RA, CondMST, diffusion) scale poorly in feature count, leaving efficient high-dimensional auditing an open systems problem.

\paragraph{From $R_{adv}$ to legal thresholds.} $R_{adv}$ captures far more exposure than classical metrics (Section~\ref{sec:disclosure_risk}), but does not yet map onto HIPAA Expert Determination or GDPR identifiability. Formalizing that relationship is the bridge regulated deployments most need.

\paragraph{Toward a standard.} The natural endpoint is a standardized, attack-based auditing benchmark: a fixed, ranked suite of reconstruction \emph{and} membership attacks with agreed reporting (mean exposure, the memorization gap, and per-subgroup incidence) that a data holder at an agency such as NIST or the Census Bureau could run before release. Our taxonomy, attacks, and metrics are a step toward it; what remains is consensus on protocol, not a question of possibility.

\begin{acks}
This research was supported by NSF \#2451163, \#2523406, NIH 1OT2OD032581, NSF NAIRR 240485 (Cloudbank AWS), and NSF NAIRR 240091 (TACC Frontera). Steven Golob is supported by an NSF CSGrad4US Fellowship. Sikha Pentyala is affiliated with the eScience Institute. We would like to thank the organizers of the NIST Privacy Collaborative Research Cycle, Christine Task and Gary Howarth, for designing and hosting the competition that motivated this work.
\end{acks}

\begin{ethics}
This work develops attacks for a defensive purpose: to measure reconstruction risk so that data holders and auditors can quantify and reduce it. All experiments use public benchmark datasets (Adult, IPUMS 1940 census microdata, California Housing, the NIST SBO extract, and the CDC BRFSS Diabetes Health Indicators); we collect no new personal data, and we attack only synthetic and de-identified releases that we generated from these datasets, never a deployed system or data product. The methodology was developed within the NIST Privacy Collaborative Research Cycle, a sanctioned red-team/blue-team program. The same techniques could in principle aid an adversary, but we judge the benefit of disclosure to outweigh this risk: we report no vulnerability in any fielded system, and every result comes with auditing guidance and the finding that differential privacy mitigates the exposure we measure.
\end{ethics}

\begin{openscience}
In keeping with PoPETs' open-science policy, all code for the attacks, generators, experiment harness, and evaluation is publicly released, along with the configuration files and hyperparameters (Appendix~\ref{sec:hyperparams}) needed to reproduce every table in this paper. All five datasets are publicly available from their original sources (Table~\ref{tab:datasets}). The repository also includes a runnable example of adding a new attack, and documents how to add a generator or a dataset.
\par\noindent\textbf{Code:}~\url{https://github.com/steveng9/Reconstruction}
\par\noindent\textbf{Archived:}~\url{https://doi.org/10.5281/zenodo.22701232}
\end{openscience}

\begin{ai}
The authors used Claude Sonnet 4.6 and Opus 5.0 (Anthropic) to revise manuscript text, install external libraries, run experiments, and query result databases. All AI-generated text was reviewed, edited for accuracy, and verified against experimental results by the authors. All scientific judgments, experimental designs, original NIST CRC participation and interpretations are the authors' own. We have manually verified and are responsible for the accuracy, originality, and integrity of the output of all AI-based tools.
\end{ai}

\bibliographystyle{ACM-Reference-Format}
\bibliography{main}

\appendix

\section{Preliminaries}

\begin{table}[t!]
  \centering
  \small
  \caption{Notation, in order of appearance. Symbols an attack lets you \emph{tune} are not repeated here; they appear, with the value we report and its alternatives, in Table~\ref{tab:hyperparams} (Appendix~\ref{sec:hyperparams}).}
  \label{tab:notation}
  \begin{tabular}{@{}p{0.28\columnwidth}p{0.68\columnwidth}@{}}
    \toprule
    \textbf{Symbol} & \textbf{Meaning} \\
    \midrule
    \multicolumn{2}{@{}l}{\emph{Problem and scoring (Sections~\ref{sec:formulation}--\ref{sec:dp})}} \\
    $D_{\text{train}}$; $D_{\text{syn}}$ & training data; release (synthetic or de-identified) \\
    $\mathcal{G}$ & generator, $D_{\text{syn}} = \mathcal{G}(D_{\text{train}})$ \\
    $\mathcal{M}$; $\varepsilon$ & DP mechanism; privacy budget \\
    $n$; $p$ & training records (the targets); features \\
    $\mathcal{F}$; $Q$; $H$ & all features; quasi-identifiers; hidden features \\
    $m$ & number of hidden features, $|H|$ \\
    $q$; $h$ & a QI feature; a hidden feature \\
    $x_i$; $x^Q_i$; $x^h_i$ & target $i$; its QI values; its value of $h$ \\
    $\hat{x}^H_i$, $\hat{x}^h_i$ & the attack's guesses \\
    $f$ & reconstruction attack, $(D_{\text{syn}}, x^Q_i) \mapsto \hat{x}^H_i$ \\
    $R_{adv}$ & rarity-weighted reconstruction advantage (Eq.~\ref{eqn:scoring}) \\
    $R_{adv}^{\mathrm{tr}}$; $R_{adv}^{\mathrm{NT}}$; $\Delta$ & $R_{adv}$ on training, non-training records; gap \\
    $C$ & number of distinct values of a feature \\
    $d_S$ & disclosure score of \citet{el_emam2020identity} \\
    $s(x)$ & membership score of record $x$ \\
    $k$ & in $k$-anonymity, $k$-NN, PrivBayes (unrelated) \\
    \multicolumn{2}{@{}l}{\emph{All pseudocode}} \\
    $a$, $b$; $v$, $w$ & features; values \\
    $\mathcal{X}_a$ & domain of feature $a$ ($C = |\mathcal{X}_a|$) \\
    $I(a;b)$ & mutual information in $D_{\text{syn}}$ \\
    $i$ & target index \\
    $\theta$ & network weights \\
    \multicolumn{2}{@{}l}{\emph{CondMST (Algorithm~\ref{alg:condmst})}} \\
    $\hat{P}$ & graphical model fit to $D_{\text{syn}}$ \\
    $\sigma$ & MST's measurement noise (set to $0$) \\
    \multicolumn{2}{@{}l}{\emph{Chaining (Appendix~\ref{sec:chaining})}} \\
    $\mathcal{A}$; $\mathcal{A}_\ell$ & supervised learner; its $\ell$-th chained predictor \\
    \multicolumn{2}{@{}l}{\emph{CondDDPM, CondRePaint (Algorithms~\ref{alg:condddpm}, \ref{alg:condrepaint})}} \\
    $c_i$ & target $i$'s encoded QI, as a vector in $\mathbb{R}^p$ \\
    $\mathbf{1}_Q$; $\odot$ & indicator of the QI coordinates; elementwise product \\
    $t$, $t'$ & diffusion timestep indices \\
    $\bar\alpha_t$ & noise schedule \\
    $\xi$; $\hat\xi_\theta$ & Gaussian noise; network that predicts it \\
    $\mathrm{noise}_t$; $\mathrm{step}_\theta$ & noise to level $t$; one reverse (denoising) step \\
    $\mathrm{decode}$ & map a vector back to feature values \\
    \multicolumn{2}{@{}l}{\emph{CoBP-RA (Algorithm~\ref{alg:cobpra})}} \\
    $\partial a$ & neighbors of feature $a$ in the graph \\
    $D^K_i$ & the $K$ records of $D_{\text{syn}}$ nearest to $x^Q_i$ \\
    $P_D$ & empirical distribution in records $D$ (smoothed) \\
    $\phi_a$; $\psi_{ab}$ & node potential; edge potential \\
    $\mu_{a\to b}$ & message from $a$ to $b$ \\
    $\beta_{a\setminus b}$; $\gamma_{a\setminus b}$ & belief of $a$ without $b$'s message; its tempering exponent \\
    $\mathrm{Ent}$ & entropy \\
    $\beta_{ih}$; $\bar\beta_h$ & belief of target $i$ about $h$; its mean over targets \\
    \multicolumn{2}{@{}l}{\emph{MultiHeadMLP, ARFFormer (Algorithms~\ref{alg:multiheadmlp}, \ref{alg:arfformer})}} \\
    $g_\theta$; $Z_\theta$ & shared layers; causal transformer \\
    $W_h$, $W_\ell$; $\pi_h$, $\pi_\ell$ & output heads; their predicted distributions \\
    $L$ & training loss \\
    $\ell$ & position in the order $h_1,\dots,h_m$ \\
    $\varnothing$ & padding token \\
    $u$; $x^{(u)}$ & guess index; one guess at a target \\
    \bottomrule
  \end{tabular}
\end{table}

\subsection{Dataset Details}\label{app:datasets}

The five datasets (summarized in Table~\ref{tab:datasets}) span the data domains and structural regimes most relevant to disclosure control; all are public benchmarks. \textbf{Adult}~\cite{adult1996} is the 1994 US Census income extract, a standard machine-learning, fairness, and privacy benchmark of mixed demographic and employment features. \textbf{NIST Arizona}~\cite{nist_arizona} is IPUMS 1940 Arizona census microdata; we use the 25-feature subset released for the NIST CRC, with the competition's two quasi-identifier sets ($\text{QI}_1$, $\text{QI}_2$; Table~\ref{tab:qi_def_arizona}). \textbf{California Housing}~\cite{california_housing} is 1990 census block-group housing data and our only fully-continuous dataset, scored by NRMSE rather than rarity-weighted accuracy. \textbf{NIST SBO}~\cite{nist_sbo} is Survey-of-Business-Owners public-use microdata, our most feature-rich dataset ($130$ mostly-binary firm-level financial and demographic attributes). \textbf{CDC Diabetes}~\cite{cdc_diabetes} is the CDC BRFSS Diabetes Health Indicators set, $22$ mostly-binary health features. Per-dataset quasi-identifier memberships are tabulated in Appendix~\ref{sec:qi_appendix} (Tables~\ref{tab:qi_def_adult}--\ref{tab:qi_def_sbo}).

\providecommand{\qirot}[1]{\rotatebox{90}{\small\texttt{#1}}}

\begin{table*}[h!]
  \centering
  \small
  \caption{%
    QI membership for the \textit{Adult} dataset.
    $\checkmark$ = known to adversary (quasi-identifier); blank = hidden (must be reconstructed).
    Features are grouped by the tier at which they first enter a QI.
    (15 features, 4 QI variants.)
  }
  \label{tab:qi_def_adult}
  \setlength{\tabcolsep}{4pt}%
  \begin{tabular}{@{}l ccc ccc cccc cc ccc@{}}
    \toprule
    & \multicolumn{3}{c}{tiny core}
    & \multicolumn{3}{c}{$+$\;$\text{QI}_{\text{demo}}$}
    & \multicolumn{4}{c}{$+$\;$\text{QI}_{\text{large}}$}
    & \multicolumn{2}{c}{beh.\ only}
    & \multicolumn{3}{c}{hidden} \\
    \cmidrule(lr){2-4}\cmidrule(lr){5-7}\cmidrule(lr){8-11}\cmidrule(lr){12-13}\cmidrule(l){14-16}
    QI ($|\text{QI}|$)
      & \qirot{age}
      & \qirot{sex}
      & \qirot{race}
      & \qirot{education}
      & \qirot{marital-status}
      & \qirot{native-country}
      & \qirot{occupation}
      & \qirot{workclass}
      & \qirot{relationship}
      & \qirot{hours-per-week}
      & \qirot{education-num}
      & \qirot{fnlwgt}
      & \qirot{capital-gain}
      & \qirot{capital-loss}
      & \qirot{income} \\
    \midrule
    $\text{QI}_{\text{tiny}}$  (3)  & $\checkmark$ & $\checkmark$ & $\checkmark$ &  &  &  &  &  &  &  &  &  &  &  &  \\
    $\text{QI}_{\text{demo}}$  (6)  & $\checkmark$ & $\checkmark$ & $\checkmark$ & $\checkmark$ & $\checkmark$ & $\checkmark$ &  &  &  &  &  &  &  &  &  \\
    $\text{QI}_{\text{large}}$ (10) & $\checkmark$ & $\checkmark$ & $\checkmark$ & $\checkmark$ & $\checkmark$ & $\checkmark$ & $\checkmark$ & $\checkmark$ & $\checkmark$ & $\checkmark$ &  &  &  &  &  \\
    $\text{QI}_{\text{beh.}}$  (6)  &  &  &  &  &  &  & $\checkmark$ & $\checkmark$ & $\checkmark$ & $\checkmark$ & $\checkmark$ & $\checkmark$ &  &  &  \\
    \bottomrule
  \end{tabular}
\end{table*}

\begin{table*}[h!]
  \centering
  \small
  \caption{%
    QI membership for the \textit{CDC Diabetes} dataset.
    $\checkmark$ = known to adversary; blank = hidden.
    \texttt{Smoker} and \texttt{PhysActivity} (under ``$+$\;demo'') and all six ``$+$\;large''
    features also appear in $\text{QI}_{\text{beh.}}$.
    \texttt{HvyAlcohol} = \texttt{HvyAlcoholConsump};
    \texttt{HeartDisease} = \texttt{HeartDiseaseorAttack}.
    (22 features, 4 QI variants.)
  }
  \label{tab:qi_def_cdc}
  \setlength{\tabcolsep}{4pt}%
  \begin{tabular}{@{}l cccc cccccc cccccc cc cccc@{}}
    \toprule
    & \multicolumn{4}{c}{tiny core}
    & \multicolumn{6}{c}{$+$\;$\text{QI}_{\text{demo}}$}
    & \multicolumn{6}{c}{$+$\;$\text{QI}_{\text{large}}$}
    & \multicolumn{2}{c}{beh.\ only}
    & \multicolumn{4}{c}{hidden} \\
    \cmidrule(lr){2-5}\cmidrule(lr){6-11}\cmidrule(lr){12-17}\cmidrule(lr){18-19}\cmidrule(l){20-23}
    QI ($|\text{QI}|$)
      & \qirot{Sex}
      & \qirot{Age}
      & \qirot{BMI}
      & \qirot{GenHlth}
      & \qirot{Education}
      & \qirot{Income}
      & \qirot{HighBP}
      & \qirot{HighChol}
      & \qirot{Smoker}
      & \qirot{PhysActivity}
      & \qirot{Fruits}
      & \qirot{Veggies}
      & \qirot{CholCheck}
      & \qirot{DiffWalk}
      & \qirot{AnyHealthcare}
      & \qirot{NoDocbcCost}
      & \qirot{HvyAlcohol}
      & \qirot{HeartDisease}
      & \qirot{Stroke}
      & \qirot{Diabetes\_binary}
      & \qirot{MentHlth}
      & \qirot{PhysHlth} \\
    \midrule
    $\text{QI}_{\text{tiny}}$  (4)  & $\checkmark$ & $\checkmark$ & $\checkmark$ & $\checkmark$ &  &  &  &  &  &  &  &  &  &  &  &  &  &  &  &  &  &  \\
    $\text{QI}_{\text{demo}}$  (10) & $\checkmark$ & $\checkmark$ & $\checkmark$ & $\checkmark$ & $\checkmark$ & $\checkmark$ & $\checkmark$ & $\checkmark$ & $\checkmark$ & $\checkmark$ &  &  &  &  &  &  &  &  &  &  &  &  \\
    $\text{QI}_{\text{large}}$ (16) & $\checkmark$ & $\checkmark$ & $\checkmark$ & $\checkmark$ & $\checkmark$ & $\checkmark$ & $\checkmark$ & $\checkmark$ & $\checkmark$ & $\checkmark$ & $\checkmark$ & $\checkmark$ & $\checkmark$ & $\checkmark$ & $\checkmark$ & $\checkmark$ &  &  &  &  &  &  \\
    $\text{QI}_{\text{beh.}}$  (10) &  &  &  &  &  &  &  &  & $\checkmark$ & $\checkmark$ & $\checkmark$ & $\checkmark$ & $\checkmark$ & $\checkmark$ & $\checkmark$ & $\checkmark$ & $\checkmark$ & $\checkmark$ &  &  &  &  \\
    \bottomrule
  \end{tabular}
\end{table*}

\begin{table*}[h!]
  \centering
  \small
  \caption{%
    QI membership for the \textit{NIST Arizona} dataset, 25-feature competition subset.
    $\checkmark$ = known to adversary; blank = hidden.
    $\text{QI}_1$ (cultural/identity attributes) and $\text{QI}_2$ (geographic/mobility) are the
    Privacy CRC competition's own sets and share \texttt{RACE} and \texttt{SEX}; the column groups
    describe these two. $\text{QI}_{\text{medium}}$, used only in the RA-as-MIA experiment
    (Section~\ref{sec:mia}), additionally knows \texttt{AGE} and \texttt{GQ}.
    (25 IPUMS features, 3 QI variants.)
  }
  \label{tab:qi_def_arizona}
  \setlength{\tabcolsep}{4pt}%
  \begin{tabular}{@{}l cc ccccc ccccc ccccccccccccc@{}}
    \toprule
    & \multicolumn{2}{c}{both QIs}
    & \multicolumn{5}{c}{$\text{QI}_1$ only}
    & \multicolumn{5}{c}{$\text{QI}_2$ only}
    & \multicolumn{13}{c}{hidden} \\
    \cmidrule(lr){2-3}\cmidrule(lr){4-8}\cmidrule(lr){9-13}\cmidrule(l){14-26}
    QI ($|\text{QI}|$)
      & \qirot{RACE} & \qirot{SEX}
      & \qirot{AGEMARR} & \qirot{GQTYPE} & \qirot{IND} & \qirot{MTONGUE} & \qirot{VETSTAT}
      & \qirot{BPL} & \qirot{FARM} & \qirot{HISPAN} & \qirot{LABFORCE} & \qirot{MIGRATE5}
      & \qirot{AGE} & \qirot{CITIZEN} & \qirot{DURUNEMP} & \qirot{EDUC} & \qirot{EMPSTAT}
      & \qirot{FAMSIZE} & \qirot{GQ} & \qirot{INCWAGE} & \qirot{MARST} & \qirot{NATIVITY}
      & \qirot{OWNERSHP} & \qirot{URBAN} & \qirot{WKSWORK1} \\
    \midrule
    $\text{QI}_1$ (7) & $\checkmark$ & $\checkmark$ & $\checkmark$ & $\checkmark$ & $\checkmark$ & $\checkmark$ & $\checkmark$ &  &  &  &  &  &  &  &  &  &  &  &  &  &  &  &  &  &  \\
    $\text{QI}_2$ (7) & $\checkmark$ & $\checkmark$ &  &  &  &  &  & $\checkmark$ & $\checkmark$ & $\checkmark$ & $\checkmark$ & $\checkmark$ &  &  &  &  &  &  &  &  &  &  &  &  &  \\
    $\text{QI}_{\text{medium}}$ (12) & $\checkmark$ & $\checkmark$ & $\checkmark$ & $\checkmark$ & $\checkmark$ & $\checkmark$ & $\checkmark$ & $\checkmark$ & & $\checkmark$ & $\checkmark$ & &  $\checkmark$ & & & & $\checkmark$\\
    \bottomrule
  \end{tabular}
\end{table*}

\begin{table*}[h!]
  \centering
  \small
  \caption{%
    QI membership for the \textit{California Housing} dataset.
    $\checkmark$ = known to adversary; blank = hidden.
    $\text{QI}_1$: geographic and locational features only.
    $\text{QI}_{\text{large}}$: adds structural housing features, hiding only the three economic outcomes.
    (9 continuous features, 2 QI variants.)
  }
  \label{tab:qi_def_california}
  \begin{tabular}{@{}l cccc cc ccc@{}}
    \toprule
    & \multicolumn{4}{c}{$\text{QI}_1$}
    & \multicolumn{2}{c}{$+\,\text{QI}_{\text{large}}$}
    & \multicolumn{3}{c}{hidden} \\
    \cmidrule(lr){2-5}\cmidrule(lr){6-7}\cmidrule(l){8-10}
    QI ($|\text{QI}|$)
      & \qirot{Latitude} & \qirot{Longitude} & \qirot{HouseAge} & \qirot{Population}
      & \qirot{AveBedrms} & \qirot{AveRooms}
      & \qirot{MedInc} & \qirot{AveOccup} & \qirot{MedHouseVal} \\
    \midrule
    $\text{QI}_1$ (4)             & $\checkmark$ & $\checkmark$ & $\checkmark$ & $\checkmark$ &              &              &  &  &  \\
    $\text{QI}_{\text{large}}$ (6) & $\checkmark$ & $\checkmark$ & $\checkmark$ & $\checkmark$ & $\checkmark$ & $\checkmark$ &  &  &  \\
    \bottomrule
  \end{tabular}
\end{table*}

\begin{table*}[h!]
  \centering
  \small
  \renewcommand{\arraystretch}{1.25}
  \caption{%
    QI membership for the \textit{NIST SBO} dataset. Under $\text{QI}_{\text{demo}}$, the only variant
    evaluated for SBO, the adversary knows 7 firm-identifier and owner-demographic features (the first
    two rows) and the remaining 123 are hidden. Features are grouped by category for readability;
    a dash (---) marks a category with nothing known.
    (130 features in 16 categories, 1 QI variant.)
  }
  \label{tab:qi_def_sbo}
  \resizebox{\textwidth}{!}{%
  \begin{tabular}{@{}l r p{2.7cm} p{10.3cm}@{}}
    \toprule
    Feature category & $N$ & Known ($\text{QI}_{\text{demo}}$, 7) & Hidden (123) \\
    \midrule
    Business identifiers &  6
      & {\small\ttfamily FIPST, SECTOR, N07\_EMPLOYER, RG}
      & {\small\ttfamily TABWGT, PCT1} \\
    Owner demographics &  4
      & {\small\ttfamily SEX1, RACE1, ETH1}
      & {\small\ttfamily VET1} \\
    \midrule
    Business acquisition &  6
      & ---
      & {\small\ttfamily FOUNDED1, PURCHASED1, INHERITED1, RECEIVED1, ACQUIRENR1, ACQYR1} \\
    Owner role \& background & 12
      & ---
      & {\small\ttfamily PROVIDE1, MANAGE1, FINANCIAL1, FNCTNABV1, FNCTNR1, HOURS1, PRMINC1, SELFEMP1, EDUC1, AGE1, BORNUS1, DISVET1} \\
    Basic business & 4
      & ---
      & {\small\ttfamily ESTABLISHED, HOMEBASED, FRANCHISE, FRANCHISER50} \\
    Customer / market type & 8
      & ---
      & {\small\ttfamily FEDERAL, STATELOCAL, OTHERBUS, INDIVIDUALS, CUSTNR, EXPORTS, OPSOUTSIDE, OUTSOURCE} \\
    Languages spoken & 18
      & ---
      & {\small\ttfamily ENGLISH, ARABIC, CHINESE, FRENCH, GERMAN, GREEK, HINDI, ITALIAN, JAPANESE, KOREAN, POLISH, PORTUGUESE, RUSSIAN, SPANISH, TAGALOG, VIETNAMESE, LANGOTHER, LANGNR} \\
    Workforce composition & 7
      & ---
      & {\small\ttfamily FULLTIME, PARTTIME, DAYLABOR, TEMPSTAFF, LEASED, CONTRACTORS, EMPNR} \\
    Digital presence & 4
      & ---
      & {\small\ttfamily WEBSITE, ECOMMERCE, ECOMMPCT, ONLINEPURCH} \\
    Operating schedule / activity & 6
      & ---
      & {\small\ttfamily LT40HOURS, LT12MONTHS, SEASONAL, OCCASIONALLY, ACTIVITYNABV, ACTIVITYNR} \\
    Ownership structure & 3
      & ---
      & {\small\ttfamily HUSBWIFE, FAMILYBUS, NUMOWNERS} \\
    Financial outcomes & 3
      & ---
      & {\small\ttfamily EMPLOYMENT\_NOISY, PAYROLL\_NOISY, RECEIPTS\_NOISY} \\
    Startup capital access (SC*) & 15
      & ---
      & {\small\ttfamily SCSAVINGS, SCASSETS, SCEQUITY, SCCREDIT, SCGOVTLOAN, SCGOVTGUAR, SCBANKLOAN, SCFAMLOAN, SCVENTURE, SCGRANT, SCOTHER, SCDONTKNOW, SCNONENEEDED, SCNOTREPORTED, SCAMOUNT} \\
    Expansion capital access (EC*) & 16
      & ---
      & {\small\ttfamily ECSAVINGS, ECASSETS, ECEQUITY, ECCREDIT, ECGOVTLOAN, ECGOVTGUAR, ECBANKLOAN, ECFAMLOAN, ECVENTURE, ECPROFITS, ECGRANT, ECOTHER, ECDONTKNOW, ECNOACCESS, ECNOEXPAND, ECNOTREPORTED} \\
    Employee benefits & 6
      & ---
      & {\small\ttfamily HEALTHINS, RETIREMENT, PROFITSHARE, HOLIDAYS, BENENABV, BENENR} \\
    Business cessation & 12
      & ---
      & {\small\ttfamily OPERATING, RETIRED, DECEASED, ONETIME, LOWSALES, NOBUSCRED, NOPERSCRED, STARTANOTHER, SOLDBUS, CEASEOTHER, CEASENR, CEASENA} \\
    \midrule
    \textit{Total} & \textit{130} & \textit{7} & \textit{123} \\
    \bottomrule
  \end{tabular}%
  }
\end{table*}

\subsection{Quasi-Identifier Definitions}\label{sec:qi_appendix}
Tables~\ref{tab:qi_def_adult}--\ref{tab:qi_def_sbo} enumerate features and quasi-identifier memberships for each of the five datasets evaluated in this paper.  For Adult and CDC~Diabetes (Tables~\ref{tab:qi_def_adult}--\ref{tab:qi_def_cdc}), features are shown as columns with a $\checkmark$ marking membership in each QI variant; blank = hidden.  NIST Arizona (Table~\ref{tab:qi_def_arizona}) uses the same format for the two competition QI variants, $\text{QI}_1$ and $\text{QI}_2$.  California Housing (Table~\ref{tab:qi_def_california}) has two QI variants.  For NIST~SBO (Table~\ref{tab:qi_def_sbo}), the 130 features are too numerous for a column-per-feature layout; instead, features are grouped into 16 semantic categories and QI membership is described per category.  $\text{QI}_{\text{demo}}$ for Adult, CDC, and SBO corresponds to \texttt{QI1} in the source code; Arizona's $\text{QI}_1$ and $\text{QI}_2$ keep the competition's names.

\subsection{De-Identification and Synthetic Data Generation Methods}\label{sec:sdg_descriptions}

The thirteen methods studied span three broad families.
\emph{De-ident\-ification (statistical disclosure limitation)} methods transform or suppress real records without generating new ones.
\emph{Statistical synthesis} methods fit explicit probabilistic models to the data and draw new records from those models; the DP variants add calibrated noise to the fitted statistics.
\emph{Deep generative synthesis} methods train neural networks to learn the data distribution and sample from the learned model.
Six methods carry a formal differential-privacy guarantee (MST, AIM, PrivBayes, PrivSyn, MWEM-PGM, and PrivateGSD, chosen for mechanistic diversity); all others are non-DP. The exact generation settings for every method (library defaults except where noted) are consolidated in Appendix~\ref{sec:hyperparams}.

\bigskip
\noindent\textbf{De-Identification (non-synthetic, non-DP)}

\paragraph{RankSwap.}
Records are sorted by a continuous key feature, partitioned into consecutive bins of fixed size, and within each bin the values of designated sensitive numerical columns are permuted among records~\cite{moore1996rankswap,templ2015sdcmicro}.
All non-swapped columns are released verbatim, so the output consists of \emph{real} records with certain attributes replaced.
RankSwap preserves marginal distributions of the swapped features but breaks individual-level linkages.
In our experiments the swapped columns are the continuous / high-cardinality numeric attributes for each dataset (e.g., for \textsc{Adult}: \texttt{age}, \texttt{fnlwgt}, \texttt{education-num}, \texttt{capital-gain}, \texttt{capital-loss}, \texttt{hours-per-week}).

\paragraph{CellSuppression.}
Records whose quasi-identifier combination falls into a group with fewer than $k$ members are removed entirely; the remaining records are released unmodified~\cite{willenborg2001elements}.
This enforces a form of $k$-anonymity on the released microdata.
The suppressed fraction can be substantial for datasets with many rare strata.
In our experiments we use $k=5$ and key variables that form the quasi-identifier of interest for each dataset (e.g., for \textsc{Adult}: \texttt{age}, \texttt{workclass}, \texttt{education}, \texttt{sex}, \texttt{race}, \texttt{native-country}).
Because suppression reduces record count, the released file is smaller than the original training set; attacks on CellSuppression synth operate on a reduced candidate pool.

\noindent\textbf{Statistical Synthesis: Differentially Private}

\paragraph{MST}
Maximum Spanning Tree~\cite{mckenna2021winning}, the winning entry in the 2018 NIST Differential Privacy Synthetic Data Challenge.
Selects a set of pairwise marginals by computing a maximum spanning tree of mutual information scores, measures those marginals with calibrated Gaussian noise to achieve $(\varepsilon, \delta)$-DP, then fits a graphical model (via Private-PGM~\cite{mckenna2019graphical}) and samples synthetic records from it.
We sweep $\varepsilon \in \{0.1, 0.3, 1, \dots, 1000\}$.

\paragraph{AIM}
Adaptive and Iterative Mechanism~\cite{mckenna2022aim}.
Extends MST with an online marginal-selection strategy: rather than fixing the marginal set upfront, AIM iteratively measures the marginal that reduces model error the most, allocating the privacy budget adaptively.
The same Private-PGM backend is used for graphical-model fitting and sampling.
AIM achieves higher utility than MST at matched $\varepsilon$ because it targets the most informative statistics first.

\paragraph{PrivBayes.}
Bayesian-network synthesis~\cite{zhang2017privbayes}. Greedily learns a network in which every attribute has at most $k$ parents, choosing each parent set with the exponential mechanism; it then releases noisy conditional distributions and samples records ancestrally through the network. We use $k{=}2$ at every budget.  

\paragraph{PrivSyn.}
Model-free marginal synthesis~\cite{zhang2021privsyn}. Measures noisy one- and two-way marginals, adjusts them so that marginals sharing a column agree on that column's counts, and then applies the Gradually Update Method (GUM): starting from records drawn from the one-way marginals, it repeatedly changes individual records' values until the record set's own marginals match the noisy targets. No graphical model is fit and none is sampled. Our implementation measures every two-way marginal under a cell-count cap.

\paragraph{MWEM-PGM}
The Multiplicative Weights Exponential Mechanism~\cite{hardt2012simple} with a Private-PGM backend~\cite{mckenna2019graphical}. Where MST fixes all its two-way marginals up front, MWEM-PGM picks them one per round: it privately selects the two-way marginal that the model fit so far gets most wrong, measures it with Gaussian noise, and refits. We run one round per column.

\paragraph{PrivateGSD}
Genetic-algorithm synthesis~\cite{liu2023generating}. Privately answers a workload of marginal queries (all two-way marginals here), then evolves a population of synthetic records by mutation and crossover to minimize the error against those noisy answers. There is no generative model; the release is the fittest population.

\bigskip
\noindent\textbf{Statistical Synthesis: Non-DP}

\paragraph{Synthpop.}
A sequential synthesis method implemented in the \texttt{synth\-pop} R package~\cite{nowok2016synthpop}.
Each feature is synthesised in turn by fitting a conditional model (default: CART) given all previously synthesised columns.
Captures complex non-linear conditional distributions.
No differential-privacy guarantee; widely used in official statistics and social-science disclosure control.

\bigskip
\noindent\textbf{Deep Generative Synthesis: Non-DP}

\paragraph{CTGAN}
Conditional Tabular GAN~\cite{xu2019modeling}.
Adversarially trained: a conditional generator (conditioned on a randomly sampled categorical column to cover the full marginal) competes against a PacGAN discriminator~\cite{lin2018pacgan} that operates on pairs of records to reduce mode collapse.
Continuous columns are transformed via mode-specific normalisation.
Training can be unstable; CTGAN tends to excel on datasets with many categorical columns and struggle on continuous-heavy ones.

\paragraph{TVAE}
Tabular Variational Autoencoder~\cite{xu2019modeling}.
An encoder maps records to a latent Gaussian, and a decoder reconstructs synthetic records by sampling and decoding.
Uses the same mode-specific normalisation as CTGAN but is not adversarially trained, yielding more stable optimisation at the cost of potentially under-representing rare modes.

\paragraph{ARF}
Adversarial Random Forests~\cite{watson2023adversarial}.
Iteratively trains a random forest to classify real versus synthetic records; the forest's leaf partitions define a density estimate from which new synthetic data are drawn.
A hybrid between ensemble methods and generative modelling (no neural networks or minimax game).
The ``adversarial'' label refers to the use of a classifier to drive density estimation, not a GAN-style training objective.

\paragraph{TabDDPM}
Denoising Diffusion Probabilistic Model for tabu\-lar da\-ta \cite{kotelnikov2023tabddpm}.
Learns to reverse a Gaussian noise process applied to continuous feature embeddings (multinomial diffusion for categorical features).
At inference, starts from noise and iteratively denoises to produce synthetic records.
Achieves state-of-the-art fidelity on many tabular benchmarks and, in our experiments, tends to produce the highest-vulnerability synthetic data.

\subsection{Membership Inference}\label{sec:mia_prelim}

Membership inference (MIA) asks whether a candidate record $x$ was in the training set $D_{\text{train}}$ used to fit the generator. An attack assigns $x$ a real-valued \emph{membership score} $s(x)$, higher meaning more member-like, and a decision is obtained by thresholding $s$. We evaluate three scores against the synthetic release $D_{\text{syn}}$. \emph{SynthDistance}~\cite{hayes2017logan} sets $s(x) = -\min_{y \in D_{\text{syn}}} d(x,y)$: members tend to lie closer to the synthetic data the generator produced. \emph{NNDR} (nearest-neighbour distance ratio)~\cite{stadler2022synthetic} normalizes that distance by the distance to the nearest leave-one-out \emph{training} neighbour, which down-weights records that are close to everything. Our \emph{RA-as-MIA} (Section~\ref{sec:mia}) instead uses reconstruction quality itself as the score: how well the full reconstruction attack recovers $x$'s hidden features from its QI.

Because every score is a ranking, we report a threshold-free metric, the ROC AUC (Table~\ref{tab:mia_comparison}).

\section{Implementation Details}\label{sec:impl}

Three attacks (CondMST, CondDDPM, CondRePaint) turn an SDG method into a reconstruction primitive by conditioning its generative process on the target's quasi-identifiers: fit (or reuse) a generator on the synthetic release, then sample the hidden features conditioned on each target's known QI values. We describe the three adaptations below, then CoBP-RA, MultiHeadMLP, and ARFFormer, with pseudocode for all six new attacks; the remaining attacks are the standard supervised models of Section~\ref{sec:attack_taxonomy}.

\label{sec:pseudocode}Algorithms~\ref{alg:condmst}--\ref{alg:arfformer} sketch the six attacks. Each listing spells out what is new about its attack and names the standard machinery it reuses (belief propagation, graphical-model fitting, network training) without writing it out. Hyperparameters that a listing uses are in Table~\ref{tab:hyperparams} (Appendix~\ref{sec:hyperparams}). Every model is fit once per release and applied to all $n$ targets; each listing treats one target at a time. Table~\ref{tab:notation} collects the symbols used in the paper and Table~\ref{tab:hyperparams} the options each attack offers, with no symbol in both.

\subsection{Conditional MST (CondMST)}\label{sec:partial_sdg_mst}

CondMST fits an MST graphical model (Private-PGM~\cite{mckenna2019graphical}) directly on the \emph{synthetic} release rather than on private data, then samples each target's hidden features conditioned on its observed QI. Fitting is \emph{noiseless}: because the synthetic data is already public, we drive the mechanism's Gaussian noise to zero ($\sigma\!\to\!0$) so the model captures the release's own pairwise-marginal structure as faithfully as possible. Continuous columns are discretized into equal-width bins over each column's synthetic range before fitting, and a hidden bin decodes to its midpoint at sampling time.

The model is fit to a set of cliques $\mathcal{C}$, each a set of features whose joint marginal it matches (Algorithm~\ref{alg:condmst}). Line~\ref{line:condmst_c1} makes every feature its own clique. The \textsc{bounded} variant we report then adds one clique of $\kappa$ features per hidden feature $h$, namely $h$ and the $\kappa{-}1$ QI features with the highest mutual information with it (line~\ref{line:condmst_bounded}), and line~\ref{line:condmst_pairs} adds two-feature cliques in decreasing mutual information, each only if it links two features not yet linked, until all features are linked. Skipping the bounded step leaves MST's own maximum spanning tree, making lines~\ref{line:condmst_c1}--\ref{line:condmst_fit} MST's fitting routine unchanged; the \emph{independent} variant instead fits a separate single-feature model per hidden feature, and \emph{hub} enlarges the cliques tying hidden features to their most informative QI columns. Sampling follows MST's own order, except that the QI are clamped to the target's values before it starts, so each hidden feature is drawn given every QI or already-drawn feature that shares a clique with it. \textsc{truncated}, the configuration we report, draws only among the most probable fraction $\eta$ of a feature's values (at least one); \textsc{sample} draws among all of them and \textsc{argmax} takes the most probable. We report the \textsc{bounded} variant as \textbf{CondMST} in Table~\ref{tab:ra_mean_adult}, the strongest configuration we found.

\begin{algorithm}[tb]
\caption{CondMST}\label{alg:condmst}
\begin{algorithmic}[1]
\Require $D_{\text{syn}}$; $Q$; $H$; targets $x^Q_1,\dots,x^Q_n$
\Statex \textbf{Parameters:} clique size $\kappa$; cliques \textsc{bounded} or \textsc{tree}; draw \textsc{truncated} ($\eta$), \textsc{sample}, or \textsc{argmax}
\Ensure $\hat{x}^H_1,\dots,\hat{x}^H_n$
\State $\mathcal{C} \gets \bigl\{\{a\} : a \in \mathcal{F}\bigr\}$ \label{line:condmst_c1}
\For{$h \in H$} \Comment{\textsc{bounded} only}
  \State add $\{h\} \cup \{$the $\kappa{-}1$ features $q \in Q$ of highest $I(q;h)\}$ to $\mathcal{C}$ \label{line:condmst_bounded}
\EndFor
\For{pairs $\{a,b\}$ in decreasing $I(a;b)$, until all features are linked}
  \State \textbf{if} no chain of cliques in $\mathcal{C}$ links $a$ and $b$: add $\{a,b\}$ to $\mathcal{C}$ \label{line:condmst_pairs}
\EndFor
\State $\hat{P} \gets$ graphical model fit to $D_{\text{syn}}$'s marginals on the cliques in $\mathcal{C}$ \label{line:condmst_fit}
\For{$i = 1,\dots,n$}
  \For{$h \in H$, in the sampling order of $\hat{P}$}
    \State draw $\hat{x}^h_i$ from $\hat{P}\bigl(h \mid x^Q_i$ and the $\hat{x}^{h'}_i$ drawn so far$\bigr)$
  \EndFor
\EndFor
\end{algorithmic}
\end{algorithm}

\subsection{Conditional TabDDPM (CondDDPM) and Conditional RePaint (CondRePaint)}\label{sec:partial_sdg_tabddpm}

\begin{algorithm}[tb]
\caption{CondDDPM}\label{alg:condddpm}
\begin{algorithmic}[1]
\Require encoded $D_{\text{syn}}$; QI indicator $\mathbf{1}_Q$; encoded targets $c_1,\dots,c_n$
\Ensure $\hat{x}^H_1,\dots,\hat{x}^H_n$
\Repeat \Comment{training}
  \State $x_0 \sim D_{\text{syn}}$;\; $t$ uniform on $\{0,\dots,T{-}1\}$;\; $\xi \sim \mathcal{N}(0,\mathrm{Id})$
  \State $\tilde{x}_t \gets \mathbf{1}_Q \odot x_0 + (1-\mathbf{1}_Q) \odot \bigl(\sqrt{\bar\alpha_t}\,x_0 + \sqrt{1-\bar\alpha_t}\,\xi\bigr)$ \Comment{\textsc{cond}}
  \State gradient step on $\bigl\lVert \hat\xi_\theta(\tilde{x}_t,t) - (1-\mathbf{1}_Q) \odot \xi \bigr\rVert^2$ \Comment{\textsc{cond}}
\Until{the training budget is spent}
\For{$i = 1,\dots,n$} \Comment{reconstruction}
  \State $z \sim \mathcal{N}(0,\mathrm{Id})$
  \For{$t = T{-}1,\dots,0$}
    \State $z \gets \mathrm{step}_\theta\bigl(\mathbf{1}_Q \odot c_i + (1-\mathbf{1}_Q) \odot z,\ t\bigr)$ \Comment{\textsc{cond}}
  \EndFor
  \State $\hat{x}^H_i \gets \mathrm{decode}\bigl((1-\mathbf{1}_Q) \odot z\bigr)$
\EndFor
\end{algorithmic}
\end{algorithm}

\begin{algorithm}[h!]
\caption{CondRePaint}\label{alg:condrepaint}
\begin{algorithmic}[1]
\Require $\hat\xi_\theta$ trained by Algorithm~\ref{alg:condddpm}; $\mathbf{1}_Q$; $c_1,\dots,c_n$
\Statex \textbf{Parameters:} rounds $r$; look-back $J$
\Ensure $\hat{x}^H_1,\dots,\hat{x}^H_n$
\For{$i = 1,\dots,n$}
  \State $z' \sim \mathcal{N}(0,\mathrm{Id})$
  \For{$t = T{-}1,\dots,0$}
    \State $t' \gets \max(0,\, t-J)$;\; $z \gets z'$
    \For{$j = 1,\dots,r$}
      \State $z' \gets \mathbf{1}_Q \odot \mathrm{noise}_{t'}(c_i) + (1-\mathbf{1}_Q) \odot \mathrm{step}_\theta(z,t')$
      \State $z \gets \mathrm{noise}_t(z')$
    \EndFor
  \EndFor
  \State $\hat{x}^H_i \gets \mathrm{decode}\bigl((1-\mathbf{1}_Q) \odot z'\bigr)$
\EndFor
\end{algorithmic}
\end{algorithm}

CondDDPM trains a TabDDPM diffusion model on the synthetic release with the QI columns supplied as conditioning, then generates the hidden columns for each target by running reverse diffusion conditioned on that target's QI. CondRePaint reuses the identical trained model but replaces ancestral sampling with RePaint inpainting~\cite{lugmayr2022repaint}: the known QI cells are held fixed (re-noised to the current timestep) while the hidden cells are denoised, with repeated resampling jumps to harmonize the two. Because both attacks share one checkpoint, whichever runs second reuses the cached model. Optionally, CondDDPM adds a second stage: a per-feature MLP that maps the diffusion model's hidden-feature ``hints'' (with the encoded QI) to a final prediction, correcting systematic biases in the raw diffusion output; our tables report CondDDPM without it.

Records are encoded as vectors in $\mathbb{R}^p$ (categories label-encoded, then every coordinate normalized; $\mathrm{decode}$ inverts this, rounding each categorical coordinate to the nearest valid code). Noise is added under a cosine schedule $\bar\alpha_t$ over $T$ steps, $\mathrm{noise}_t(x) = \sqrt{\bar\alpha_t}\,x + \sqrt{1-\bar\alpha_t}\,\xi$ with a fresh $\xi \sim \mathcal{N}(0,\mathrm{Id})$ at every use; the network $\hat\xi_\theta$ learns to predict that noise, and $\mathrm{step}_\theta(z,t)$ calls it to take one reverse step of the standard DDPM sampler~\cite{ho2020denoising}. CondDDPM departs from unconditional diffusion only in the lines marked \textsc{cond} (Algorithm~\ref{alg:condddpm}): in training the QI coordinates stay clean, so the network learns that they carry no noise, and in sampling they are reset to the target's values before every step. CondRePaint (Algorithm~\ref{alg:condrepaint}) reuses the trained $\hat\xi_\theta$ and changes only the sampler: at each timestep $t$ it looks back $J$ steps, to $t' = \max(0, t{-}J)$, and runs $r$ rounds, each of which denoises one step at level $t'$, overwrites the QI with the target's values noised to level $t'$, and re-noises the whole vector to level $t$.

\begin{algorithm}[tb]
\caption{CoBP-RA}\label{alg:cobpra}
\begin{algorithmic}[1]
\Require $D_{\text{syn}}$; $Q$; $H$; targets $x^Q_1,\dots,x^Q_n$
\Statex \textbf{Parameters:} neighbors $K$; column-step strength $\lambda$; graph \textsc{tree}, \textsc{complete}, or \textsc{top}-$M$
\Ensure $\hat{x}^H_1,\dots,\hat{x}^H_n$
\State $f_h \gets$ classifier of $h$ from $Q$, trained on $D_{\text{syn}}$, \textbf{for} $h \in H$
\State $G \gets$ maximum spanning tree on $Q \cup H$, weighted by $I$
\For{$i = 1,\dots,n$}
  \State $\phi_h \gets f_h(\cdot \mid x^Q_i)$ \textbf{for} $h \in H$;\; $\phi_q \gets \mathbf{1}[\,\cdot = x^q_i]$ \textbf{for} $q \in Q$ \label{line:cobp_unary}
  \State $D^K_i \gets$ the $K$ records of $D_{\text{syn}}$ nearest to $x^Q_i$
  \State $\psi_{ab}(v,w) \gets \dfrac{P_{D^K_i}(a{=}v,\ b{=}w)}{P_{D^K_i}(a{=}v)\,P_{D^K_i}(b{=}w)}$ \textbf{for} $(a,b) \in G$ \label{line:cobp_pmi}
  \State belief propagation on $G$, with messages \label{line:cobp_msg}
  \Statex \hspace{\algorithmicindent}\hspace{\algorithmicindent}$\mu_{a\to b}(w) \propto \bigl[\sum_{v} \beta_{a\setminus b}(v)\,\psi_{ab}(v,w)\bigr]^{\gamma_{a\setminus b}}$, \ $\beta_{a\setminus b} \propto \phi_a \prod_{a' \in \partial a \setminus b} \mu_{a'\to a}$
  \State $\beta_{ih} \propto \phi_h \prod_{a' \in \partial h} \mu_{a'\to h}$ \textbf{for} $h \in H$
\EndFor
\State $\beta_{ih}(v) \propto \beta_{ih}(v)\,\bigl[P_{D_{\text{syn}}}(h{=}v)\,/\,\bar{\beta}_h(v)\bigr]^{\lambda}$, where $\bar{\beta}_h = \frac{1}{n}\sum_{i} \beta_{ih}$ \label{line:cobp_col}
\State $\hat{x}^h_i \gets \operatorname*{arg\,max}_{v}\ \beta_{ih}(v)$ \textbf{for all} $i$, $h \in H$
\end{algorithmic}
\end{algorithm}

\subsection{Belief-Propagation Correction (CoBP-RA)}\label{sec:marginal_rf}

 CoBP-RA starts from independent per-feature prediction posteriors (one classifier per hidden feature, trained on the release) and corrects them with the dependencies among features that the release exposes. It places the features on a graph, by default the maximum spanning tree of pairwise mutual information; attaches to each edge a table of how much more often pairs of values co-occur than chance among the target's nearest synthetic neighbors; and runs sum--product belief propagation, so that a confident feature can shift an uncertain neighbor. Algorithm~\ref{alg:cobpra} gives the configuration we report.

Line~\ref{line:cobp_unary} sets each node's potential: for a hidden feature, the classifier's posterior for the target; for a QI feature, $\mathbf{1}[\,\cdot = x^q_i]$, which puts all its probability on the target's known value, so QI nodes act as fixed evidence. Line~\ref{line:cobp_pmi} sets each edge's potential: $\psi_{ab}(v,w)$ says how many times more often values $v$ and $w$ occur together among the target's $K$ nearest synthetic neighbors, $D^K_i$, than they would if $a$ and $b$ were independent. Neighbors are Euclidean on the ordinal-encoded QI, and $P_{D^K_i}$ falls back to $P_{D_{\text{syn}}}$ when $D^K_i$ holds too few records per cell of the $(a,b)$ table. A message $\mu_{a\to b}(w)$ is $a$'s evidence that its neighbor $b$ takes value $w$: $a$'s belief without $b$'s own message, $\beta_{a\setminus b}$, carried across the edge by $\psi_{ab}$. \emph{Tempering} raises it to the power $\gamma_{a\setminus b} = 1-\mathrm{Ent}(\beta_{a\setminus b})/\log|\mathcal{X}_a|$, which is $1$ when that belief is certain and $0$ when it is uniform, so an unsure sender barely moves its neighbor. Line~\ref{line:cobp_col} is the \emph{column step}: it rescales the beliefs $\beta_{ih}$ of feature $h$ by one common factor per value, moving their average over all targets a fraction $\lambda$ (on the log scale) toward $h$'s marginal in $D_{\text{syn}}$. Features with many values stay off the graph, since their edge tables have too many cells to estimate from $K$ neighbors; they keep the classifier's posterior, followed by the column step. On the default \textsc{tree} graph, one sweep up and one down computes every message; the \textsc{complete} and \textsc{top}-$M$ graphs contain cycles, so there all messages are updated in parallel, each averaged with its previous value, until they stop changing or a round limit is reached, and the beliefs are then approximate.

When the synthetic features are independent (all $\mathrm{PMI}\approx 0$) the pairwise messages vanish and CoBP-RA reduces to the underlying random forest, up to the column step; its gains come entirely from joint structure the release happens to preserve. The continuous variant quantile-bins each hidden feature, runs the same message passing on bin labels, and decodes to bin midpoints (Appendix~\ref{sec:cal_memorization_appendix}).

\subsection{MultiHeadMLP and ARFFormer}\label{sec:neural_attacks}

\begin{algorithm}[tb]
\caption{MultiHeadMLP}\label{alg:multiheadmlp}
\begin{algorithmic}[1]
\Require $D_{\text{syn}}$; $Q$; $H$; targets $x^Q_1,\dots,x^Q_n$
\Ensure $\hat{x}^H_1,\dots,\hat{x}^H_n$
\State $g_\theta \gets$ MLP with shared hidden layers, on the one-hot QI
\State $\pi_h(\cdot \mid x^Q) \gets \operatorname{softmax}\bigl(W_h\, g_\theta(x^Q)\bigr)$ \textbf{for} $h \in H$ \Comment{one head per feature}
\State train $\theta$ and all $W_h$ with Adam on $D_{\text{syn}}$, minimizing $L = -\sum_{h \in H} \log \pi_h(x^h \mid x^Q)$,
\Statex \hspace{\algorithmicindent}early-stopped on validation $L$
\State $\hat{x}^h_i \gets \operatorname*{arg\,max}_{v}\ \pi_h(v \mid x^Q_i)$ \textbf{for all} $i$, $h \in H$
\end{algorithmic}
\end{algorithm}

\begin{algorithm}[tb]
\caption{ARFFormer}\label{alg:arfformer}
\begin{algorithmic}[1]
\Require $D_{\text{syn}}$; $Q = (q_1,\dots,q_{|Q|})$; $H = (h_1,\dots,h_m)$; targets $x^Q_1,\dots,x^Q_n$
\Statex \textbf{Parameters:} token-dropout rate $\tau$; guesses per target $U$
\Ensure $\hat{x}^H_1,\dots,\hat{x}^H_n$
\State $Z_\theta \gets$ causal transformer on tokens $(x^{q_1},\dots,x^{q_{|Q|}},x^{h_1},\dots,x^{h_m})$
\State $\pi_\ell(\cdot \mid x) \gets \operatorname{softmax}\bigl(W_\ell\,[Z_\theta(x)]_{|Q|+\ell-1}\bigr)$ \textbf{for} $\ell \le m$ \Comment{sees $x^Q$ and $x^{h_1},\dots,x^{h_{\ell-1}}$}
\State train with AdamW on $D_{\text{syn}}$, minimizing $-\sum_{\ell} \log \pi_\ell(x^{h_\ell} \mid \tilde{x})$,
\Statex \hspace{\algorithmicindent}where $\tilde{x}$ is $x$ with each hidden value set to $\varnothing$ with probability $\tau$;
\Statex \hspace{\algorithmicindent}early-stopped on mean validation accuracy
\For{$i = 1,\dots,n$}
  \State $x^{(u)} \gets (x^Q_i, \varnothing, \dots, \varnothing)$ \textbf{for} $u \le U$
  \For{$\ell = 1,\dots,m$}
    \State $\hat{x}^{h_\ell}_i \gets \operatorname*{arg\,max}_{v}\ \frac{1}{U}\sum_{u=1}^{U} \pi_\ell(v \mid x^{(u)})$
    \State draw $x^{(u),h_\ell} \sim \pi_\ell(\cdot \mid x^{(u)})$ \textbf{for} $u \le U$ \Comment{extend each guess}
  \EndFor
\EndFor
\end{algorithmic}
\end{algorithm}

\begin{table*}[t]
  \centering
  \small
  \caption{Attack hyperparameters and options. Symbols are those of Algorithms~\ref{alg:condmst}--\ref{alg:arfformer} (all defined in Table~\ref{tab:notation}); where an attack offers alternatives, the value we report comes first and the others follow in parentheses. Parameters not listed are at their scikit-learn, LightGBM, or PyTorch defaults.}
  \label{tab:hyperparams}
  \begin{tabular}{@{}llll@{}}
    \toprule
    \textbf{Attack} & \textbf{Parameter} & \textbf{Symbol} & \textbf{Value (alternatives)} \\
    \midrule
    KNN & neighbors (distance-weighted) & & $1$ \\
    Random forest & trees; maximum depth & & $25$; $25$ \\
    LightGBM & boosting rounds & & $100$ \\
    Logistic regression & iterations & & $100$ \\
    SVM & kernel; regularization strength & & RBF; $1$ \\
    MLP & hidden layer width; dropout & & $300$; $0.2$ \\
        & batch size; Adam learning rate & & $264$; $3\times10^{-4}$ \\
        & maximum epochs; patience & & $500$; $60$ \\
    TabPFN & training rows (subsampled); ensemble size & & $1024$; $32$ \\
           & values above which a feature falls back to the random forest & & $10$ \\
    Linear program & marginal queries (solved with Gurobi) & & $3$-way \\
    \midrule
    CoBP-RA & classifier & $f_h$ & random forest, as above (LightGBM, naive Bayes, MLP) \\
            & graph & $G$ & \textsc{tree} (\textsc{complete}, \textsc{top}-$M$) \\
            & QI nodes; tempering & & on; on (each can be off) \\
            & neighbors for edge tables & $K$ & $100$ \\
            & Laplace smoothing & & $10^{-6}$ \\
            & minimum records per edge-table cell, else global table & & $5$ \\
            & maximum values of a graph feature & & $50$ \\
            & column-step strength & $\lambda$ & $0.5$ ($0$ = off) \\
            & column-step target & & marginal in $D_{\text{syn}}$ (marginal in $D^K_i$) \\
            & loopy graphs: edges; rounds; damping; tolerance & $M$ & $2\times$nodes; $20$; $0.5$; $10^{-4}$ \\
            & continuous variant: quantile bins (global tables) & & $20$ \\
    \midrule
    CondMST & equal-width bins per continuous column & & $20$ \\
            & cliques & $\mathcal{C}$ & \textsc{bounded} (\textsc{tree}, independent, hub) \\
            & clique size & $\kappa$ & $3$ \\
            & draw & & \textsc{truncated} (\textsc{sample}, \textsc{argmax}) \\
            & kept fraction of values & $\eta$ & $10\%$ \\
            & Private-PGM fitting iterations & & $10{,}000$ \\
    \midrule
    CondDDPM, & denoiser MLP widths & & $[512,1024,1024,1024,1024,512]$ \\
    CondRePaint & dropout; weight decay & & $0.1$; $10^{-5}$ \\
            & training steps; batch size; learning rate & & $200{,}000$; $4096$; $6\times10^{-4}$ \\
            & diffusion timesteps & $T$ & $2000$ \\
            & CondRePaint rounds; look-back & $r$; $J$ & $10$; $10$ \\
            & CondDDPM second-stage MLP & & off (on) \\
    \midrule
    MultiHeadMLP & shared hidden-layer widths; dropout & & $(1000, 500)$; $0$ \\
            & batch size; Adam learning rate & & $256$; $5\times10^{-4}$ \\
            & maximum epochs; patience & & $300$; $125$ \\
    \midrule
    ARFFormer & encoder blocks; heads; embedding width & & $3$; $4$; $128$ \\
            & feed-forward width; dropout & & $256$; $0.1$ \\
            & batch size; AdamW learning rate; weight decay & & $256$; $5\times10^{-4}$; $0.01$ \\
            & learning rate halved after epochs without improvement & & $5$ \\
            & maximum epochs; patience & & $200$; $20$ \\
            & token-dropout rate & $\tau$ & $0.15$ \\
            & decoding; guesses per target & $U$ & average over guesses (\textsc{greedy}); $32$ \\
    \bottomrule
  \end{tabular}
\end{table*}

 Both attacks predict every hidden feature with one network, trained on an $80/20$ train/validation split of the release with early stopping. MultiHeadMLP (Algorithm~\ref{alg:multiheadmlp}) is an MLP whose hidden layers, [linear, ReLU, batch-norm] blocks, are shared by all hidden features, with one output layer (a \emph{head}) per feature; summing the heads' losses trains the shared layers on every hidden feature at once. ARFFormer, the autoregressive feature transformer (Algorithm~\ref{alg:arfformer}), predicts the hidden features one at a time, each given the QI and the hidden features before it. It embeds each value as a token (with a learned position embedding and an extra padding token $\varnothing$), and $Z_\theta$ is a stack of causally masked transformer encoder blocks, so the output at position $|Q|{+}\ell{-}1$, which predicts $h_\ell$, sees only the QI and $h_1,\dots,h_{\ell-1}$. In training, token dropout replaces each hidden value by $\varnothing$ with probability $\tau$, so that the model does not over-trust earlier hidden values, which at inference are its own guesses. ARFFormer is early-stopped on mean per-feature validation accuracy rather than on its loss, which high-cardinality features dominate. At inference it does not commit to one guess for the earlier features. For each target it samples $U$ complete guesses of $h_1,\dots,h_m$ from the model, feature by feature, and predicts each $h_\ell$ as the value with the highest probability averaged over those $U$ guesses. This picks the most likely value of each feature on its own, which is what per-feature accuracy rewards. A \textsc{greedy} option instead feeds back a single best guess per feature.

\section{Hyperparameters}\label{sec:hyperparams}

 Unless noted, every attack and generator runs at fixed defaults. Table~\ref{tab:hyperparams} gives the attacks' values; the artifact's \texttt{attack\_defaults.py} lists every parameter, including those left at library defaults.

\paragraph{Generators.} All thirteen SDG methods run at their library defaults, with these exceptions. The DP synthesizers receive continuous columns pre-binned into $20$ equal-width ordinal bins, so that the full privacy budget is spent on synthesis rather than on private bound estimation. RankSwap permutes the continuous / high-cardinality columns within rank blocks (Appendix~\ref{sec:sdg_descriptions} lists the swapped columns per dataset), and CellSuppression enforces $k{=}5$-anonymity on each dataset's QI key variables. TabDDPM uses the same denoiser as the diffusion attacks (widths, $200{,}000$ steps, $T{=}2000$; Table~\ref{tab:hyperparams}), except that for the $130$-column NIST SBO data it is widened to $[1024,2048,2048,2048,1024]$ and trained for $300{,}000$ steps.

\paragraph{Generator implementations.} For reproducibility we note the underlying libraries: MST and AIM use SmartNoise-Synth (\texttt{snsynth}); TVAE and CTGAN use the Synthetic Data Vault (SDV)~\cite{patki2016synthetic}; ARF uses Synthcity's \texttt{arf} plugin; TabDDPM uses the official tabular-diffusion implementation~\cite{kotelnikov2023tabddpm}. PrivBayes uses the DataSynthesizer package (correlated-attribute mode); PrivateGSD uses its authors' \texttt{genetic\_sd} library; MWEM-PGM is our implementation on Private-PGM; and PrivSyn pairs the GUM record-update routine of the SynMeter reference code with our own zCDP-calibrated marginal measurement. The de-identification and statistical-synthesis baselines call R through \texttt{rpy2}: Synthpop uses the \texttt{synthpop} package (CART engine), and RankSwap and CellSuppression use \texttt{sdcMicro} (\texttt{rankSwap} and \texttt{lo\-cal\-Sup\-press\-ion}, respectively).

\section{Attack Enhancements: Ensembling and Chaining}\label{sec:chain_and_ensemble}

Two natural strategies promise to strengthen reconstruction attacks: ensembling the predictions of several attacks for the same feature, and, when several hidden features must be predicted, chaining predictions auto-regressively. Both played a role in our NIST CRC winning submission. In extensive controlled experiments against fixed SDG configurations (spanning more than 200 distinct (sample, SDG, QI) combinations), however, neither enhancement reliably improves upon the best individual attack. We report this as an informative negative finding. The oracle analysis below reveals that the information required to improve does \emph{exist} in the data; it simply cannot be accessed by any of the schemes we tried. This locates the problem precisely: the bottleneck is not attack sophistication but a structural property of reconstruction on synthetic data. Synthetic releases preserve population-level distributional structure; they are blind to which specific records are hardest to reconstruct. No voting or prediction scheme can overcome the fundamental ceiling set by how much joint structure the release retains. This theme runs through both ensembling and chaining, and we develop it in each subsection below.

\subsection{Ensembling}\label{sec:ensembling}

The \emph{Best ensemble} row of Table~\ref{tab:ra_mean_adult} is a single fixed configuration applied to every generator: \textbf{hard (majority) voting} over five base attacks: CoBP-RA (the primary), RandomForest, LightGBM, MLP, and KNN, each at its default hyperparameters (Appendix~\ref{sec:hyperparams}). We selected it as the best-averaging combination among the configurations we swept; it nonetheless lands below CoBP-RA alone, as the oracle analysis below explains.

\begin{figure*}[t]
  \centering
  \includegraphics[width=.8\textwidth]{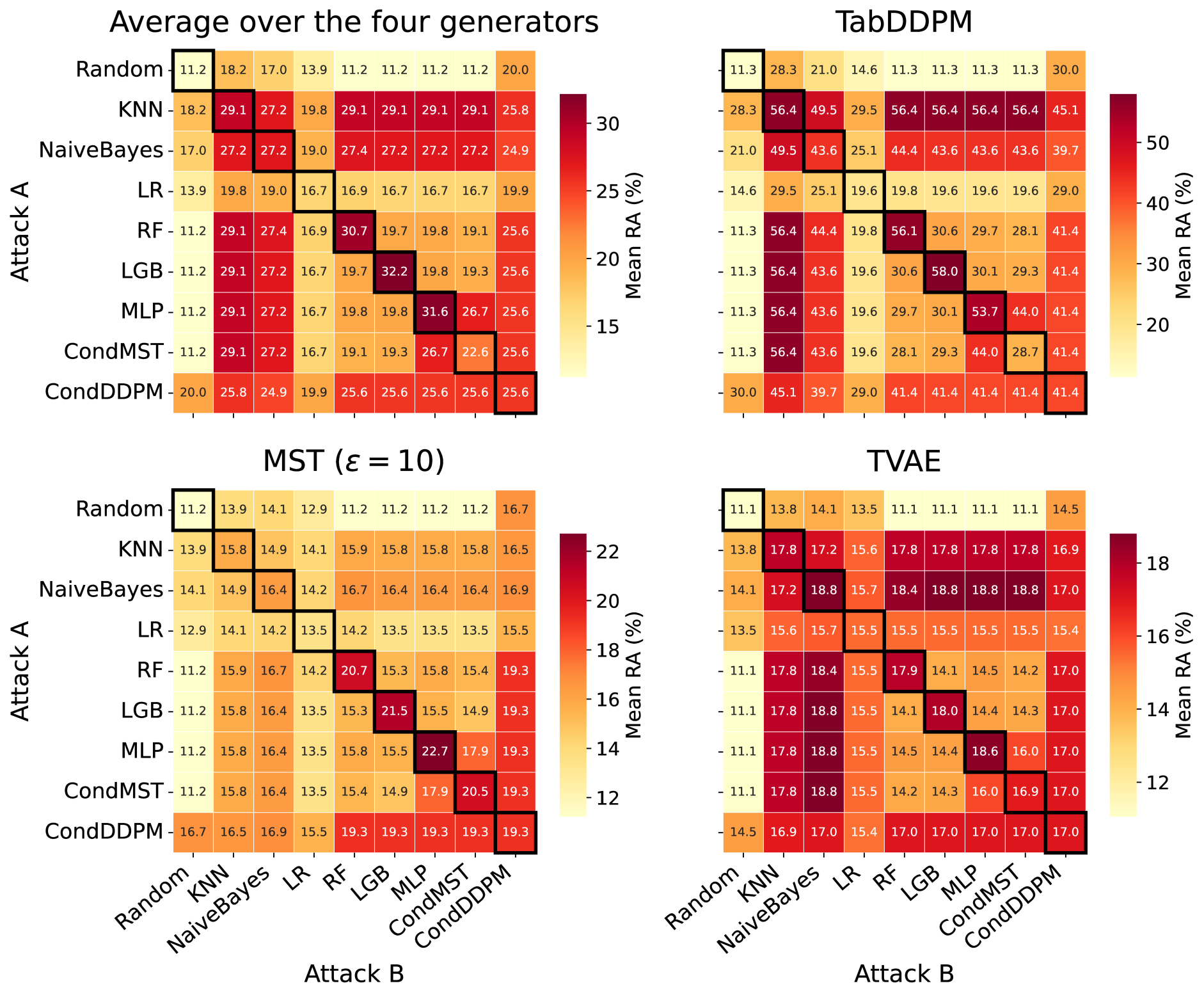}
  \caption{Pairwise ensemble reconstruction accuracy ($R_{adv}$ mean): one panel averaging the four generators swept (Synthpop, TabDDPM, MST $\varepsilon{=}10$, TVAE) and one for each but Synthpop. Rows and columns index the individual attacks: diagonal cells (black border) give each attack alone, off-diagonal cells the soft-voting ensemble of that pair; colour encodes $R_{adv}$. No off-diagonal pair improves meaningfully on the better of its two attacks. (Adult, 1k rows, $\text{QI}_{\text{demo}}$.)}
  \Description{Four heatmap panels in a two-by-two grid, titled ``Average over the four generators'', ``TabDDPM'', ``MST ($\varepsilon = 10$)'', and ``TVAE''. Each panel is a nine-by-nine matrix whose rows and columns index the same nine attacks, in order: Random, KNN, NaiveBayes, LR, RF, LGB, MLP, CondMST, and CondDDPM. Every cell is shaded and annotated with a mean reconstruction advantage in percent, against a per-panel colour bar at its right; attack names label the rows of the left panels and the columns of the bottom panels. Diagonal cells, outlined in black, give each attack's own score; off-diagonal cells give the soft-voting ensemble of the corresponding row and column attacks. In all four panels the highest values sit on or immediately beside the diagonal: in the TabDDPM panel, for instance, the best individual scores are LGB at 58.0 and KNN at 56.4, and no off-diagonal pair exceeds either. Pairing a strong attack with a weaker one yields a value between the two rather than above both.}
  \label{fig:heatmap_ensemble}
\end{figure*}

We exhaustively evaluated all $\binom{9}{2} = 36$ two-attack pairwise soft-voting ensembles, covering the nine leading attacks from Table~\ref{tab:ra_mean_adult} across four SDG methods (Synthpop, TabDDPM, TVAE, MST $\varepsilon{=}10$). Figure~\ref{fig:heatmap_ensemble} visualizes the pairwise grids for four panels (the average over those four generators plus separately TabDDPM, MST $\varepsilon{=}10$, and TVAE) with the diagonal giving each attack's individual score and the off-diagonal its pairwise ensemble. For each of the 5 training samples and 4 SDG methods, we record the best pairwise ensemble RA against the best individual RA for that (sample, SDG) pair; across these 20 evaluation jobs, the mean gain was $-1.2$ RA points and only 5 of the 20 jobs showed any positive gain at all, with the maximum positive improvement being $+0.12$~pp. We also evaluated a meta-learning ensemble (stacking): a per-feature logistic regression trained on held-out synthetic-data predictions (out-of-fold cross-validation on the synthetic dataset, where some synthetic records are withheld when training each base model, and then those withheld predictions are used to train the meta-learner) augmented with QI features. Stacking outperformed soft voting but still fell below the best individual attack. The reason is distribution shift: the meta-model learns reliability weights (which base model to trust in which QI region) from synthetic data, but real training targets occupy different regions of feature space than synthetic records, so the learned confidence estimates do not transfer.

A third approach, \emph{oracle-tuned weighted averaging}, grid-searched all 816 convex weight combinations (step size 0.1) over five attacks (CoBP-RA, LightGBM, MLP, KNN, NaiveBayes) across five samples and five SDG methods. The optimal weights on this grid were CoBP-RA=0.20, MLP=0.50, KNN=0.20, NaiveBayes=0.10, LightGBM=0.00, achieving $26.7$\% RA averaged over all SDG methods. This is identical to CoBP-RA alone ($26.7$\%). Even with weights tuned directly on the test distribution (an oracle advantage no real attacker has), the optimal weighted combination provides no improvement over the single best attack. This closes the question of whether cleverer weighting schemes could rescue ensembling.

 An oracle shows why. An attacker that could pick, for each record, whichever attack happens to be right would gain $+11.4$ points over the best single attack, so the attacks do err on different records. Picking the best attack per \emph{feature} instead gains only $+2.7$~pp. The gain is thus available only record by record, and to collect it an ensemble must tell, from a target's QI, which attack to trust on that target. Stacking learns exactly such trust weights from the QI and still falls below the best single attack, so the release does not carry that signal. The pairwise heatmaps (Figure~\ref{fig:heatmap_ensemble}) show the outcome: no off-diagonal pair improves meaningfully on the better of its two constituents.

\subsection{Chaining}\label{sec:chaining}

The \emph{Best chain} row of Table~\ref{tab:ra_mean_adult} is also a single fixed configuration: \textbf{CoBP-RA} as the base attack with \textbf{hard chaining} under a \textbf{mutual-information ordering} (hidden features predicted in decreasing mutual information with the QI), each predicted value fed back as an additional known feature. This order predicts first the features the QI determines best, so later steps condition on the most reliable predictions. It was the strongest of the orderings and feedback modes we swept, and like the best ensemble it does not surpass CoBP-RA alone.

Chaining~\cite{read2011classifier} predicts hidden features sequentially, feeding each predicted value back as an additional input for the next step. Formally, for hidden features ordered as $H = \{h_1, \dots, h_m\}$, the $\ell$-th predictor is trained by a supervised learner $\mathcal{A}$ as
\begin{equation}\label{eqn:chain_train}
\mathcal{A}_\ell = \mathcal{A}\!\left(D_{\text{syn}}[Q \cup \{h_1,\dots,h_{\ell-1}\}],\; D_{\text{syn}}[h_\ell]\right),
\end{equation}

and at inference, predicted values are substituted for earlier features:
\begin{equation}\label{eqn:chain_infer}
\hat{h}_\ell = \mathcal{A}_\ell\!\left(D_{\text{train}}[Q \cup \{\hat{h}_1,\dots,\hat{h}_{\ell-1}\}]\right).
\end{equation}
Besides the mutual-information order, we tried decreasing correlation with $Q$ and random orders. Hard chaining (feeding back the most-likely predicted label) was essentially neutral across all configurations: the best result was $+0.3$~pp (RF, mutual-information ordering on Adult) and the MLP degraded by $1.9$~pp. No ordering strategy produced consistent improvements.

\begin{table*}[t]
\centering
\resizebox{\textwidth}{!}{%
\begin{tabular}{c | c c | c c c c c c c c c}
& \multicolumn{2}{c|}{Mean over all SDG} & \multicolumn{9}{c}{Per-SDG score ($\text{QI}_1$)} \\
    \hline
    \multirow{2}{*}{Team} & $\text{QI}_1$ & $\text{QI}_2$ & Cell & Rank & \multirow{2}{*}{ARF} & \multirow{2}{*}{TVAE} & Synth- & MST & MST & AIM & AIM \\
    & (avg.) & (avg.) & Supp. & Swap & & & pop & $(\varepsilon{=}1)$ & $(\varepsilon{=}10)$ & $(\varepsilon{=}1)$ & $(\varepsilon{=}10)$ \\
    \hline
    (ours) & \textbf{654} & \textbf{598} & 415 & \textbf{1078} & \textbf{346} & \textbf{343} & 363 & \textbf{349} & \textbf{338} & \textbf{351} & \textbf{349} \\
    (ours) & 569 & 506 & 414 & 438 & 317 & 342 & \textbf{365} & 314 & 307 & 290 & 314 \\
    (ours) & 573 & 499 & \textbf{430} & 426 & 324 & \textbf{343} & \textbf{365} & 325 & 306 & 293 & 325 \\
    ? & 381 & 373 & 248 & 245 & 243 & 246 & 245 & 245 & 242 & 244 & 245 \\
    ? & 370 & 359 & 246 & 245 & 241 & 243 & 246 & 228 & 222 & 220 & 228 \\
\end{tabular}
}
\caption{Official NIST Privacy CRC scoreboard. Rows = competing teams; our three submissions appear in the first three rows. $R_{adv}$ (Eq.~\ref{eqn:scoring}) is \emph{summed} across all hidden features per the official scoring protocol. The \textbf{Mean over all SDG} columns give each submission's average over all nine SDG methods, under known-QI sets $\text{QI}_1$ (7 features, 18 hidden) and $\text{QI}_2$ (12 features, 10 hidden); the nine right-hand columns are per-SDG scores under $\text{QI}_1$. Feature-level QI membership is in Table~\ref{tab:qi_def_arizona}. (NIST Arizona, 25 features, 10k rows.)}\label{tab:nist_results}
\end{table*}

Three additional mechanisms designed to reduce error propagation also failed: \emph{soft chaining} (passing full posterior vectors instead of hard labels) degraded RF by $25.9$~pp as noisy probabilities corrupted downstream conditioning; \emph{confidence-gated chaining} (replacing low-confidence predictions with a uniform prior) recovered no performance at any threshold; and \emph{Gibbs chaining} (iterating the chain over multiple passes) settled slightly below the non-chained score for RF (locking in incorrect early assignments) and progressively degraded MLP.

The oracle bound provides the definitive perspective: conditioning on \emph{true} hidden-feature values at each step yields only a few points: $+4.0$~pp for RF and $+5.8$~pp for MLP on TabDDPM, and at most $+2.9$~pp on MST $\varepsilon{=}10$ and Synthpop (Adult 1k, $\text{QI}_{\text{demo}}$, 5 samples). Even \emph{perfect} previous predictions thus add little signal beyond the QI. This rules out error propagation as the primary limitation. The synthetic joint distribution simply does not preserve enough higher-order conditional structure (beyond what QI captures) for chaining to exploit. Notably, this holds for both MST (whose model is fit to pairwise marginals alone; best hard-chaining result on MST $\varepsilon{=}1000$: $+0.1$~pp over non-chained) and TabDDPM (which models the full joint via diffusion; best hard-chaining result on TabDDPM: $+0.2$~pp). The near-zero gains on TabDDPM are especially instructive: despite learning a high-fidelity joint generative model, the synthetic data TabDDPM produces does not encode enough higher-order \emph{conditional} structure (hidden$|$QI$+$previous-hidden) to make auto-regressive conditioning useful beyond the direct QI signal.

\begin{table}[t]
  \centering
  \small
  \caption{Per-feature $R_{adv}$ (\%) averaged over five attacks (NaiveBayes, KNN, RF, CoBP-RA, MLP) under MST. Bold = maximum per row. (Adult, 10k rows, $\text{QI}_{\text{demo}}$.)}
  \label{tab:feature_eps_breakdown}
  \begin{tabular}{lrrrrr}
    \toprule
    Feature & $\varepsilon{=}0.1$ & $\varepsilon{=}1$ & $\varepsilon{=}10$ & $\varepsilon{=}100$ & $\varepsilon{=}1000$ \\
    \midrule
    income         & 59.5 & 58.6 & 58.8 & \textbf{59.6} & 59.3 \\
    relationship   & 39.5 & 54.2 & 55.4 & \textbf{55.5} & 54.9 \\
    education-num  & 22.0 & 56.3 & 79.6 & \textbf{83.9} & 83.4 \\
    workclass      & 11.8 & 11.8 & 11.8 & \textbf{12.4} & 12.3 \\
    occupation     &  8.4 & 11.9 & \textbf{13.2} & 12.5 & 12.9 \\
    hours-per-week &  \textbf{1.2} &  1.1 & \textbf{1.2} &  \textbf{1.2} &  1.1 \\
    capital-loss   &  0.0 &  0.0 &  0.0 &  0.0 &  0.0 \\
    capital-gain   &  0.0 &  0.0 &  0.0 &  0.0 &  0.0 \\
    fnlwgt         &  0.0 &  0.0 &  0.0 &  0.0 &  0.0 \\
    \bottomrule
  \end{tabular}
\end{table}

\begin{table*}[t!]
  \centering
  \small
  \caption{Reconstruction accuracy ($R_{adv}$, \%) on CDC Diabetes. Mode baseline $\approx42.5\%$. Bold = best per column among ML attacks. $\dagger$~= novel attack. (1k rows, $\text{QI}_{\text{demo}}$.)}
  \label{tab:ra_mean_cdc}
  \resizebox{\textwidth}{!}{%
  \begin{tabular}{lrrrrrrrrrrrrrrrr}
    \toprule
    Attack & \rotatebox{60}{RankSwap} & \rotatebox{60}{Cell Supp.} & \rotatebox{60}{Synthpop} & \rotatebox{60}{TVAE} & \rotatebox{60}{CTGAN} & \rotatebox{60}{ARF} & \rotatebox{60}{TabDDPM} & \rotatebox{60}{MST $(\varepsilon{=}0.1)$} & \rotatebox{60}{MST $(\varepsilon{=}1)$} & \rotatebox{60}{MST $(\varepsilon{=}10)$} & \rotatebox{60}{MST $(\varepsilon{=}100)$} & \rotatebox{60}{MST $(\varepsilon{=}1000)$} & \rotatebox{60}{AIM $(\varepsilon{=}1)$} & \rotatebox{60}{AIM $(\varepsilon{=}3)$} & \rotatebox{60}{AIM $(\varepsilon{=}10)$} & \rotatebox{60}{Avg.} \\
    \midrule
    Mode & 42.5 & 42.5 & 42.5 & 42.5 & 42.5 & 42.5 & 42.5 & 42.5 & 42.5 & 42.5 & 42.5 & 42.5 & 42.5 & 42.5 & 42.5 & 42.5 \\
    \midrule
    KNN & \textbf{83.5} & \textbf{59.4} & 46.7 & 45.1 & 42.6 & 46.7 & \textbf{75.7} & 42.4 & 42.2 & 44.1 & 44.2 & 44.2 & 43.1 & 43.9 & 45.1 & \textbf{49.9} \\
    Random Forest & 75.6 & 58.6 & 46.9 & 45.2 & 42.4 & 45.0 & 75.1 & 42.3 & 42.5 & 43.9 & 44.1 & 44.2 & 42.6 & 44.2 & 45.0 & 49.2 \\
    MLP & 61.5 & 57.3 & 48.4 & 46.2 & \textbf{42.8} & 46.6 & 71.0 & 42.2 & \textbf{43.1} & 44.9 & 45.4 & 45.2 & \textbf{43.2} & 44.7 & 45.7 & 48.5 \\
    CondDDPM$^\dagger$ & 81.9 & 47.6 & 46.8 & 45.1 & 42.3 & 46.3 & 73.0 & 42.2 & 42.3 & 43.8 & 44.2 & 44.3 & 42.8 & 43.9 & 45.0 & 48.8 \\
    CoBP-RA$^\dagger$ & 74.6 & 59.0 & \textbf{48.6} & \textbf{46.3} & 42.6 & \textbf{46.8} & 74.1 & \textbf{42.5} & 42.5 & \textbf{45.6} & \textbf{45.8} & \textbf{45.7} & 43.1 & \textbf{45.0} & \textbf{46.1} & \textbf{49.9} \\
    \midrule
    \textbf{Avg.} & \textbf{75.4} & \textbf{56.4} & \textbf{47.5} & \textbf{45.6} & \textbf{42.5} & \textbf{46.3} & \textbf{73.8} & \textbf{42.3} & \textbf{42.5} & \textbf{44.5} & \textbf{44.7} & \textbf{44.7} & \textbf{43.0} & \textbf{44.3} & \textbf{45.4} & \\
    \bottomrule
  \end{tabular}
  }
\end{table*}

\begin{table*}[h]
  \centering
  \small
  \caption{Reconstruction accuracy ($R_{adv}$, \%) on CDC Diabetes at 100k rows. Mode baseline $=42.2\%$ throughout; bold = best per column among ML attacks. KNN, CondDDPM, and TabPFN were not run at this scale. $\dagger$~= novel attack. ($\text{QI}_{\text{demo}}$.)}
  \label{tab:cdc_100k}
  \resizebox{\textwidth}{!}{%
  \begin{tabular}{lrrrrrrrrrrrrrr}
    \toprule
    Attack & \rotatebox{60}{RankSwap} & \rotatebox{60}{Cell Supp.} & \rotatebox{60}{Synthpop} & \rotatebox{60}{TVAE} & \rotatebox{60}{CTGAN} & \rotatebox{60}{ARF} & \rotatebox{60}{TabDDPM} & \rotatebox{60}{MST $(\varepsilon{=}0.1)$} & \rotatebox{60}{MST $(\varepsilon{=}1)$} & \rotatebox{60}{MST $(\varepsilon{=}10)$} & \rotatebox{60}{MST $(\varepsilon{=}100)$} & \rotatebox{60}{MST $(\varepsilon{=}1000)$} & \rotatebox{60}{AIM $(\varepsilon{=}1)$} & \rotatebox{60}{Avg.} \\
    \midrule
    Mode & 42.2 & 42.2 & 42.2 & 42.2 & 42.2 & 42.2 & 42.2 & 42.2 & 42.2 & 42.2 & 42.2 & 42.2 & 42.2 & 42.2 \\
    \midrule
    Random Forest & \textbf{62.9} & \textbf{87.1} & 47.3 & 45.1 & 45.2 & 45.8 & 47.7 & 43.6 & 43.8 & 43.3 & 43.2 & 43.3 & 45.6 & 49.5 \\
    \textsc{mlp} & 45.9 & 46.9 & 45.6 & 45.0 & 45.2 & 45.0 & 45.9 & 43.7 & 43.2 & 43.2 & 43.2 & 43.2 & 46.0 & 44.8 \\
    Naive Bayes & 49.2 & 49.5 & \textbf{48.4} & \textbf{48.1} & \textbf{48.2} & \textbf{48.8} & \textbf{49.6} & \textbf{45.1} & \textbf{46.5} & \textbf{47.2} & \textbf{47.3} & \textbf{48.1} & \textbf{48.9} & 48.1 \\
    CoBP-RA$^\dagger$ & \textbf{62.9} & 85.7 & 47.8 & 45.7 & 46.1 & 46.5 & 48.1 & 44.4 & 44.0 & 44.1 & 44.0 & 44.0 & 46.0 & \textbf{49.9} \\
    \midrule
    \textbf{Avg.} & \textbf{55.2} & \textbf{67.3} & \textbf{47.3} & \textbf{46.0} & \textbf{46.2} & \textbf{46.5} & \textbf{47.8} & \textbf{44.2} & \textbf{44.4} & \textbf{44.4} & \textbf{44.4} & \textbf{44.6} & \textbf{46.6} & \\
    \bottomrule
  \end{tabular}}
\end{table*}

\begin{table*}[h]
  \centering
  \small
  \caption{Reconstruction accuracy ($R_{adv}$, \%) on NIST SBO. Mode baseline $= 32.1\%$ throughout; bold = best per column among ML attacks. Avg.: mean over generators (column) or over the attacks below Mode (row). $\dagger$~= novel attack. (1k rows, $\text{QI}_{\text{demo}}$.)}
  \label{tab:ra_mean_nist_sbo}
  \resizebox{\textwidth}{!}{%
  \begin{tabular}{lrrrrrrrrrrrrr}
    \toprule
    Attack & \rotatebox{60}{RankSwap} & \rotatebox{60}{Cell Supp.} & \rotatebox{60}{Synthpop} & \rotatebox{60}{TVAE} & \rotatebox{60}{CTGAN} & \rotatebox{60}{ARF} & \rotatebox{60}{TabDDPM} & \rotatebox{60}{MST $(\varepsilon{=}0.1)$} & \rotatebox{60}{MST $(\varepsilon{=}1)$} & \rotatebox{60}{MST $(\varepsilon{=}10)$} & \rotatebox{60}{MST $(\varepsilon{=}100)$} & \rotatebox{60}{MST $(\varepsilon{=}1000)$} & \rotatebox{60}{Avg.} \\
    \midrule
    Mode & 32.1 & 32.1 & 32.1 & 32.0 & 32.1 & 32.1 & 32.1 & 32.1 & 32.1 & 32.1 & 32.1 & 32.1 & 32.1 \\
    \midrule
    KNN & 72.2 & 39.9 & 33.9 & 33.0 & 33.1 & 33.3 & 32.1 & \textbf{32.5} & 32.0 & 32.4 & 32.6 & 32.7 & 36.6 \\
    Naive Bayes & 39.6 & 36.4 & 33.3 & \textbf{33.2} & 32.0 & \textbf{34.3} & 32.1 & 32.1 & \textbf{32.3} & 32.4 & 32.6 & 32.4 & 33.6 \\
    Random Forest & 66.0 & 35.5 & 34.1 & 33.0 & 32.6 & 33.0 & 32.1 & 32.2 & 32.0 & 32.6 & 32.6 & 32.7 & 35.7 \\
    CoBP-RA$^\dagger$ & \textbf{80.8} & \textbf{42.3} & \textbf{34.6} & 33.1 & \textbf{34.3} & 33.5 & 32.1 & 32.2 & 32.0 & \textbf{32.9} & \textbf{33.6} & \textbf{33.4} & \textbf{37.9} \\
    \midrule
    \textbf{Avg.} & \textbf{64.7} & \textbf{38.5} & \textbf{34.0} & \textbf{33.1} & \textbf{33.0} & \textbf{33.5} & \textbf{32.1} & \textbf{32.2} & \textbf{32.1} & \textbf{32.6} & \textbf{32.9} & \textbf{32.8} & \\
    \bottomrule
  \end{tabular}
  }
\end{table*}

\begin{table}[ht!]
  \centering
  \small
  \caption{Memorization test on California Housing. NRMSE: lower = better reconstruction; $\Delta = \mathrm{NRMSE}_{\mathrm{tr}} - \mathrm{NRMSE}_{\mathrm{NT}}$, so \emph{negative} $\Delta$ signals memorization (bold: $\Delta\le-0.015$). $\ddagger$\,At MST $\varepsilon{=}0.1$ the implied holdout NRMSE ($\mathrm{tr}-\Delta$) exceeds the mean-imputation baseline (${\approx}0.24$) for all three attacks, indicating attack failure on very noisy data rather than memorization. $\dagger$~= novel attack. (1k rows, $\text{QI}_{\text{large}}$.)}
  \label{tab:memorization_california}
  \begin{tabular}{l rr rr rr}
    \toprule
    & \multicolumn{2}{c}{RandomForest} & \multicolumn{2}{c}{BayesianRidge} & \multicolumn{2}{c}{CoBP-RA$^\dagger$} \\
    \cmidrule(lr){2-3}\cmidrule(lr){4-5}\cmidrule(l){6-7}
    SDG method & $\mathrm{tr}$ & $\Delta$ & $\mathrm{tr}$ & $\Delta$ & $\mathrm{tr}$ & $\Delta$ \\
    \midrule
    TabDDPM   & 0.067 & $\mathbf{-0.034}$ & 0.119 & $-0.001$ & 0.158 & $\mathbf{-0.112}$ \\
    Synthpop  & 0.107 & $\mathbf{-0.033}$ & 0.121 & $-0.001$ & 0.191 & $\mathbf{-0.115}$ \\
    RankSwap  & 0.092 & $\mathbf{-0.017}$ & 0.120 & $-0.002$ & 0.245 & $-0.007$ \\
    ARF       & 0.120 & $-0.005$          & 0.132 & $-0.001$ & 0.204 & $+0.001$ \\
    TVAE      & 0.136 & $-0.002$          & 0.147 & $+0.001$ & 0.194 & $-0.001$ \\
    CTGAN     & 0.172 & $+0.001$          & 0.165 & $+0.001$ & 0.245 & $+0.022$ \\
    \midrule
    MST $\varepsilon{=}0.1$$^\ddagger$ & 0.264 & $\mathbf{-0.026}$ & 0.232 & $-0.008$ & 0.317 & $\mathbf{-0.123}$$^\ddagger$ \\
    MST $\varepsilon{=}1$    & 0.183 & $\phantom{-}0.000$ & 0.184 & $+0.001$ & 0.182 & $\phantom{-}0.000$ \\
    MST $\varepsilon{=}10$   & 0.154 & $-0.002$           & 0.156 & $-0.004$ & 0.169 & $\phantom{-}0.000$ \\
    MST $\varepsilon{=}100$  & 0.154 & $\phantom{-}0.000$ & 0.160 & $-0.003$ & 0.166 & $+0.002$ \\
    MST $\varepsilon{=}1000$ & 0.158 & $-0.007$           & 0.153 & $-0.007$ & 0.252 & $\mathbf{-0.039}$ \\
    \bottomrule
  \end{tabular}
\end{table}

\section{Additional Results}

\subsection{NIST Privacy CRC Scoreboard}\label{sec:nist_appendix}

 Table~\ref{tab:nist_results} gives the official NIST CRC scoreboard: our submissions placed first under both quasi-identifier sets.

\paragraph{Attack choice in the competition setting.} Our winning $\text{QI}_2$ submissions used KNN and NaiveBayes rather than the tree ensembles that lead on Adult, a reminder that the best attack is scenario-dependent (Section~\ref{sec:taxonomy}). With a small training set ($n{=}10\,000$) and a high-dimensional ordinal QI ($12$ features), neighbor lookup effectively finds demographic twins, and NaiveBayes exploits strong independent QI--target marginals, whereas a random forest needs more data per leaf to learn the interactions that grow sparse as QI dimensionality rises. As a rule of thumb, small high-dimensional-QI settings favor KNN and NaiveBayes; larger datasets with moderate QIs favor RF and MLP.

\begin{table*}[h]
  \centering
  \small
  \caption{Reconstruction accuracy by quasi-identifier setting; the CDC counterpart of Table~\ref{tab:qi_analysis_adult}. Within each section, cells average over the \emph{intersection} of hidden features across that section's QI variants, so columns are comparable within a section (per-variant QI memberships in Table~\ref{tab:qi_def_cdc}); bold = row maximum within a section. $\Delta_{\text{size}}{=}\text{QI}_{\text{large}}{-}\text{QI}_{\text{tiny}}$, $\Delta_{\text{comp}}{=}\text{QI}_{\text{beh.}}{-}\text{QI}_{\text{demo}}$, and $\Delta^{*}_{t{\to}d}$ is the gain from tiny to demographic QI, scored on the broader 12-feature two-way intersection. (CDC Diabetes, 1k rows, mean over all SDG methods.)}
  \label{tab:qi_analysis_cdcdiabetes}
  \resizebox{\textwidth}{!}{%
  \begin{tabular}{lrrrrr|rrr}
    \toprule
     & \multicolumn{5}{c}{QI size sweep ($\nearrow$ adversary knowledge)} & \multicolumn{3}{c}{Composition contrast (matched $|\text{QI}|$)} \\
    \cmidrule(lr){2-6}\cmidrule(l){7-9}
    Attack & $\text{QI}_{\text{tiny}}$ & $\text{QI}_{\text{demo}}$ & $\Delta^{*}_{t{\to}d}$ & $\text{QI}_{\text{large}}$ & $\Delta_{\text{size}}$ & $\text{QI}_{\text{demo}}$ & $\text{QI}_{\text{beh.}}$ & $\Delta_{\text{comp}}$ \\
     & (4 kn.) & (10 kn.) &  & (16 kn.) &  & (10 kn.) & (10 kn.) &  \\
    \midrule
    Mode & \textbf{34.91} & \textbf{34.91} & $+$0.00 & \textbf{34.91} & $+$0.00 & \textbf{27.36} & \textbf{27.36} & $+$0.00 \\
    Random & 34.85 & \textbf{34.88} & $+$0.02 & 34.85 & $+$0.01 & \textbf{27.32} & 27.32 & $-$0.00 \\
    \midrule
    \textsc{knn} & 40.32 & 42.20 & $+$2.03 & \textbf{42.23} & $+$1.91 & \textbf{34.49} & 29.97 & $-$4.53 \\
    Naive Bayes & 36.38 & 38.68 & $+$2.36 & \textbf{38.87} & $+$2.49 & \textbf{31.18} & 29.61 & $-$1.57 \\
    Random Forest & 39.71 & 41.19 & $+$1.73 & \textbf{41.33} & $+$1.62 & \textbf{33.64} & 29.56 & $-$4.08 \\
    LightGBM & 38.87 & 40.95 & $+$2.31 & \textbf{41.08} & $+$2.21 & \textbf{33.43} & 29.14 & $-$4.29 \\
    \textsc{mlp} & 38.99 & 40.46 & $+$1.71 & \textbf{41.05} & $+$2.06 & \textbf{33.03} & 29.59 & $-$3.44 \\
    \bottomrule
  \end{tabular}
  }
\end{table*}

\subsection{Extended Linear Reconstruction Results}\label{sec:lp_appendix}

For the 5-category \texttt{race} target (Adult, 1k rows), the LP achieves $28.0$\% RA on average across SDG methods, essentially matching RF's $27.1$\%; gains concentrate against TabDDPM (LP: $65.5$\%, RF: $64.2$\%). At 10k rows, LP scores $34.5$\% vs.\ RF's $33.9$\%, a marginal difference that does not justify the substantially higher compute cost. For joint binary prediction (Adult 10k, jointly predicting \texttt{income} and \texttt{sex} as a $2^2{=}4$-class problem), LP scores $72.1$ vs.\ RF $74.0$, with LP not winning against any individual SDG method. On CDC Diabetes (jointly predicting \texttt{Diabetes\_binary}, \texttt{Stroke}, \texttt{HighBP}), RF beats LP at both 1k ($59.4$ vs.\ $58.1$) and 100k ($59.1$ vs.\ $57.8$) rows. In no setting does the LP surpass simple ML baselines.

\subsection{Feature-Level Reconstruction Accuracy by MST $\varepsilon$}\label{sec:feature_eps_appendix}

Table~\ref{tab:feature_eps_breakdown} reports per-feature $R_{adv}$ averaged over five attacks (NaiveBayes, KNN, RandomForest, CoBP-RA, MLP) across MST $\varepsilon \in \{0.1, 1,$ $10, 100, 1000\}$ on Adult 10k ($\text{QI}_{\text{demo}}$). The aggregate plateau of Section~\ref{sec:conditions} conceals distinct per-feature dynamics.

 \emph{Settled early}: \texttt{relationship} jumps from $39.5\%$ to $54.2\%$ at $\varepsilon{=}1$ and then freezes, while \texttt{income}, \texttt{workclass}, and \texttt{hours-per-week} barely move across the whole range. \emph{Climbing to $\varepsilon{=}10$}: \texttt{occupation} ($15$ categories) rises from $8.4\%$ to $13.2\%$ and then holds, its many conditional cells needing more budget to resolve.

 \emph{Climbing furthest}: \texttt{education-num} rises from $22.0\%$ to $83.9\%$ by $\varepsilon{=}100$. It is a one-to-one recoding of the known QI \texttt{education}, a maximal-information edge that MST's spanning tree selects and measures ever more cleanly as the noise falls. The signal is not equally available to every attack, and the split runs along attack \emph{family}, not strength: at MST $\varepsilon{=}1000$ the discriminative attacks recover the feature almost completely (MultiHeadMLP $100.0\%$, Random Forest $87.1\%$), whereas the diffusion-based conditional attacks, which sample a plausible value from a learned joint instead of learning the mapping, do not (CondDDPM $30.7\%$, CondRePaint $19.0\%$). One redundant column thus inflates the discriminative attacks specifically, and cross-family comparisons on Adult should be read with that in mind.

\emph{Always zero}: \texttt{fnlwgt} (8,477 distinct values) is unrecoverable at all $\varepsilon$.

\subsection{NIST SBO Reconstruction Results}\label{sec:sbo_appendix}

Table~\ref{tab:ra_mean_nist_sbo} reports reconstruction on NIST SBO at 1k rows under $\text{QI}_{\text{demo}}$ (the per-feature structure behind the aggregate is discussed in Section~\ref{sec:conditions}). De-identification is again the exposed regime: rank swapping leaks heavily (CoBP-RA $80.8\%$, KNN $72.2\%$), because small firms form near-singleton QI groups that pass through the swap unchanged, and cell suppression follows ($42.3\%$). Every generative and DP release, by contrast, sits at or within a point of the $32.1\%$ Mode baseline (TabDDPM matches Mode exactly for all attacks, and the full MST $\varepsilon$ sweep barely moves), so this $130$-feature financial data leaves little structure reconstructable beyond the marginal. CoBP-RA is best or tied in nearly every column, consistent with the main benchmark.

\subsection{CDC Diabetes Reconstruction}\label{sec:cdc_appendix}
Table~\ref{tab:ra_mean_cdc} gives the full per-generator breakdown for CDC Diabetes (1k rows, $\text{QI}_{\text{demo}}$) summarized in Section~\ref{sec:whatdata}. The dataset is the most exposed categorical setting we test under de-identification (KNN $83.5\%$ on rank swapping), and TabDDPM nearly matches it ($75.7\%$) despite being fully synthetic. DP is correspondingly tight: across the entire MST sweep KNN moves only $42.2\!\to\!44.2\%$. CoBP-RA leads or ties most columns. Because the features are mostly binary the Mode baseline is high ($42.5\%$), so it is the \emph{gap above baseline}, not the raw value, that measures attack advantage.

\subsection{Dataset Size: CDC Diabetes at Scale}\label{sec:cdc_100k_appendix}
Table~\ref{tab:cdc_100k} reports reconstruction on CDC Diabetes at $100{,}000$ rows, the same attacks and generators as the 1k results in Table~\ref{tab:ra_mean_cdc}. The size effect is strongly mechanism-dependent (Section~\ref{sec:conditions}): de-identification grows \emph{more} exposed with scale (Cell Suppression: RF $58.6\%\!\to\!87.1\%$, as more QI groups clear the $k$-anonymity threshold and ship verbatim), while high-fidelity diffusion grows \emph{less} exposed (TabDDPM: RF $75.1\%\!\to\!47.7\%$, as more training data dilutes per-record memorization). MST remains pinned near its baseline at every budget.

\subsection{QI Analysis: CDC Diabetes}\label{sec:cdc_qi_appendix}

Table~\ref{tab:qi_analysis_cdcdiabetes} repeats on CDC Diabetes the QI analysis run for Adult (Section~\ref{subsec:qi}), and the headline holds: composition outweighs size. Enlarging the known set from 4 to 16 features lifts the random forest only $+1.6$~pp on the shared hidden-feature intersection, and every attack has saturated by the 10-feature $\text{QI}_{\text{demo}}$. Composition moves risk more: at a matched $|\text{QI}|{=}10$ the demographic QI beats the behavioral one by $+4.0$~pp for the random forest and $+4.5$ for KNN: the \emph{opposite} sign to Adult, where behavioral knowledge dominated. On this mostly-binary health data the demographic attributes carry the predictive structure, whereas on Adult the employment and hours features did; in both, \emph{which} features the adversary knows matters more than \emph{how many}.

\subsection{Continuous-Domain Memorization}\label{sec:cal_memorization_appendix}

The memorization test of Section~\ref{sec:memorization} carries over to continuous data by replacing rarity-weighted accuracy with normalized RMSE (lower is better) and a regression attack for the classifier. We run three regressors (a random forest, Bayesian ridge regression, and the continuous variant of CoBP-RA, Appendix~\ref{sec:marginal_rf}) on California Housing and report, for each, the training-target error $\mathrm{tr}$ and the train-minus-holdout gap $\Delta$ (Table~\ref{tab:memorization_california}). The same three-regime structure as the categorical datasets reappears, with the memorization sign reversed by the change of metric: resynthesizing generators that memorize (TabDDPM, Synthpop) show negative $\Delta$, while the DP generators sit at $\Delta\approx0$.

One pattern is worth isolating, because it shows that \emph{reconstruction accuracy and memorization sensitivity are not the same quantity}. On TabDDPM and Synthpop, the continuous CoBP-RA has by far the \emph{worst} training error ($\mathrm{tr}=0.158$ and $0.191$, versus $0.067$ and $0.107$ for the random forest) yet by far the \emph{largest} memorization gap ($\Delta=-0.112$ and $-0.115$, versus $-0.034$ and $-0.033$). A smooth global regressor like Bayesian ridge, conversely, reconstructs reasonably but registers essentially no gap ($\Delta\approx-0.001$). The explanation is what each attack keys on: CoBP-RA's per-record neighbourhood couplings (local PMI from a target's nearest synthetic rows) latch onto exactly the idiosyncratic joint structure a memorizing generator copies from individual training records, so they fire much harder on members than on holdouts even though the resulting point estimate is noisier overall. Bayesian ridge, fitting one global linear map, cannot express that record-specific structure and so neither reconstructs it nor detects it. The practical lesson for auditing: the best memorization \emph{detector} need not be the best \emph{reconstructor}; an attack that exploits local joint structure can expose memorization that a lower-error but globally-smooth attack completely misses.

\subsection{Disparate Impact: Per-Attack and Per-Feature}\label{sec:disparity_perfeature_appendix}

These tables support the claim in Section~\ref{sec:disparate} that disparate exposure is set by the \emph{generator}, not by the attack, and that it concentrates on features whose values are individually identifying. Both use the same setting as Table~\ref{tab:disparate_impact} (Adult, 10k, $\text{QI}_{\text{demo}}$); outliers are the top-$10\%$ by QI-space Isolation Forest.

Table~\ref{tab:disparity_perfeature} localizes the penalty. Under both leaking generators the gap is positive on individually identifying features (\texttt{occupation}, \texttt{hours-per-week}) and negative on features dominated by one common value (\texttt{education-num}, \texttt{relationship}). \texttt{fnlwgt} is the sharpest mechanism contrast: record retention exposes it for outliers (Cell Supp.\ ${+}36$) but synthesis cannot reproduce a near-unique value even for them (TabDDPM ${+}2$). \texttt{income} is the one feature whose gap is a property of the data rather than the release: it favors outliers under both (about ${+}10$), and the Mode baseline shows it too.

\begin{table}[ht!]
  \centering
  \small
  \caption{Table~\ref{tab:disparate_impact}'s outlier penalty broken out by attack $\times$ generator: each cell is the \emph{ratio} of mean row-level $R_{adv}$ on the outlier $10\%$ to that on the typical $90\%$ (${>}1$ = atypical individuals reconstructed more accurately), with the absolute mean row $R_{adv}$ (\%) in parentheses. The ratio is meaningful only where the attack reconstructs: cells at or above $1.6\%$ mean row $R_{adv}$ are bold, and ratios on near-zero cells (noise over noise) should be ignored. $\dagger$~= novel attack. (Adult, 10k rows, $\text{QI}_{\text{demo}}$.)}
  \label{tab:disparity_perattack}
  \resizebox{\columnwidth}{!}{%
  \begin{tabular}{l cccc}
    \toprule
    \textbf{Attack} & CellSupp.\ & TabDDPM & Synthpop & MST $(\varepsilon{=}1)$ \\
    \midrule
    Mode          & 1.05 \tiny(0.2)          & 0.73 \tiny(0.1)          & 0.67 \tiny(0.1)          & 0.76 \tiny(0.1) \\
    KNN           & \textbf{2.18} \tiny(34.5) & \textbf{2.88} \tiny(1.6) & 0.79 \tiny(1.5)          & 0.93 \tiny(0.2) \\
    NaiveBayes    & \textbf{2.18} \tiny(32.1) & 3.61 \tiny(1.0)          & \textbf{1.31} \tiny(1.7) & 1.04 \tiny(0.1) \\
    RandomForest  & \textbf{2.17} \tiny(34.7) & \textbf{2.56} \tiny(1.7) & \textbf{0.92} \tiny(1.9) & 1.00 \tiny(0.2) \\
    CoBP-RA$^\dagger$ & \textbf{2.17} \tiny(34.7) & \textbf{2.62} \tiny(1.7) & \textbf{1.15} \tiny(1.8) & 1.14 \tiny(0.3) \\
    MLP           & 2.97 \tiny(0.6)          & 2.79 \tiny(0.5)          & \textbf{1.14} \tiny(1.9) & 1.30 \tiny(0.3) \\
    LightGBM      & 2.96 \tiny(0.9)          & 1.51 \tiny(0.2)          & 1.65 \tiny(0.4)          & 1.24 \tiny(0.3) \\
    \bottomrule
  \end{tabular}}
\end{table}

\subsection{Data Qualities}\label{sec:quality_appendix}

Table~\ref{tab:quality_overview} gives the fidelity/utility/vulnerability profile summarized graphically by Figure~\ref{fig:quality_overview} in the main text (Adult). Tables~\ref{tab:quality_arizona}--\ref{tab:quality_california} extend it to the remaining four datasets, using the metric definitions of Table~\ref{tab:quality_overview} together with one additional column, \textbf{Prop.\ AUC}~\cite{snoke2018general} (propensity AUC: the AUC of a classifier trained to distinguish real from synthetic records; ${\downarrow}$, $0.5$ = indistinguishable), and the same favorable-direction arrows; they let the reconstruction results of Section~\ref{sec:conditions} be read against the quality of the release each attack faced. The recurring pattern is that the de-identification methods (RankSwap, CellSuppression) lead most fidelity metrics because they retain real records (which is also what makes them the most exposed), while the DP generators trade fidelity for their $\varepsilon$ guarantee. The four datasets sharpen this differently. On \textbf{NIST Arizona} (Table~\ref{tab:quality_arizona}) TabDDPM is the strongest \emph{synthetic} generator (TSTR ratio $1.02$) and MST fidelity climbs steadily with $\varepsilon$. On \textbf{CDC Diabetes} (Table~\ref{tab:quality_cdc}), whose features are mostly binary, fidelity is high across the board and three generators saturate the TSTR-ratio cap. On \textbf{NIST SBO} (Table~\ref{tab:quality_sbo}) the $130$ noisy financial columns invert the usual order: TabDDPM collapses (Col.\ pairs $0.35$) while RankSwap and the higher-$\varepsilon$ MST models fare best. And on \textbf{California Housing} (Table~\ref{tab:quality_california}), fully continuous, the DP marginal models lose most utility (TSTR $R^2\!\to\!0$ at small $\varepsilon$) whereas TabDDPM and RankSwap retain it.

\begin{table}[t]
  \centering
  \small
  \caption{Per-feature outlier gap for the two leaking generators: each cell is mean row-level $R_{adv}$ on the outlier $10\%$ \emph{minus} that on the typical $90\%$, in percentage points, averaged over KNN, NaiveBayes, Random Forest, and CoBP-RA, so a positive value means the feature is reconstructed \emph{better} for atypical individuals. We take a difference rather than Table~\ref{tab:disparity_perattack}'s ratio because many per-feature scores are near zero for everyone, where a ratio is unstable or undefined. (Adult, 10k rows, $\text{QI}_{\text{demo}}$.)}
  \label{tab:disparity_perfeature}
  \small
  \resizebox{\columnwidth}{!}{%
  \begin{tabular}{l l rr}
    \toprule
    \textbf{Feature} & \textbf{Type} & CellSupp.\ & TabDDPM \\
    \midrule
    \texttt{fnlwgt}         & cont.\ (near-unique) & $+36$ & $+2$  \\
    \texttt{occupation}     & categorical          & $+22$ & $+21$ \\
    \texttt{hours-per-week} & numeric              & $+22$ & $+18$ \\
    \texttt{income}         & binary               & $+11$ & $+11$ \\
    \texttt{workclass}      & categorical          & $+7$  & $+6$  \\
    \texttt{capital-gain}   & cont.\ (sparse)      & $+7$  & $+4$  \\
    \texttt{capital-loss}   & cont.\ (sparse)      & $+4$  & $+5$  \\
    \midrule
    \texttt{education-num}  & ordinal              & $-5$  & $-4$  \\
    \texttt{relationship}   & categorical          & $-8$  & $-7$  \\
    \bottomrule
  \end{tabular}}
\end{table}

\begin{table}[t]
  \centering
  \caption{Memorization gaps of Table~\ref{tab:memorization_and_ds_risk} with their $95\%$ Student-$t$ intervals and one-sample $t$-test $p$-values against zero, over the $5$ disjoint samples. (Adult, 1k rows, $\text{QI}_{\text{linear}}$; RF+KNN+MLP.)}
  \label{tab:memo_stats}
  \small
  \begin{tabular}{@{}lrcr@{}}
    \toprule
    SDG method & $\Delta$ (pp) & 95\% CI & $p$ \\
    \midrule
    \multicolumn{4}{@{}l}{\emph{De-identification}} \\
    Cell Supp. & $+20.0$ & $[+17.6, +22.4]$ & $<0.001$ \\
    RankSwap & $+13.1$ & $[+10.8, +15.3]$ & $<0.001$ \\
    \midrule
    \multicolumn{4}{@{}l}{\emph{Non-DP synthesizers}} \\
    TabDDPM & $+13.5$ & $[+10.9, +16.1]$ & $<0.001$ \\
    Synthpop & $+1.6$ & $[-2.4, +5.7]$ & $0.32$ \\
    CTGAN & $+0.5$ & $[-1.9, +2.9]$ & $0.61$ \\
    TVAE & $-0.1$ & $[-1.6, +1.4]$ & $0.89$ \\
    ARF & $+2.3$ & $[-3.2, +7.8]$ & $0.31$ \\
    \midrule
    \multicolumn{4}{@{}l}{\emph{DP synthesizers}} \\
    AIM $(\varepsilon{=}1)$ & $+0.3$ & $[-2.3, +3.0]$ & $0.73$ \\
    MST $(\varepsilon{=}1)$ & $+0.2$ & $[-1.6, +2.1]$ & $0.77$ \\
    MST $(\varepsilon{=}10)$ & $+1.0$ & $[-3.3, +5.3]$ & $0.55$ \\
    MST $(\varepsilon{=}1000)$ & $+1.0$ & $[-4.5, +6.5]$ & $0.64$ \\
    PrivBayes $(\varepsilon{=}1)$ & $-0.1$ & $[-1.7, +1.6]$ & $0.93$ \\
    PrivBayes $(\varepsilon{=}1000)$ & $+0.8$ & $[-0.9, +2.5]$ & $0.25$ \\
    PrivSyn $(\varepsilon{=}1)$ & $+0.6$ & $[-0.7, +1.9]$ & $0.26$ \\
    PrivSyn $(\varepsilon{=}1000)$ & $+0.5$ & $[-4.0, +5.0]$ & $0.77$ \\
    MWEM-PGM $(\varepsilon{=}1)$ & $+0.4$ & $[-1.6, +2.3]$ & $0.63$ \\
    MWEM-PGM $(\varepsilon{=}1000)$ & $+0.7$ & $[-1.3, +2.7]$ & $0.39$ \\
    PrivateGSD $(\varepsilon{=}1)$ & $+2.1$ & $[-0.4, +4.5]$ & $0.08$ \\
    PrivateGSD $(\varepsilon{=}1000)$ & $+1.2$ & $[-0.4, +2.9]$ & $0.11$ \\
    \bottomrule
  \end{tabular}
\end{table}

\begin{table*}[t]
  \centering
  \caption{Paired $t$-tests on the privacy-budget curve: change in $R_{adv}$ (percentage points, mean of Random Forest, Naive Bayes, and CoBP-RA) between two budgets, paired on the disjoint training sample. Shading marks significance: \sigkeyA, \sigkeyB, \sigkeyC, unshaded not significant; `---' is not evaluated. (Adult, 10k rows, all three QIs.)}
  \label{tab:eps_stats}
  \small
  \setlength{\tabcolsep}{4pt}
  \begin{tabular}{@{}l rrr rrr rrr@{}}
    \toprule
     & \multicolumn{3}{c}{$\text{QI}_{\text{demo}}$} & \multicolumn{3}{c}{$\text{QI}_{\text{large}}$} & \multicolumn{3}{c}{$\text{QI}_{\text{behavioral}}$} \\
    \cmidrule(lr){2-4}\cmidrule(lr){5-7}\cmidrule(l){8-10}
    Generator \hspace{4pt} $\varepsilon$: & $0.1{\to}10$ & $1{\to}10$ & $10{\to}1000$ & $0.1{\to}10$ & $1{\to}10$ & $10{\to}1000$ & $0.1{\to}10$ & $1{\to}10$ & $10{\to}1000$ \\
    \midrule
    MST & \sigA{$+8.6$} & \sigB{$+3.4$} & \sigB{$+0.8$} & \sigA{$+10.9$} & \sigB{$+6.2$} & \sigC{$+1.2$} & \sigA{$+7.5$} & \sigA{$+3.3$} & \sigB{$+1.1$} \\
    AIM & \sigA{$+7.8$} & \sigA{$+2.5$} & --- & \sigA{$+9.7$} & \sigB{$+4.2$} & --- & \sigA{$+8.3$} & \sigB{$+3.1$} & --- \\
    PrivBayes & \sigA{$+8.3$} & \sigA{$+2.8$} & $-0.8$ & \sigA{$+11.4$} & \sigA{$+4.5$} & $-0.9$ & \sigA{$+8.9$} & \sigA{$+3.2$} & $-0.6$ \\
    PrivSyn & \sigA{$+6.4$} & \sigA{$+3.1$} & $+0.1$ & \sigA{$+5.7$} & \sigB{$+2.5$} & $+0.6$ & \sigA{$+6.6$} & \sigA{$+2.7$} & $+0.4$ \\
    MWEM-PGM & \sigA{$+6.6$} & $+0.3$ & $0.0$ & \sigA{$+7.0$} & $+0.8$ & $+0.1$ & \sigA{$+8.6$} & \sigC{$+1.5$} & $-0.2$ \\
    PrivateGSD & \sigA{$+7.8$} & \sigA{$+3.7$} & \sigA{$+2.2$} & \sigA{$+7.7$} & \sigB{$+3.6$} & \sigB{$+3.4$} & \sigA{$+7.7$} & \sigB{$+3.5$} & \sigB{$+2.3$} \\
    \bottomrule
  \end{tabular}
\end{table*}

\begin{table*}[ht]
  \centering
  \small
  \caption{Synthetic-data quality profile and reconstruction vulnerability for Adult: the per-release values behind Figure~\ref{fig:quality_overview}, where the metrics are defined. Arrows (${\uparrow}$/${\downarrow}$) in the column headers give the favorable direction, and \textbf{RF $R_{adv}$} is averaged as in Table~\ref{tab:ra_mean_adult}. \textit{Train--Train} (real vs.\ real) is a fidelity benchmark; best per column excluding it in \textbf{bold}. All six DP synthesizers appear at $\varepsilon{=}1$ and $\varepsilon{=}1000$, so fidelity and vulnerability can be read against the budget. The same profile for the other four datasets follows below (Tables~\ref{tab:quality_arizona}--\ref{tab:quality_california}). (15 features, 10k rows.)}
  \label{tab:quality_overview}
  \resizebox{\textwidth}{!}{%
  \begin{tabular}{lrrrrrrr|r}
    \toprule
    \textbf{SDG method} & \textbf{TSTR ratio} $\uparrow$ & \textbf{Mean JSD} $\downarrow$ & \textbf{Col.\ shapes} $\uparrow$ & \textbf{Pairwise TVD} $\downarrow$ & \textbf{Col.\ pairs} $\uparrow$ & \textbf{Corr.\ diff} $\downarrow$ & \textbf{Wass.\ OHE} $\downarrow$ & \textbf{RF $R_{adv}$} \\
    \midrule
MST $(\varepsilon{=}0.1)$ & 0.651 & 0.101 & 0.766 & 0.158 & 0.840 & 0.068 & 2.939 & 16.0 \\
MST $(\varepsilon{=}1)$ & 0.764 & 0.028 & 0.820 & 0.058 & 0.943 & 0.055 & 0.451 & 22.1 \\
MST $(\varepsilon{=}10)$ & 0.837 & 0.007 & 0.826 & 0.039 & 0.959 & 0.052 & 0.067 & 24.1 \\
MST $(\varepsilon{=}1000)$ & 0.839 & 0.002 & 0.827 & 0.037 & 0.961 & 0.051 & 0.024 & 25.4 \\
AIM $(\varepsilon{=}1)$ & 0.983 & 0.035 & 0.819 & 0.049 & 0.946 & 0.041 & 0.512 & 20.9 \\
AIM $(\varepsilon{=}10)$ & 0.989 & 0.011 & 0.824 & 0.020 & 0.972 & 0.034 & 0.161 & 21.9 \\
AIM $(\varepsilon{=}1000)$ & 0.991 & 0.007 & 0.825 & 0.014 & 0.980 & 0.015 & 0.152 & 22.2 \\
PrivBayes $(\varepsilon{=}1)$ & 0.823 & 0.258 & 0.607 & 0.444 & 0.577 & 0.083 & 10.919 & 20.8 \\
PrivBayes $(\varepsilon{=}1000)$ & 0.913 & 0.012 & 0.821 & 0.055 & 0.951 & 0.036 & 0.345 & 19.3 \\
PrivSyn $(\varepsilon{=}1)$ & 0.917 & 0.041 & 0.907 & 0.075 & 0.901 & 0.117 & 0.730 & 16.4 \\
PrivSyn $(\varepsilon{=}1000)$ & 0.971 & 0.004 & 0.918 & 0.025 & 0.951 & 0.022 & 0.137 & 19.5 \\
MWEM-PGM $(\varepsilon{=}1)$ & 0.719 & 0.037 & 0.812 & 0.063 & 0.890 & 0.083 & 0.767 & 21.6 \\
MWEM-PGM $(\varepsilon{=}1000)$ & 0.831 & 0.006 & 0.820 & 0.037 & 0.909 & 0.059 & 0.258 & 22.2 \\
PrivateGSD $(\varepsilon{=}1)$ & 0.973 & 0.037 & 0.816 & 0.067 & 0.930 & 0.148 & 0.647 & 18.0 \\
PrivateGSD $(\varepsilon{=}1000)$ & 0.988 & 0.001 & 0.827 & 0.003 & \textbf{0.991} & \textbf{0.005} & \textbf{0.007} & 21.8 \\
\midrule
RankSwap & 0.972 & \textbf{0.000} & 0.987 & \textbf{0.000} & 0.845 & 0.012 & 0.608 & 25.7 \\
Cell Supp. & \textbf{1.021} & 0.016 & \textbf{0.993} & 0.015 & 0.986 & \textbf{0.005} & 0.216 & 52.3 \\
\midrule
Synthpop & 1.002 & 0.011 & \textbf{0.993} & 0.024 & 0.978 & 0.011 & 0.272 & 28.8 \\
TVAE & 0.982 & 0.103 & 0.871 & 0.139 & 0.789 & 0.045 & 3.103 & 24.1 \\
CTGAN & 0.964 & 0.089 & 0.862 & 0.168 & 0.823 & 0.035 & 2.753 & 21.2 \\
ARF & 0.973 & 0.092 & 0.894 & 0.187 & 0.841 & 0.014 & 3.016 & 20.5 \\
TabDDPM & 1.018 & 0.016 & 0.990 & 0.022 & 0.977 & 0.014 & 0.332 & 38.9 \\
\midrule
\textit{Train--Train} & --- & 0.014 & 0.990 & 0.026 & 0.970 & 0.010 & --- & --- \\
    \bottomrule
  \end{tabular}
  }
\end{table*}

\begin{table*}[ht]
  \centering
  \small
  \caption{Synthetic-data quality profile for NIST Arizona; metrics as defined in Figure~\ref{fig:quality_overview}, best per column excluding \textit{Train--Train} in \textbf{bold}. (25 features, 10k rows.)}
  \label{tab:quality_arizona}
  \resizebox{\textwidth}{!}{%
  \begin{tabular}{lrrrrrrr}
    \toprule
    \textbf{SDG method} & \textbf{TSTR ratio} $\uparrow$ & \textbf{Mean JSD} $\downarrow$ & \textbf{Pairwise TVD} $\downarrow$ & \textbf{Wass.\ OHE} $\downarrow$ & \textbf{Corr.\ diff} $\downarrow$ & \textbf{Col.\ pairs} $\uparrow$ & \textbf{Prop.\ AUC} $\downarrow$ \\
    \midrule
    MST $(\varepsilon{=}0.1)$ & 0.711 & 0.101 & 0.140 & 4.301 & 0.157 & 0.829 & 0.869 \\
    MST $(\varepsilon{=}1)$ & 0.778 & 0.036 & 0.048 & 0.859 & 0.061 & 0.943 & 0.920 \\
    MST $(\varepsilon{=}10)$ & 0.719 & 0.010 & 0.026 & 0.138 & 0.046 & 0.963 & 0.918 \\
    MST $(\varepsilon{=}100)$ & 0.770 & 0.005 & 0.023 & 0.058 & 0.042 & 0.966 & 0.905 \\
    MST $(\varepsilon{=}1000)$ & 0.774 & 0.004 & 0.023 & \textbf{0.054} & 0.043 & 0.966 & 0.924 \\
    AIM $(\varepsilon{=}1)$ & 0.796 & 0.054 & 0.055 & 5.637 & 0.064 & 0.890 & 0.588 \\
    \midrule
    RankSwap & 1.041 & \textbf{0.000} & \textbf{0.000} & 0.245 & 0.035 & 0.966 & \textbf{0.451} \\
    Cell Supp. & \textbf{1.045} & 0.012 & 0.013 & 0.361 & \textbf{0.007} & \textbf{0.977} & 0.517 \\
    \midrule
    Synthpop & 0.986 & 0.010 & 0.033 & 0.303 & 0.009 & 0.919 & 0.499 \\
    TVAE & 0.897 & 0.094 & 0.129 & 3.566 & 0.039 & 0.827 & 0.599 \\
    CTGAN & 0.712 & 0.090 & 0.133 & 4.223 & 0.053 & 0.829 & 0.655 \\
    ARF & 0.961 & 0.064 & 0.128 & 3.504 & 0.036 & 0.798 & 0.561 \\
    TabDDPM & 1.019 & 0.020 & 0.024 & 0.620 & 0.022 & 0.963 & 0.506 \\
    \midrule
    \textit{Train--Train} & -- & 0.013 & 0.019 & -- & 0.009 & 0.977 & -- \\
    \bottomrule
  \end{tabular}}
\end{table*}

\begin{table*}[ht]
  \centering
  \small
  \caption{Synthetic-data quality profile for CDC Diabetes; metrics as defined in Figure~\ref{fig:quality_overview}, best per column excluding \textit{Train--Train} in \textbf{bold}. (22 features, 1k rows.)}
  \label{tab:quality_cdc}
  \resizebox{\textwidth}{!}{%
  \begin{tabular}{lrrrrrrr}
    \toprule
    \textbf{SDG method} & \textbf{TSTR ratio} $\uparrow$ & \textbf{Mean JSD} $\downarrow$ & \textbf{Pairwise TVD} $\downarrow$ & \textbf{Wass.\ OHE} $\downarrow$ & \textbf{Corr.\ diff} $\downarrow$ & \textbf{Col.\ pairs} $\uparrow$ & \textbf{Prop.\ AUC} $\downarrow$ \\
    \midrule
    MST $(\varepsilon{=}0.1)$ & 0.763 & 0.210 & 0.386 & 12.952 & 0.193 & 0.568 & 0.983 \\
    MST $(\varepsilon{=}1)$ & 0.902 & 0.057 & 0.105 & 4.176 & 0.126 & 0.793 & 0.823 \\
    MST $(\varepsilon{=}10)$ & 0.916 & 0.007 & 0.033 & 4.103 & 0.133 & 0.939 & 0.575 \\
    MST $(\varepsilon{=}100)$ & 0.910 & 0.002 & 0.025 & 4.250 & 0.087 & 0.955 & 0.539 \\
    MST $(\varepsilon{=}1000)$ & 0.886 & 0.001 & 0.026 & 4.364 & 0.069 & 0.955 & 0.548 \\
    AIM $(\varepsilon{=}1)$ & 0.898 & 0.063 & 0.099 & 4.077 & 0.193 & 0.770 & 0.785 \\
    AIM $(\varepsilon{=}10)$ & 0.941 & 0.009 & 0.028 & 4.128 & 0.169 & 0.948 & 0.569 \\
    \midrule
    RankSwap & 1.100 & \textbf{0.000} & \textbf{0.000} & \textbf{0.169} & \textbf{0.017} & \textbf{0.989} & 0.485 \\
    Cell Supp. & \textbf{1.100} & 0.083 & 0.141 & 3.593 & 0.087 & 0.834 & 0.749 \\
    \midrule
    Synthpop & 1.053 & 0.014 & 0.031 & 0.710 & 0.034 & 0.958 & \textbf{0.473} \\
    TVAE & 0.994 & 0.137 & 0.198 & 5.706 & 0.073 & 0.720 & 0.848 \\
    CTGAN & 0.892 & 0.069 & 0.141 & 4.494 & 0.197 & 0.779 & 0.778 \\
    ARF & 0.987 & 0.023 & 0.047 & 1.760 & 0.035 & 0.891 & 0.524 \\
    TabDDPM & 1.100 & 0.019 & 0.034 & 0.950 & 0.031 & 0.953 & 0.501 \\
    \midrule
    \textit{Train--Train} & -- & 0.018 & 0.037 & -- & 0.039 & 0.948 & -- \\
    \bottomrule
  \end{tabular}}
\end{table*}

\begin{table*}[ht]
  \centering
  \small
  \caption{Synthetic-data quality profile for NIST SBO; metrics as defined in Figure~\ref{fig:quality_overview}, best per column excluding \textit{Train--Train} in \textbf{bold}. Pairwise TVD is omitted (not computed for ${>}30$ categorical columns); the large Wasserstein values reflect SBO's noisy continuous financial variables dominating the OHE-space distance. (130 features, 1k rows.)}
  \label{tab:quality_sbo}
  \resizebox{\textwidth}{!}{%
  \begin{tabular}{lrrrrrr}
    \toprule
    \textbf{SDG method} & \textbf{TSTR ratio} $\uparrow$ & \textbf{Mean JSD} $\downarrow$ & \textbf{Wass.\ OHE} $\downarrow$ & \textbf{Corr.\ diff} $\downarrow$ & \textbf{Col.\ pairs} $\uparrow$ & \textbf{Prop.\ AUC} $\downarrow$ \\
    \midrule
    MST $(\varepsilon{=}0.1)$ & 0.979 & 0.407 & 131.4 & 0.263 & 0.310 & 0.999 \\
    MST $(\varepsilon{=}1)$ & 1.013 & 0.103 & 21.1 & 0.275 & 0.847 & 0.953 \\
    MST $(\varepsilon{=}10)$ & 0.992 & 0.029 & 4.7 & 0.177 & 0.946 & 0.801 \\
    MST $(\varepsilon{=}100)$ & 0.990 & 0.006 & 1.9 & 0.175 & 0.969 & 0.918 \\
    MST $(\varepsilon{=}1000)$ & 0.997 & 0.004 & 3.5 & 0.141 & 0.971 & 0.958 \\
    \midrule
    RankSwap & \textbf{1.100} & \textbf{0.000} & \textbf{0.368} & 0.073 & \textbf{1.000} & \textbf{0.480} \\
    Cell Supp. & 1.004 & 0.059 & 9.572 & 0.087 & 0.959 & 0.612 \\
    \midrule
    Synthpop & 1.011 & 0.019 & 3.399 & \textbf{0.051} & 0.975 & 0.486 \\
    TVAE & 0.989 & 0.160 & 29.370 & 0.144 & 0.852 & 0.946 \\
    CTGAN & 0.994 & 0.118 & 25.342 & 0.270 & 0.837 & 0.799 \\
    ARF & 1.003 & 0.027 & 6.713 & 0.073 & 0.966 & 0.603 \\
    TabDDPM & 1.027 & 0.423 & 118.2 & 0.486 & 0.353 & 0.997 \\
    \midrule
    \textit{Train--Train} & -- & 0.021 & -- & 0.088 & 0.978 & -- \\
    \bottomrule
  \end{tabular}}
\end{table*}

\begin{table*}[ht]
  \centering
  \small
  \caption{Synthetic-data quality profile for California Housing; metrics as defined in Figure~\ref{fig:quality_overview}, best per column excluding \textit{Train--Train} in \textbf{bold}. TSTR ratio uses regression $R^2$ (not F1-macro); the categorical-marginal metrics (mean JSD, pairwise TVD) are undefined for fully continuous data and omitted, and Col.\ pairs is computed over discretized columns. (9 features, 1k rows.)}
  \label{tab:quality_california}
  \resizebox{\textwidth}{!}{%
  \begin{tabular}{lrrrrr}
    \toprule
    \textbf{SDG method} & \textbf{TSTR ratio} $\uparrow$ & \textbf{Wass.\ OHE} $\downarrow$ & \textbf{Corr.\ diff} $\downarrow$ & \textbf{Col.\ pairs} $\uparrow$ & \textbf{Prop.\ AUC} $\downarrow$ \\
    \midrule
    MST $(\varepsilon{=}0.1)$ & 0.000 & 7.313 & 0.174 & 0.610 & 0.977 \\
    MST $(\varepsilon{=}1)$ & 0.000 & 10.046 & -- & 0.692 & 0.947 \\
    MST $(\varepsilon{=}10)$ & 0.484 & 8.319 & 0.124 & 0.760 & 0.870 \\
    MST $(\varepsilon{=}100)$ & 0.610 & 9.238 & 0.121 & 0.795 & 0.815 \\
    MST $(\varepsilon{=}1000)$ & 0.586 & 11.354 & 0.101 & 0.903 & 0.933 \\
    AIM $(\varepsilon{=}1)$ & 0.000 & 9.981 & -- & 0.595 & 0.901 \\
    AIM $(\varepsilon{=}10)$ & 0.550 & 8.282 & 0.136 & 0.691 & 0.837 \\
    \midrule
    RankSwap & \textbf{1.100} & 0.795 & 0.047 & 0.914 & 0.488 \\
    \midrule
    Synthpop & 1.046 & 0.668 & \textbf{0.036} & \textbf{0.955} & 0.495 \\
    TVAE & 0.708 & 5.548 & 0.121 & 0.706 & 0.913 \\
    CTGAN & 0.076 & 7.066 & 0.167 & 0.617 & 0.945 \\
    ARF & 0.854 & 2.118 & 0.050 & 0.888 & 0.547 \\
    TabDDPM & 1.100 & \textbf{0.630} & 0.043 & 0.953 & \textbf{0.475} \\
    \midrule
    \textit{Train--Train} & -- & -- & 0.043 & 0.965 & -- \\
    \bottomrule
  \end{tabular}}
\end{table*}


\subsection{Full \texorpdfstring{$\varepsilon$}{epsilon} Sweep, per Attack and QI}\label{sec:eps_appendix}

\begin{figure*}[h]
  \centering
  \includegraphics[width=.87\textwidth]{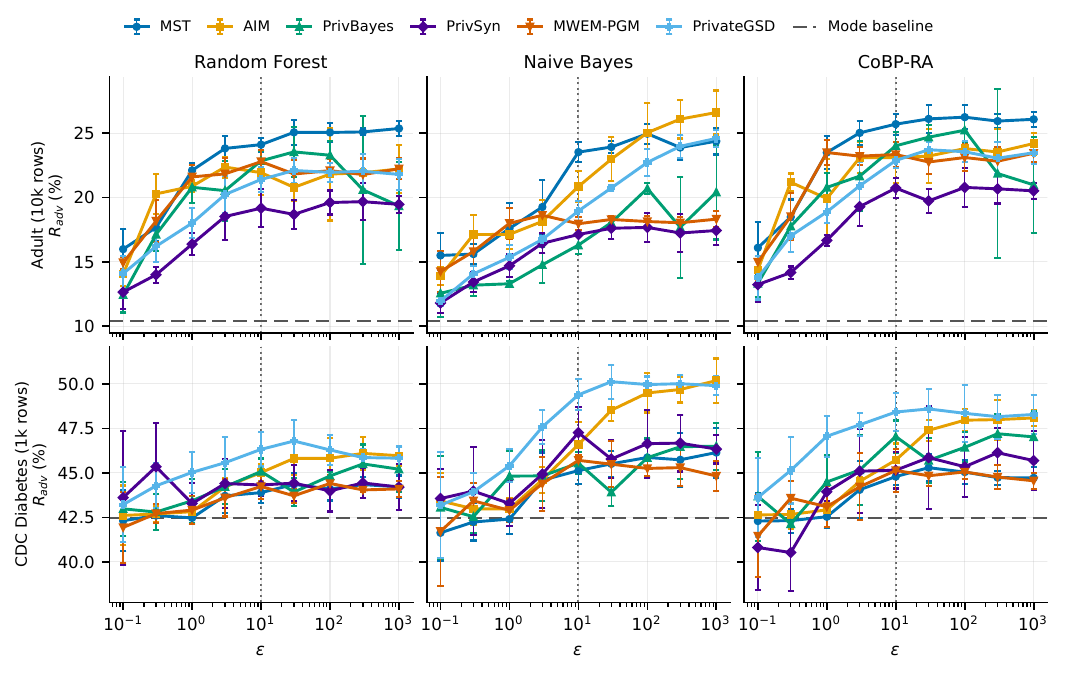}
  \caption{Per-attack view of Figure~\ref{fig:eps_curves}: reconstruction risk against privacy budget (log scale), one column per attack and one row per dataset. Bars are $95\%$ confidence intervals, the dotted line marks $\varepsilon{=}10$, and the dashed rule is the Mode baseline. (Adult 10k and CDC Diabetes 1k, $\text{QI}_{\text{demo}}$; Random Forest, Naive Bayes and CoBP-RA.)}
  \Description{Six line plots in two rows of three. Columns are the attacks Random Forest, Naive Bayes and CoBP-RA; the top row is Adult 10k rows and the bottom row CDC Diabetes 1k rows, each row sharing one vertical scale. In every panel the horizontal axis is the privacy budget epsilon on a logarithmic scale from 0.1 to 1000 and the vertical axis is reconstruction advantage, with one coloured line per DP synthesizer (MST, AIM, PrivBayes, PrivSyn, MWEM-PGM and PrivateGSD), error bars for 95 percent confidence intervals, a dotted vertical line at epsilon equals 10, and a dashed horizontal Mode-baseline rule. Across all six panels the curves rise steeply up to roughly epsilon equals 10 and are nearly flat above it; the main exceptions are Naive Bayes on AIM in both rows and on PrivateGSD in the Adult row, which keep climbing. On Adult, MST is the highest curve at moderate and large budgets; on CDC, PrivateGSD and AIM are highest. The Adult panels sit some fifteen points above their Mode baseline, whereas the CDC panels sit only a few points above theirs.}
  \label{fig:eps_curves_perattack}
\end{figure*}

\begin{figure*}[h]
  \centering
  \includegraphics[width=.7\textwidth]{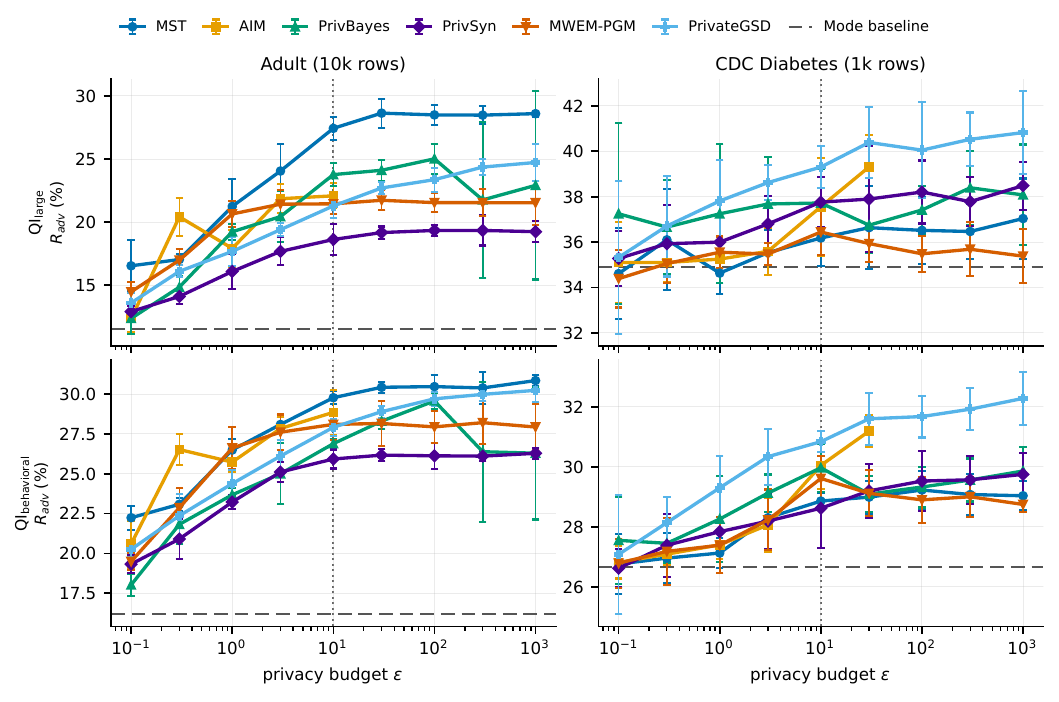}
  \caption{Figure~\ref{fig:eps_curves} for the two further QI sets (defined in Tables~\ref{tab:qi_def_adult} and~\ref{tab:qi_def_cdc}): reconstruction risk against privacy budget (log scale), mean of Random Forest, Naive Bayes, and CoBP-RA, one row per QI and one column per dataset. Bars are $95\%$ confidence intervals, the dotted line marks $\varepsilon{=}10$, and the dashed rule is the Mode baseline. AIM's runtime grows steeply with the budget, so on these QIs it is swept only to $\varepsilon{=}10$ (Adult) and $\varepsilon{=}30$ (CDC). (Adult 10k and CDC Diabetes 1k, $\text{QI}_{\text{large}}$ and $\text{QI}_{\text{behavioral}}$.)}
  \Description{Four line plots in a two-by-two grid. Rows are the quasi-identifier sets QI-large (top) and QI-behavioral (bottom); columns are Adult with 10k rows (left) and CDC Diabetes with 1k rows (right). Each panel plots reconstruction advantage against the privacy budget epsilon on a logarithmic axis from 0.1 to 1000, with one coloured line and marker per DP synthesizer (MST, AIM, PrivBayes, PrivSyn, MWEM-PGM, PrivateGSD), 95 percent error bars, a dotted vertical line at epsilon 10 and a dashed horizontal Mode baseline; the AIM line stops at epsilon 10 on Adult and 30 on CDC. On Adult the curves rise from 12 to 17 at epsilon 0.1 to between 19 and 27 at epsilon 10 (QI-large), and from 18 to 22 to between 26 and 30 (QI-behavioral), then flatten, with MST highest, PrivSyn lowest, and PrivBayes dipping at epsilon 300. On CDC every curve stays within about six points of its Mode baseline, rising gently, with PrivateGSD highest.}
  \label{fig:eps_curves_otherqi}
\end{figure*}

\begin{figure*}[h]
  \centering
  \includegraphics[width=.87\textwidth]{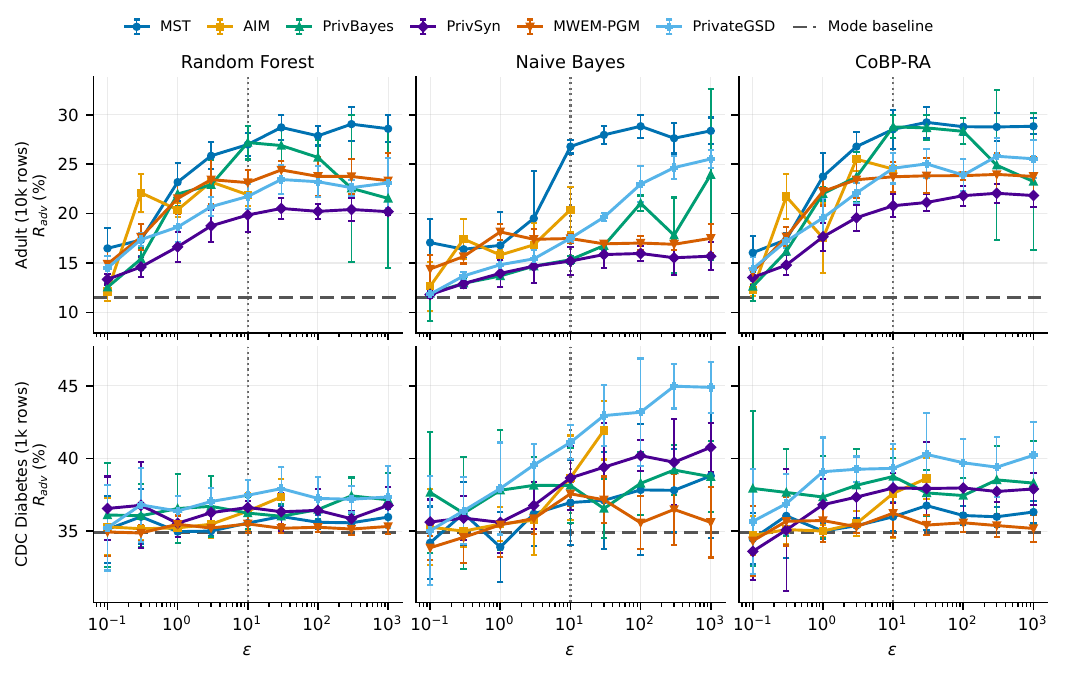}
  \caption{Per-attack view of Figure~\ref{fig:eps_curves_otherqi}, drawn as in Figure~\ref{fig:eps_curves_perattack}. (Adult 10k and CDC Diabetes 1k, $\text{QI}_{\text{large}}$; Random Forest, Naive Bayes and CoBP-RA.)}
  \Description{Six line plots in two rows of three, laid out as the QI-demo per-attack figure: columns Random Forest, Naive Bayes and CoBP-RA; top row Adult 10k, bottom row CDC Diabetes 1k; each panel plots reconstruction advantage against epsilon (log scale) with one line per DP synthesizer, error bars, a dotted line at epsilon 10 and a dashed Mode baseline. On Adult, Random Forest and CoBP-RA curves rise until epsilon 10 and flatten, with MST highest; under Naive Bayes, MST stays low until epsilon 3 and then jumps, while PrivateGSD and PrivBayes keep climbing to epsilon 1000. On CDC the curves sit a few points above the Mode baseline, with PrivateGSD highest under Naive Bayes.}
  \label{fig:eps_curves_perattack_qilarge}
\end{figure*}

\begin{figure*}[h]
  \centering
  \includegraphics[width=.87\textwidth]{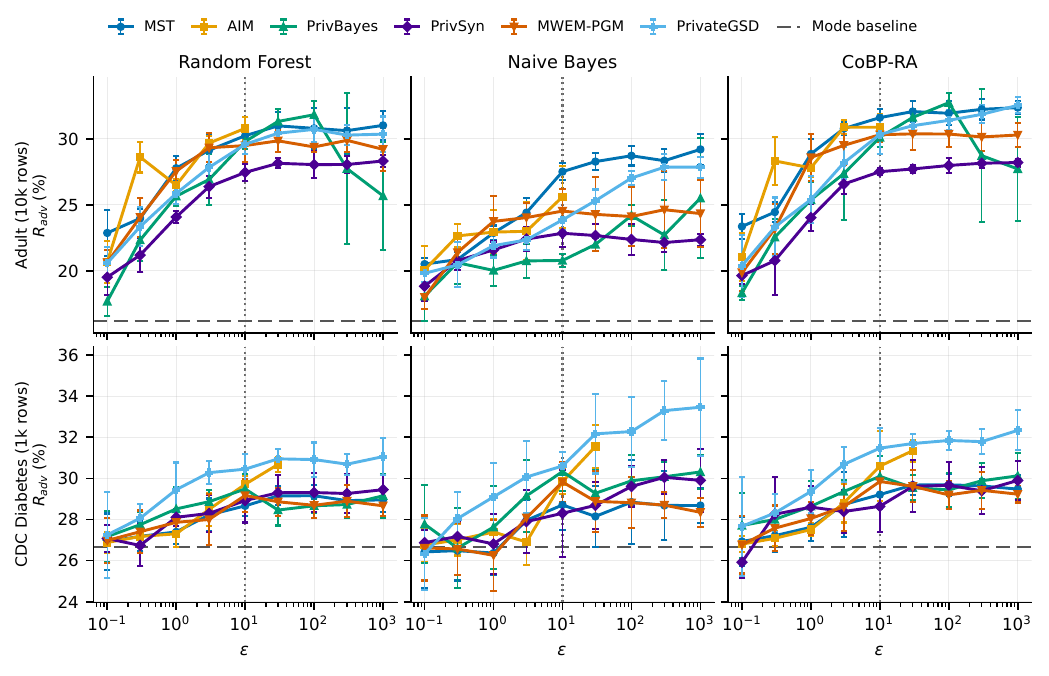}
  \caption{Per-attack view of Figure~\ref{fig:eps_curves_otherqi}, drawn as in Figure~\ref{fig:eps_curves_perattack}. (Adult 10k and CDC Diabetes 1k, $\text{QI}_{\text{behavioral}}$; Random Forest, Naive Bayes and CoBP-RA.)}
  \Description{Six line plots in two rows of three: columns Random Forest, Naive Bayes and CoBP-RA; top row Adult 10k, bottom row CDC Diabetes 1k; one line per DP synthesizer against epsilon on a log scale, with error bars, a dotted line at epsilon 10 and a dashed Mode baseline. On Adult the Random Forest and CoBP-RA curves climb from about 20 to 30 by epsilon 10 and then flatten, while PrivBayes drops sharply at epsilon 300 and 1000; Naive Bayes curves are lower and flatter. On CDC the curves climb gently from the Mode baseline near 27 to between 28 and 33, with PrivateGSD highest under Naive Bayes.}
  \label{fig:eps_curves_perattack_qibehavioral}
\end{figure*}

 Figure~\ref{fig:eps_curves_perattack} breaks the $\text{QI}_{\text{demo}}$ curves of Figure~\ref{fig:eps_curves} out by attack; Figure~\ref{fig:eps_curves_otherqi} repeats Figure~\ref{fig:eps_curves} for the two further QIs, and Figures~\ref{fig:eps_curves_perattack_qilarge} and~\ref{fig:eps_curves_perattack_qibehavioral} break those out by attack. Every view keeps the shape of Section~\ref{sec:whichgenerator}: a steep climb to $\varepsilon{\approx}10$, then little movement. Three departures need explaining.

\paragraph{The curves that keep climbing.} Past $\varepsilon{=}10$ the curves that still rise are Naive Bayes curves: on AIM (Adult $20.8\!\to\!25.0$ by $\varepsilon{=}100$; CDC $46.6\!\to\!50.2$ by $\varepsilon{=}1000$) and on Adult's PrivateGSD ($18.9\!\to\!24.6$), while Random Forest on the same releases moves by at most a point. This does not contradict the plateau. The cap at $\varepsilon{\approx}10$ is on the joint structure a release can represent, which Random Forest and CoBP-RA exploit; Naive Bayes reads only the pairwise association of each hidden feature with each QI feature, so it keeps gaining while a mechanism keeps sharpening its pairwise marginals. PrivateGSD and PrivBayes do: from $\varepsilon{=}1$ to $1000$ their pairwise TVD (lower is more faithful) falls ${\approx}20\times$ and ${\approx}8\times$, against $3\times$ for PrivSyn and under $2\times$ for MST and MWEM-PGM (Table~\ref{tab:quality_overview}). AIM gets there differently: as $\varepsilon$ grows, its selection rule stops penalizing large marginals, so it measures ever larger ones, and its runtime grows with them. PrivBayes is the one mechanism that is unstable rather than flat at high budgets. Its greedy structure search learns a different network from each training sample, and at $\varepsilon{=}300$ and $1000$ one sample each yields a much less reconstructable release, widening its $95\%$ interval to $\pm5.4$~pp at $\varepsilon{=}300$, against under $\pm0.5$~pp for MST.

\paragraph{Why the rankings differ between datasets.} On Adult, MST is the most exposed DP generator from $\varepsilon{=}3$ up, and CoBP-RA is the strongest attack against every DP generator at $\varepsilon{=}10$. On CDC, PrivateGSD and AIM are the most exposed ($48.0$ and $48.1$ at $\varepsilon{=}1000$, against MST's $45.0$), and Naive Bayes leads against five of the six generators, at $\varepsilon{=}10$ and at $\varepsilon{=}1000$. Both flips follow the data, not the sample size (summarized below): on Adult at 1k rows CoBP-RA still leads against every DP generator from $\varepsilon{=}10$ up, and on CDC at 100k rows Naive Bayes still leads against every synthetic release (Table~\ref{tab:cdc_100k}). Adult has one dominant, deterministic link, \texttt{education-num} as a recoding of the known \texttt{education} (Appendix~\ref{sec:feature_eps_appendix}); MST's spanning tree always keeps that edge, and this one feature accounts for essentially all of MST's lead over PrivateGSD. CDC's signal is instead spread thinly across many weak dependencies of binary features on the QI. A tree keeps few of them, so the mechanisms with the broadest coverage leak most, and Naive Bayes, which adds up many small pairwise signals, is the attack that fits.

\begin{center}
\small
\begin{tabular}{@{}lcc@{}}
  \toprule
  Strongest attack & 1k rows & larger \\
  \midrule
  Adult & CoBP-RA & CoBP-RA (10k)  \\
  CDC Diabetes & Naive Bayes & Naive Bayes (100k) \\
  \bottomrule
\end{tabular}
\end{center}
Along a row (same data, more rows) the leader stays the same; down the 1k column (same size, other data) it changes (--- = not evaluated).

\paragraph{PrivSyn: more private, or less coherent?} On Adult, PrivSyn has the lowest aggregate exposure of any DP generator, at every budget and on all three QIs (Figures~\ref{fig:eps_curves} and~\ref{fig:eps_curves_otherqi}), and its releases are nonetheless faithful: it has the best column shapes of any DP generator at every budget we sweep, and at $\varepsilon{=}1000$ only AIM and PrivateGSD beat it on utility, pairwise fidelity, and correlation error (Table~\ref{tab:quality_overview}). Yet a random forest recovers less from it than from any other DP release ($19.5$, against MST's $25.4$).

Every fidelity metric in that table audits one- or two-way statistics, whereas reconstruction asks whether a row's hidden values agree with all of its QI values at once. PrivSyn never fits a joint model: it edits records drawn independently per column until their pairwise counts match. Its one- and two-way statistics are therefore as faithful as the other DP releases', and we conjecture that agreement among three or more columns of a row is what it lacks. Two observations fit this reading. PrivateGSD, also model-free but fit an order of magnitude more tightly to the pairwise marginals, leaks more, which suggests that a tighter pairwise fit brings some higher-order agreement with it. And PrivSyn's low exposure disappears on CDC, where the leading attack, Naive Bayes, needs only pairwise structure. PrivSyn's lower exposure is thus not a stronger guarantee (the formal guarantee is identical at matched $\varepsilon$) but a release whose rows agree less beyond pairs. For data consumed as low-order statistics or standard prediction tasks that is an attractive trade; for analyses that need interactions beyond pairs it is a cost that one- and two-way fidelity metrics, ours included, do not reveal.

\section{Statistical Reproducibility}\label{sec:statistics}

 Every $R_{adv}$ we report is a mean over 5 \emph{disjoint} training samples drawn from the same source dataset. The samples share no records, so the five scores behind a cell are independent replicates of the whole pipeline (sampling, synthesis, attack). Where we give a $95\%$ confidence interval it is $\bar{x} \pm t_{0.975,4}\,s/\sqrt{5}$, the usual interval for a mean of five values, with $t_{0.975,4}{\approx}2.78$ in place of the normal $1.96$ because the spread $s$ is itself estimated from five values. Two claims in the main text rest on tests. \emph{Memorization} (Table~\ref{tab:memorization_and_ds_risk}): a gap $\Delta$ counts as memorization only if a one-sample $t$-test over the five samples rejects $\Delta{=}0$; Table~\ref{tab:memo_stats} gives every row's interval and $p$-value, and only the three gaps of $13$~pp or more clear it. \emph{The privacy-budget plateau} (Section~\ref{sec:whichgenerator}): Table~\ref{tab:eps_stats} compares two budgets of one mechanism with a paired $t$-test, pairing each training sample's release at one budget with its release at the other, which removes the variation between samples.

 Table~\ref{tab:eps_stats} reads as follows. From $\varepsilon{=}0.1$ to $10$ every mechanism gains $5.7$--$11.4$~pp on every QI, and every one of these gains is significant ($p{\le}0.001$). From $\varepsilon{=}10$ to $1000$ the same mechanisms move by $-0.9$ to $+3.4$~pp. For PrivBayes, PrivSyn, and MWEM-PGM that change is indistinguishable from zero on all three QIs; MST's is detectable but at most $1.2$~pp. PrivateGSD keeps climbing ($+2.2$ to $+3.4$~pp), as does AIM in the sweep itself, almost entirely through Naive Bayes (Appendix~\ref{sec:eps_appendix}), so we claim that risk flattens past $\varepsilon{\approx}10$, not that it stops.

One caveat bounds how finely these numbers should be read. Re-running the full scoring pipeline on identical inputs moves a cell by ${\approx}1\%$ of its value (MST $\varepsilon{=}1000$ under RF: $0.00289$ vs.\ $0.00285$), and small subgroups move considerably more: race \emph{Other} ($n{\approx}70$) moved from $0.19$ to $0.16$ across two runs of the same configuration. Differences below roughly half a percentage point, and any subgroup claim resting on a few tens of records, should be treated as within the resolution of the measurement.

\FloatBarrier

\end{document}